D I S S E R T A T I O N

# CARBON MONOXIDE ADSORBTION AND CARBON MONOXIDE OXIDATION STUDIES ON THIOLATE SUPPORTED GOLD NANOCLUSTERS DISPENSED ON METALLIC OXIDES

ausgeführt zum Zwecke der Erlangung des akademischen Grades einer Doktorin der Naturwissenschaften unter der Leitung von

Univ. Prof. Mag. rer. nat. Dr. rer. nat. Guenther RUPPRECHTER
und
Prof. Ass. Dr. rer. nat. Karin FOETTINGER

E 165
Institut für Materialchemie

eingereicht an der Technischen Universität Wien
Technischen Chemie

von

Mag. rer. nat. Dilek DEMIR

0600046
Wurmbrand – Stuppach Siedlung 9
A-7111 Parndorf

Wien, am 06. 03. 2017

# CARBON MONOXIDE ADSORBTION AND CARBON MONOXIDE OXIDATION STUDIES ON THIOLATE SUPPORTED GOLD NANOCLUSTERS DISPENSED ON METALLIC OXIDES

# ABSTRACT

Thiolate-protected gold nanoclusters $Au_n(SR)_m$ (n = 38, 40; m = 24) represent atomically well-defined active centers for catalysis with resolved crystal structures, enabling the creation of a homogeneous and defined surface.

In this work, $Au_n(SR)_m$ (n = 38, 40; m = 24) supported by $Al_2O_3$ and $CeO_2$ are studied and characterized using techniques such as Thermogravimetric Analysis, Fourier Transform Infrared Spectroscopy and Kinetic Tests.

The results from these measurements answered open questions related to the effects of pretreatment conditions, support material and the number of gold atoms.

Thermogravimetric Analysis on bare gold nanoclusters showed that the thermal removal in oxidative and inert atmosphere is a three-step procedure for $Au_{40}(SR)_{24}$. For $CeO_2$ supported gold nanoclusters ($Au_n(SR)_m$ (n = 38, 40; m = 24)) showed gold nanocluster characteristic changes, whereas for $Al_2O_3$ supported gold nanoclusters this was not observed.

Through the theoretical calculations mass loss of the gold nanocluster and mass loss of the gold nanocluster, the Thermogravimetric Analysis gave information that at 473 K thiolates are removed.

For the Fourier-Transformed Infrared Spectroscopy and Kinetic Tests 473 K was taken as the pretreatment temperature. For the pretreatment atmosphere either oxidative, inert or reductive atmosphere was taken pretreatment and the pretreatment temperature was set to 473 K, in order to investigate and validate observations from literature.

Spectroscopic studies were performed for $Au_{40}(SR)_{24}/CeO_2$ after oxidative pretreatment at 473 K reveal that at lower temperatures $Au^{\delta-}$ and $Ce^{3+}$ sites are active to carbon monoxide adsorbed at room temperature on the catalysts surface. At higher temperatures $Au^{\delta+}$ is the most active sites for carbon monoxide adsorption on the gold nanocluster catalysts surface. For $Au_{38}(SR)_{24}/CeO_2$ after oxidative pretreatment at 473 K only $Au^{\delta-}$ and $Ce^{3+}$ sites are active to carbon monoxide adsorbed on the catalysts surface.

For $Au_{40}(SR)_{24}/Al_2O_3$ and $Au_{38}(SR)_{24}/Al_2O_3$ no CO adsorption are observed.

Kinetics prove that after the partial removal of thiolate ligands in oxidative pretreatment atmosphere at 473 K pretreatment temperature, $Au_{40}(SR)_{24}/CeO_2$ is more efficient and faster in the conversion of carbon monoxide into carbon dioxide, even at moderate temperatures, e.g. 300 K to 425 K, compared to $Au_{38}(SR)_{24}/CeO_2$.

By contrast, gold nanoclusters supported on $Al_2O_3$ start to convert from 425 K on.

The kinetic results are compared to gold nanoparticles Au/$Al_2O_3$ and Au/$CeO_2$. The correlation of characterization via FT-IR spectroscopy with kinetics studies gives insight into the gold clusters' activity and tests their utilization to bridge the gap between organometallic complex-based homogenous catalysis and nanoparticle based heterogeneous catalysis as well as the gap between the bulk crystal structure model catalysis and real-world catalysis.

**CONTENTS**

# 1.Chapter: INTRODUCTION

The primary objective of medieval alchemists was to transmute various substances into gold. To date, this objective has never been achieved. Gold, the “noblest” of all metals, has a low affinity with any reactive gas such as hydrogen ($H_2$) or oxygen ($O_2$). The obstacle here is the repulsion between the orbitals of the adsorbate and the filled shell, which diminishes the reaction probability [1]–[59].

In Haruta’s pioneering work, it was documented that gold nanoparticles displayed a better catalytic activity for CO adsorption and CO oxidation reactions on the surface, even at low temperatures, compared to existing commercial catalysts [60]–[62].

Haruta et al. [62] showed that when gold is deposited on a metal oxide supported by a co-precipitation method, the metal exhibits high catalytic activity for CO oxidation at low temperatures. Relevant factors for catalytic activity are the interaction between the gold nanoparticle catalysts as well as the support material, the interface between catalyst and oxide surface and the size of the gold catalysts.

The catalytic activity of gold depends on particle size, with reactivity increasing as the particle’s diameter decreases. The catalytic reaction mechanism was studied theoretically, and it was shown that while isolated gold atoms cannot activate $O_2$, small gold clusters are excellent catalysts for $O_2$ activation [1], [63].

## 1.1 Cluster Science

Research on metal clusters of Manganese was triggered by the idea of adding a third dimension to the periodic table. The third dimension implies either the cluster size or the number of atoms in clusters [64].

In the late 1970s, groups led by Schumacher [65] and Knight [66] focused on the accounting of magic numbers (n = 2, 8, 18, 20, 34, 40, 58, 92, 138, ……). These numbers were often observed as mass spectroscopy (MS) peaks of unusually high intensity.

The appropriate Aufbau principle [[64]] of delocalized “superatomic orbitals” of metal clusters is $|1S^2| 1P^6 | 1D^{10} \quad 2S^2 | 1F^{14} \quad 2P^6 | 1G^{18} \quad 2D^{10} \quad 3S^2 |$……, wherein S – P – D – F – G – denotes the angular momentum quantum number, and 1 – 2 – 3 – corresponds to the radial nodes.

As mentioned above, exceptional stability is present for a total count of *2, 8, 18, 20, ......* for ligand-protected gold clusters as predicted with the spherical model [4], [67]–[109]. The magic cluster size corresponds to the filling of electron shells of these energy levels.
In [43], Walther and colleagues employed the counting equation for free clusters to determine the total number of electrons ne for thiolate-protected gold nanoclusters

$$n_e = N\ V^{\#}\,A - M - z \qquad \textbf{(1)}$$

In equation ***(1)*** $n_e$ is the total number of electrons, N corresponds to the number of metal atoms, A is Avogadro's number, $V^{\#}$ is the atomic valence, M is the number of electrons withdrawn by the ligands and z is the overall charge of the cluster complex. The cluster size can be classified into one of the three following terms: "very small clusters" consist of 2–20 atoms, "small clusters" consist of 20–500 atoms and "large clusters" contain 500–$10^7$ atoms. This classification system was introduced by the groups led by Knight and Schumacher [110].
A comparison of the packing electronic structure, magnetism, electrochemical and charge transport properties shows that subnanometer to 2 nm gold nanoparticles are very different from larger Au nanoparticles (>2 nm to 100 nm) [111].
The clusters described above are protected by ligands such as thiolates, amines and phosphines. The ligands form a monolayer on the cluster surface (see Figure 1), and accordingly are referred to as monolayer–protected clusters [112].

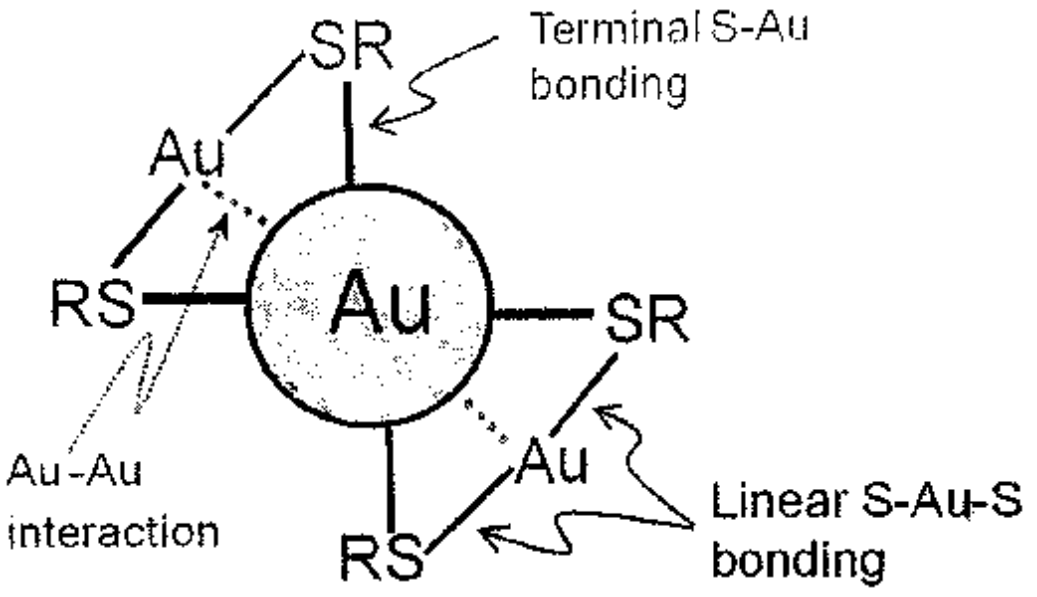

*Figure 1 Simple sketch of a thiolate -protected gold nanoclusters ($Au_n(SR)_m$). Taken from [84].*

## 1.2 THIOLATE LIGANDS

Ligands [113] have two notable features (see Figure 2). One is their tendency to prevent clusters from aggregation in a solvent or in the solid phase. Second, the ligand influences the physical and chemical properties of the metal cluster, prominently affecting both chirality and photoluminescence and optical properties. Ligand-protected Au clusters are atomically precise and written as $Au_n(SR)_m$, where n corresponds to the number of gold atoms and m to the number of thiolate ligands SR.

SR [112], [114]–[119] corresponds to thiolate ligands and R represents the alkyl group. To be precise, in this work SR refers to the ligand $SCH_2$ $CH_2$ Ph. Possible bending patterns for thiolate ligands are shown in Figure 3 [120].

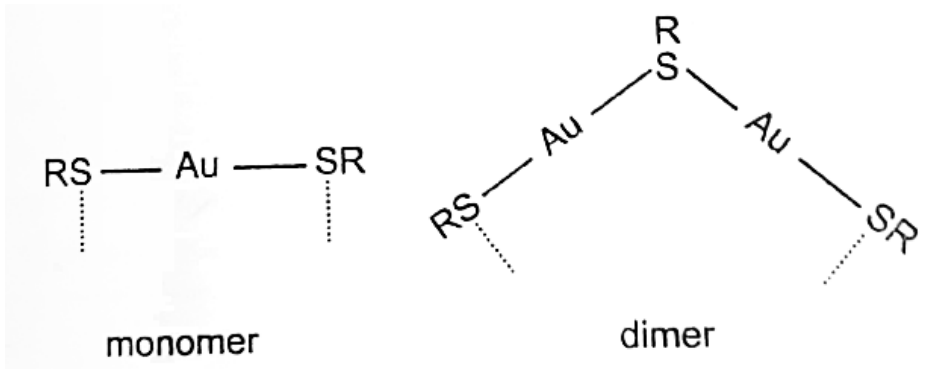

*Figure 2 Staple monomer (RS-Au-SR) motif (left) and dimer (RS-Au-SR-Au-SR) motif. Taken from [84].*

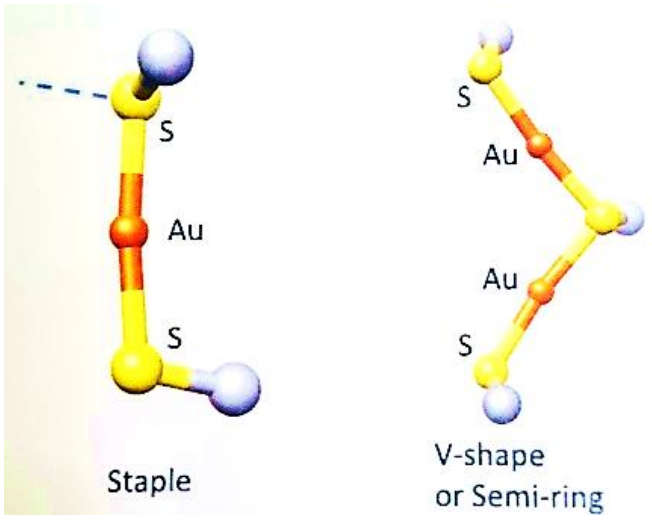

*Figure 3 Possible binding patterns between gold and thiolate ligands. Taken from [120].*

The gold nanocluster catalysts (Figure 1) are protected by ligands (e.g. thiols, phosphines, amines etc.), which typically are located as a monolayer on the nanocluster surface as monomer and dimer motifs, as shown in Figure 3 [112], [121]–[123].

The thiolate ligands protect the adsorption of any other atoms with the core gold atoms and stabilize the gold cluster system. The presence of the thiolate ligands makes it increasingly difficult to understand, if the ligands desorb and more gold atoms become accessible, after rising the surrounding temperature higher than 300K.

Therefore, ligand construction of gold nanocluster catalysts must be understood and controlled better. This all plays a role to understand the catalytic activity/performance, depending on the size.

## 1.3 LITERATURE REVIEW

Structure prediction is currently one of the most challenging issues in the field of cluster science. Numerous groups and researchers in the field of cluster science are working on the development of a suitable theoretical method using computational calculations.

Tsukuda, et al. [84], [124], relying on DFT-powered basin hopping, suggested that the very unique structure of $Au_{40}(SR)_{24}$ (Figure 4) is a twisted pyramid with a missing corner. This idea was developed by Tsukuda and colleagues, particularly regarding images of triangular shaped clusters developed through transmission electron microscopy.
According to models, two atoms differentiate the structure of the $Au_{38}(SR)_{24}$ from $Au_{40}(SR)_{24}$, but both possess the same number of RS-groups. In this review, two structure models for $Au_{40}(SR)_{24}$ were featured (see table from [84]).

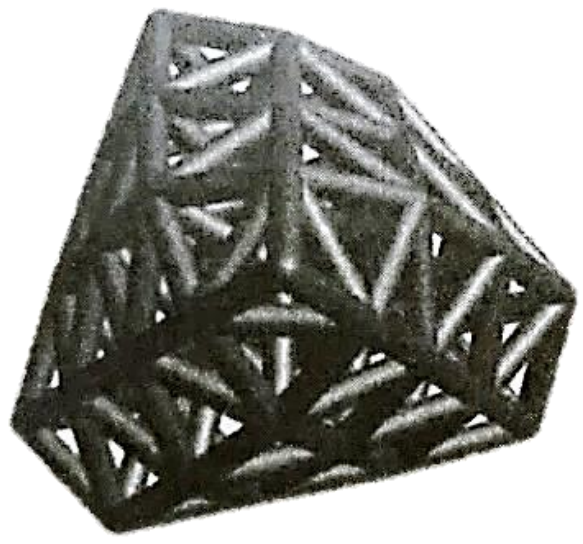

*Figure 4 A twisted golbal minimum of $Au_{40}(SR)_{24}$. Another model for the $Au_{40}(SR)_{24}$ gold nanocluster, this was found by DFT-powered basin hopping. Taken from [84].*

*Table 1 Comparison of the two different systems presented in [84] and [104].*

| Model | Malola[104] | Jiang[84] |
|---|---|---|
| core | $Au_{26}$ | $Au_{25}$ |
| Number of RS-Au-SR | 6 | 3 |
| Number of RS-Au-SR-Au-SR | 4 | 6 |
| DFT-PBE (eV)[a] | 0 | -0.09 |
| DFT-TPSS (eV)[a] | 0 | -0.53 |
| HOMO-LUMO Gap (eV)[a, b] | 0.76 | 0.86 |

[a] From TURBOMOLE V6.0 with def2-TZVP basis sets.

[b] From DFT-PBE.

One is predicted by Malola et. al. [104] as a cluster with $Au_{26}$, six monomers and four dimer staples (Figure 5, table from [84]). The second is from De-Jin, in which $Au_{25}$ gold atoms are stabilized via three monomer staples and six dimer staples (Figure 2 and Figure 3). In the table below, the detailed numbers are given (adapted from [84]).

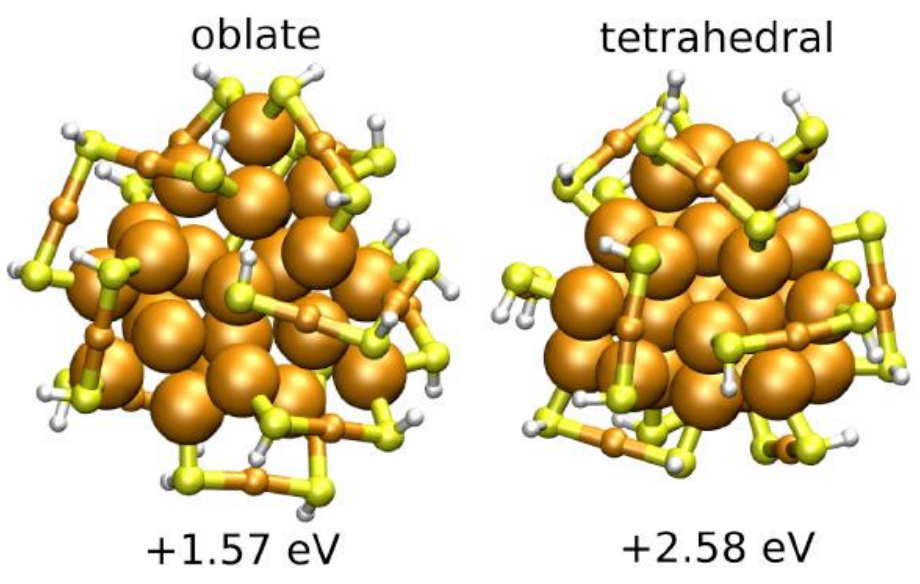

*Figure 5 Additional structures for $Au_{40}(SR)_{24}$ with alternative core shapes. Models are taken from [104] – Supporting Information.*

The latest structure model published recently is [125]. In this publication, the authors presented structure models with experimental data by X-ray crystallography on $Au_{40}(SR)_{24}$ and $Au_{52}(SR)_{32}$ gold nanoclusters (Figure 6). They found that the anatomy of the structures reveals a $Au_{25}$ atom kernel resembling a snowflake and nine surface protecting staples [67], [126], [127]. The kernel is segregated into eight tetrahedral $Au_{40}$ units, evidenced by Au-Au bond length differences. Two of the tetrahedral $Au_{40}$ units form a bi-tetrahedral antiprism. The remaining six tetrahedra from a Kekulé-like $Au_{40}$ structure with a $D_{3d}$ symmetry, which is due to the rotative arrangement of the staple motifs (Figure 6).

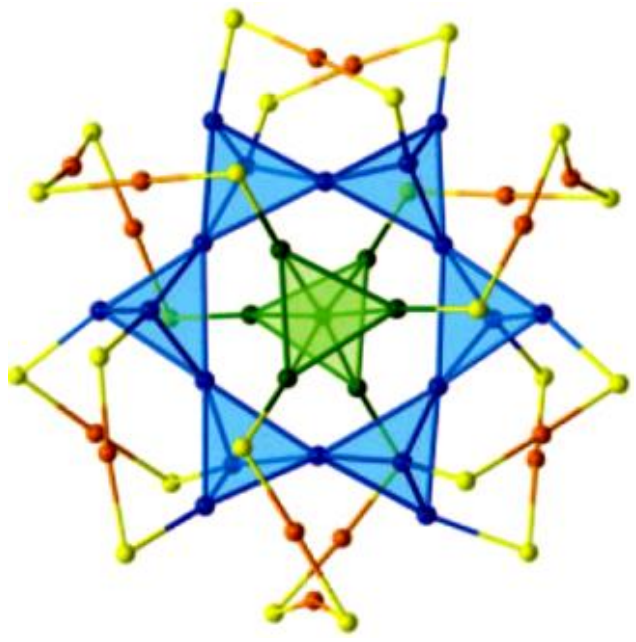

*Figure 6* *$Au_{40}(SR)_{24}$ proposed model. Taken from* [125].

## 1.4 THESIS OUTLINE

The investigation of gold nanocluster catalysts is at the heart of the current work. The focus is on $Au_{40}(SR)_{24}$ tested under high temperature CO oxidation and $Au_{38}(SR)_{24}$ tested under low temperature CO oxidation. These two new catalysts are supported by $Al_2O_3$ and $CeO_2$, which yield different results. As an introduction to the actual scientific work, an in-depth literature review on existing relevant research is presented.

The tools used for the current investigations were Thermogravimetric Analysis, FT-Infrared Spectroscopy and kinetic tests conducted in a fixed bed flow reactor.

The following chapter "Materials and Methods" clearly presents the experimental tools and the employed pretreatment procedure. It also gives information about the used samples and the reference catalysts as well as data treatment.

In "Results", the measurement outcomes of TGA measurement of $Au_n(SR)_m$, (for which n equals 38 and 40 and m equals 24) are shown.

In the first subsection, results with gold nanocluster catalysts supported by $Al_2O_3$ and $CeO_2$ are presented, and results of the available support material, $CeO_2$, are plotted.

The next subsection includes the results of FT-Infrared Spectroscopy on the reference catalysts support material $CeO_2$ as well as the results for $Au_{38}(SR)_{24}$ and $Au_{40}(SR)_{24}$ supported by $Al_2O_3$ and $CeO_2$.

The last part of "Results" combines the CO oxidation ($CO + O_2 \rightarrow CO_2$) kinetic tests of

2 wt.% $Au_{38}(SR)_{24}/CeO_2$, 1 wt. % $Au_{40}(SR)_{24}/CeO_2$, 2 wt. % $Au_{38}(SR)_{24}/ Al_2O_3$ and 1 wt. % $Au_{40}(SR)_{24}/Al_2O_3$ after pretreatment in different atmospheres, such as reductive, inert and oxygen after different pretreatment procedures at 473 K.

The gold nanocluster catalysts are compared to gold nanoparticle catalysts $Au/M_xO_y$ supported by $Al_2O_3$, $CeO_2$.

The final subsection of the "Results" chapter demonstrates the values for calculated kinetic parameters, such as activation energy and reaction rate.

The results are then discussed in detail in the chapter "Discussion" and linked to previous results in the field of gold nanocluster catalysts.

The chapter "Conclusions and Summary" summarizes the main breakthroughs of this work.

The work closes with the chapter "Outlook", in which unexplored and interesting questions in the field of gold nanocluster catalysis are presented.

# 2.Chapter: METHODS AND MATERIALS

In this chapter, the technical equipment, pretreatment procedures, sample properties, experimental settings and details about the data analysis. This description is provided for the Thermogravimetric Analysis, the Fourier – Transformation Infrared Spectroscopy and kinetic tests.

## 3.1 THERMOGRAVIMETRIC ANALYSIS

Thermogravimetric analysis is an essential laboratory tool for the characterization of solid samples. This analytical tool detects changes in mass and measures the chemical or physical processes that occur upon heating a sample [128].
To be precise, information on the quantification of loss of water, loss of solvent, loss of plasticizer, decarboxylation, thermal stability, pyrolysis, oxidation, decomposition, the amount of metallic catalytic residue remaining on any support material, and the determination of the purity of any material can be achieved with the help of this analysis.

The first step was to discover at which temperatures the different molecules are desorbed and the second was to prove the thermal stability of the unsupported and supported gold nanocluster catalysis.

### 3.1.1 EXPERIMENTAL SETUP

TGA (Figure 7) was performed on a STA 449 F3 Jupiter (NETZSCH) thermal analyzer within an aluminium oxside pan (Figure 9).

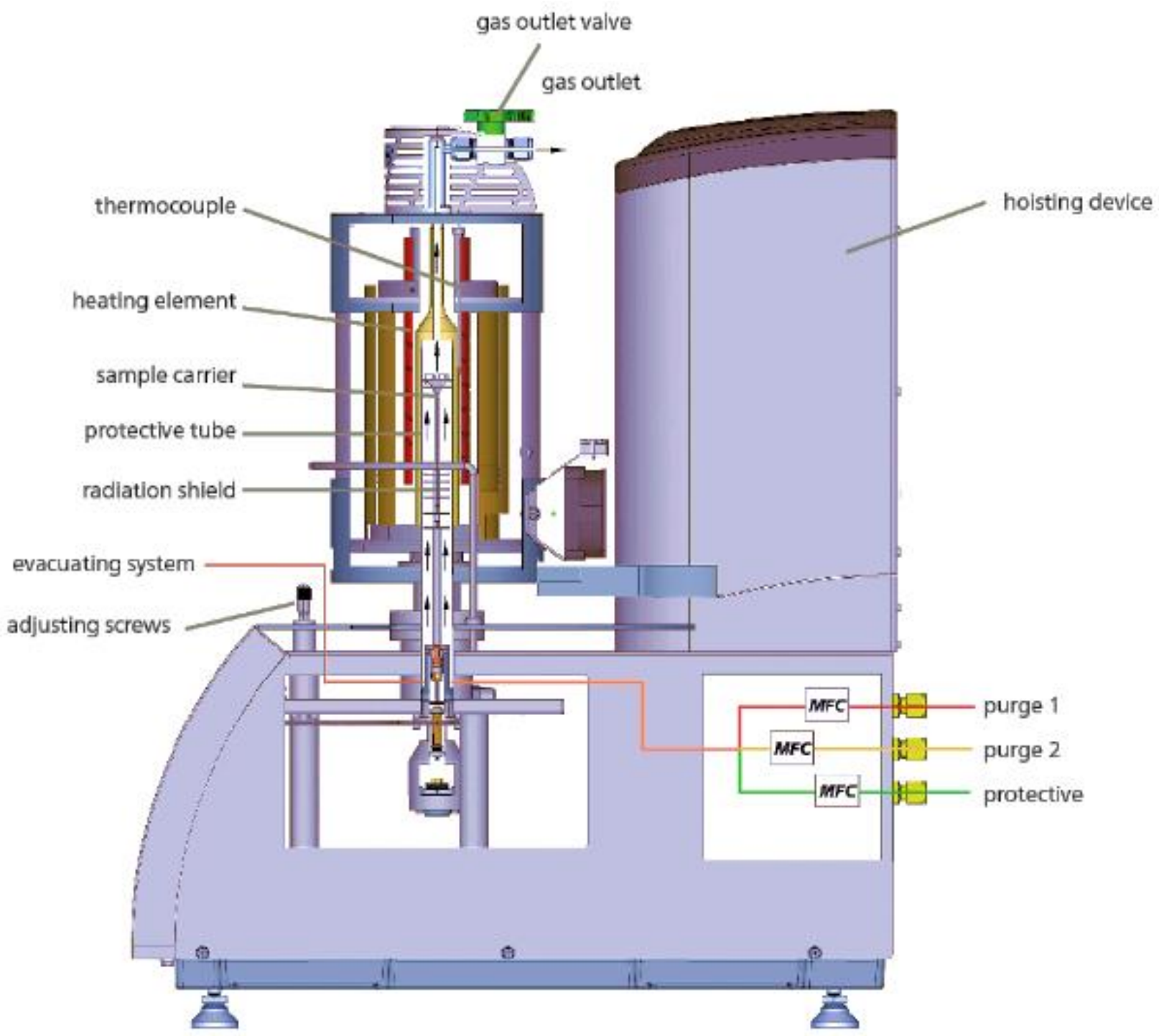

*Figure 7 A side picture of the STA 449 F3 Jupiter (NETZSCH) thermal analyzer with detailed description of the different parts. Taken from* [129].

The TGA apparatus consisted of a sample pan supported by a precision balance. The pan resided in a furnace and was heated or cooled during the experiment. The mass loss of the sample in percentage was monitored during the experiment. A sample purge gas controlled the sample environment. This gas may be inert or a reactive gas that flows over the sample and exits through an exhaust. [129].

For TGA measurement, samples were loaded into aluminum oxide pans (as shown in Figure 9) in order to prevent any reactions between the samples (such as unsupported\supported gold nanoclusters catalysts) and crucible in the course of the TGA analysis. The sample was installed after reaching room temperature to avoid prior desorption processes.

### 3.1.1 THERMOGRAVIMETRIC ANALYSIS MEASUREMENTS DOSING PROCEDURE

TGA (Figure 7) thermal analysis was taken under two different atmospheres (inert: 30mL/min $N_2$ and oxidative: 20 mL/min $O_2$). The temperature was increased from 300 K to 970 K at a ramp rate of 5 K/min. Shown in Figure 8.

For the $Au_{40}(SR)_{24}$ catalysts, the ramp rate was set at 10 K/min because this measurement was done already by Zhang et. al. [130] and the aim was to reproduce the TGA result.

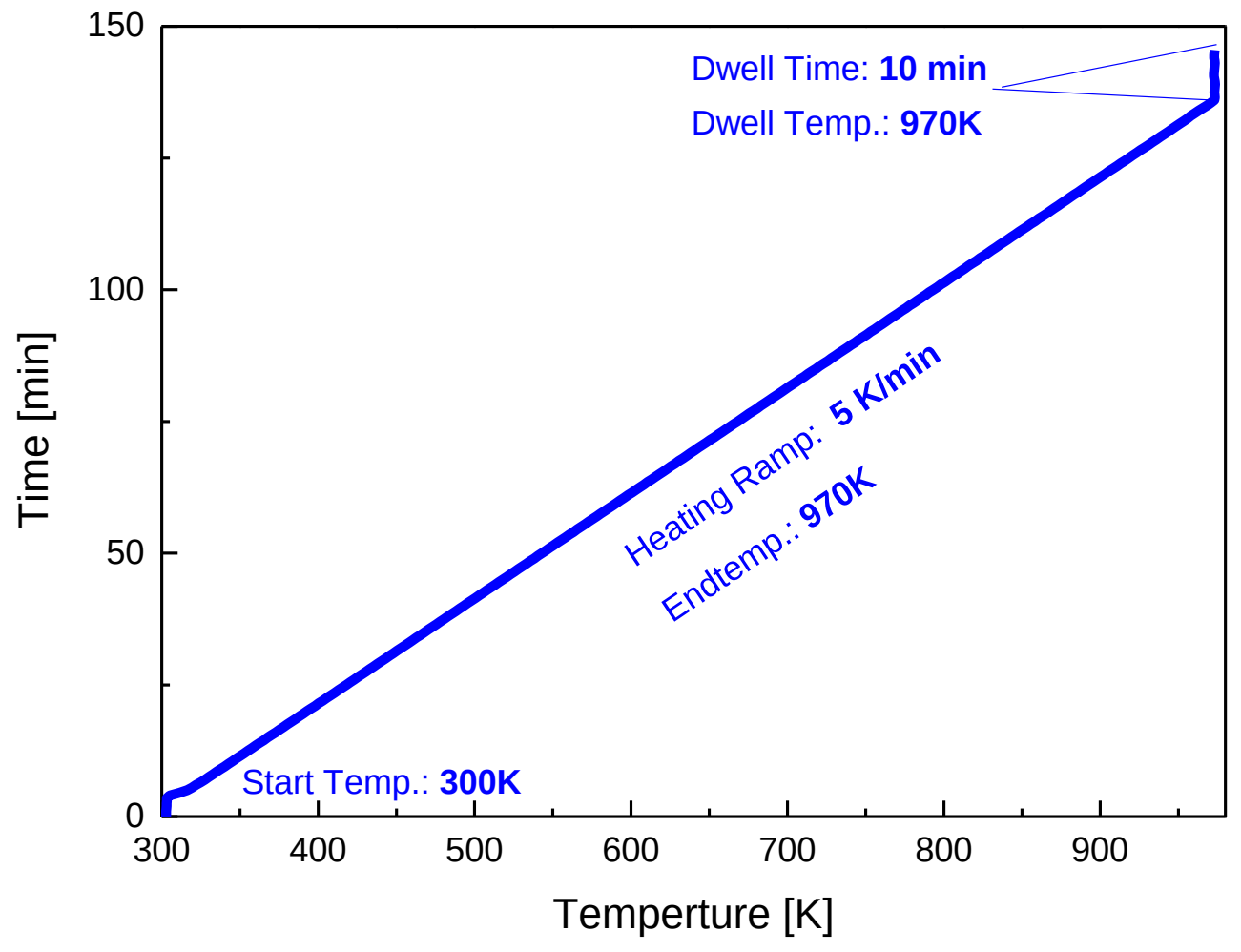

*Figure 8 Heating ramp of the TGA measurement showing temperature programmed desorption (TPD) setting of the TGA analysis, as was written out for every step.*

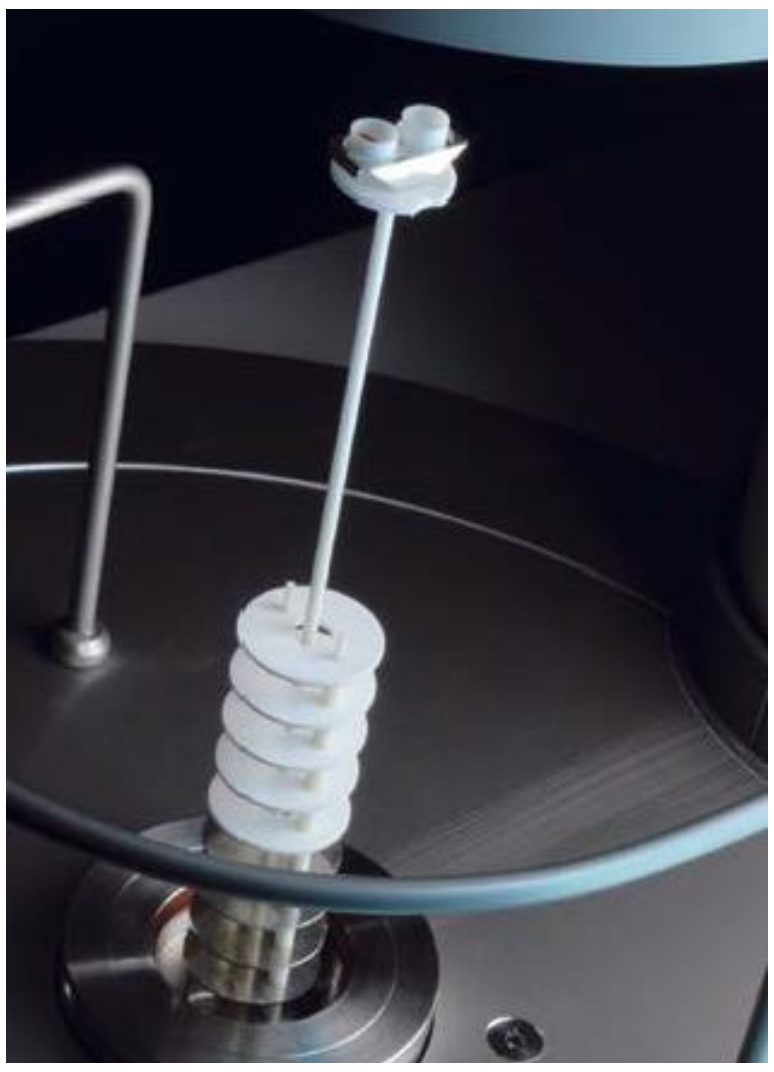

*Figure 9 Side view of the aluminum oxide pans positioned on the oven. Taken from* [129].

### 3.1.2 SAMPLE PREPARATION

The samples analyzed in both inert and oxidative atmospheres are summarized in Table 2. The measurements are shown and compared to previous work in the chapter "Results" (subsection "TGA Measurements").

Samples 2)–6) listed in Table 2 in the subsection "Measured Samples" were available as powders. The sample (10 mg) was loaded into aluminum oxide pans without any further treatment of the sample. The weight was determined by an external balance and a second time with the in-build balance inside the TGA apparatus.

$Au_{40}(SR)_{24}$ (sample 1) was dissolved in Dichloromethane and dropped into the crucible. The crucible was then left to evaporate the dichloromethane for 30–60 minutes at room temperature under atmospheric pressure in air. The TGA measurement began after the evaporation.

As an indicator, all moisture inside the crucible was removed.

*Table 2 TGA tested samples overview, where ✓ (Green Tick) indicates that the sample written on the left of the table was tested and* **X** *(cross) is placed into the line to show that the sample written on the left side of the table was not tested.*

| | ***Sample*** | ***Purge Gas*** | |
|---|---|---|---|
| | | oxidative | inert |
| *1* | **$Au_{40}(SR)_{24}$** | ✓ | ✓ |
| *2* | **$CeO_2$** | ✓ | ✓ |
| *3* | **1 wt. % $Au_{40}(SR)_{24}/CeO_2$** | ✓ | ✓ |
| *4* | **1 wt. % $Au_{40}(SR)_{24}/Al_2O_3$** | ✓ | ✓ |
| *5* | **2 wt. % $Au_{38}(SR)_{24}/CeO_2$** | ✓ | **X** |
| *6* | **2 wt. % $Au_{38}(SR)_{24}/Al_2O_3$** | ✓ | **X** |

### 3.1.3 DATA ANALYSIS

The TGA curve is displayed (Figure 10) on the abscissa (X-axis) as time or temperature. The ordinate (Y-axis) can be presented as sample weight (mg) or as sample weight percent (%).

For bare gold nanocluster $Au_{40}(SR)_{24}$ results, the ordinate is given in sample weight (mg).
For supported gold nanoclusters $Au_{40}(SR)_{24}/CeO_2$, $Au_{38}(SR)_{24}/CeO_2$, $Au_{40}(SR)_{24}/Al_2O_3$ and $Au_{38}(SR)_{24}/Al_2O_3$, the ordinate is always in weight percent (%).
In general, the value for the ordinate is kept the same but during the saving process this information got lost.

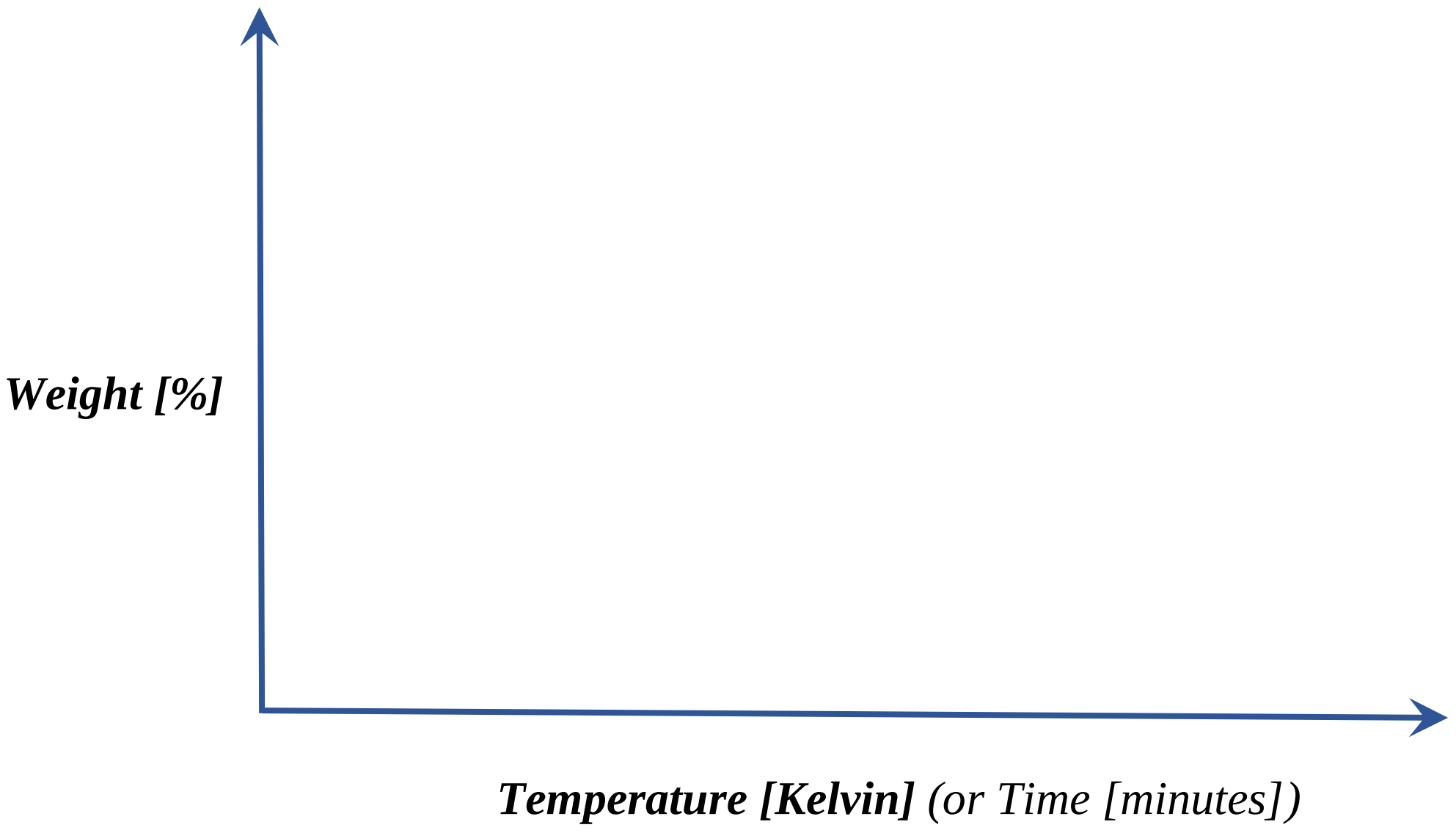

*Figure 10 TGA Thermal Curve. The example graph shows the way in which the TGA values are displayed in the following section (chapter "Results").*

## 3.2 FOURIER-TRANSFORM INFRARED SPECTROSCOPY

Pioneering work of adsorption on metallic support materials were done by Eischens et. al. [131] in the 1950s. In the mid-IR range (from 50 µm to 2.5 µm), absorbed photons were measured due to the excitation of intramolecular vibration. In catalysis, the FT-IR tool is employed primarily to answer questions related to the absorbed species and (in-)active sites to the adsorption on the catalyst's surface. The vibrational energy of molecules is given in discrete levels. Absorption photons in the mid-IR range (from 50 µm to 2.5 µm) lead to transitions between different vibrational energy levels.

### 3.2.1 EXPERIMENTAL SETUP

Infrared spectra were recorded in transmission mode using a Brucker Vertex 70 spectrometer (Figure 11) with a globar as the radiation source and a HeNe laser as an internal wavelength standard. A mercury cadmium telluride (MCT) detector with a resolution of 4 $cm^{-1}$ was installed as part of the FT-IR spectrometer. Figure 11 shows a schematic of the operative setup.
The infrared apparatus was arranged with a vacuum cell such that samples could be implanted after pressing them into small discs. The vacuum cell was linked to a turbomolecular pump (Pfeiffer HiEco Cube), so pressures of $10^{-6}$ mbar could be achieved. A rotary vane pump was used to reach the fine vacuum (<1 mbar), to prime the turbomolecular pump. The measured background pressure at room temperature was $<10^{-5}$ mbar.

Gases were introduced into the chamber via a leak valve. This part was feeding the dosing system. The pressure in the chamber was controlled via a MKS type 626 pressure gauge for pressures above 1 mbar, by a Pfeiffer Vacuum type CMR 264 compact capacitance gauge for the fine vacuum range, and by a Balzers type IKR 260 compact cold cathode gauge for pressures in the high vacuum range.

A type K thermocouple positioned around the edges of the sample holder containing the disc monitored the sample temperature. This temperature thermocouple was connected to a Eurotherm temperature controller. This construction was useful to power the heating wire applied to the sample, following a specific reductive, oxidative and inert pretreatment arranged in the same cell. The detailed pretreatment and CO adsorption procedure can be found in the subsections “Pretreatment” and “CO Dosing Procedures”.
The adsorption experiments were carried out under static pressures at 300 K utilizing a compatible IR cell.

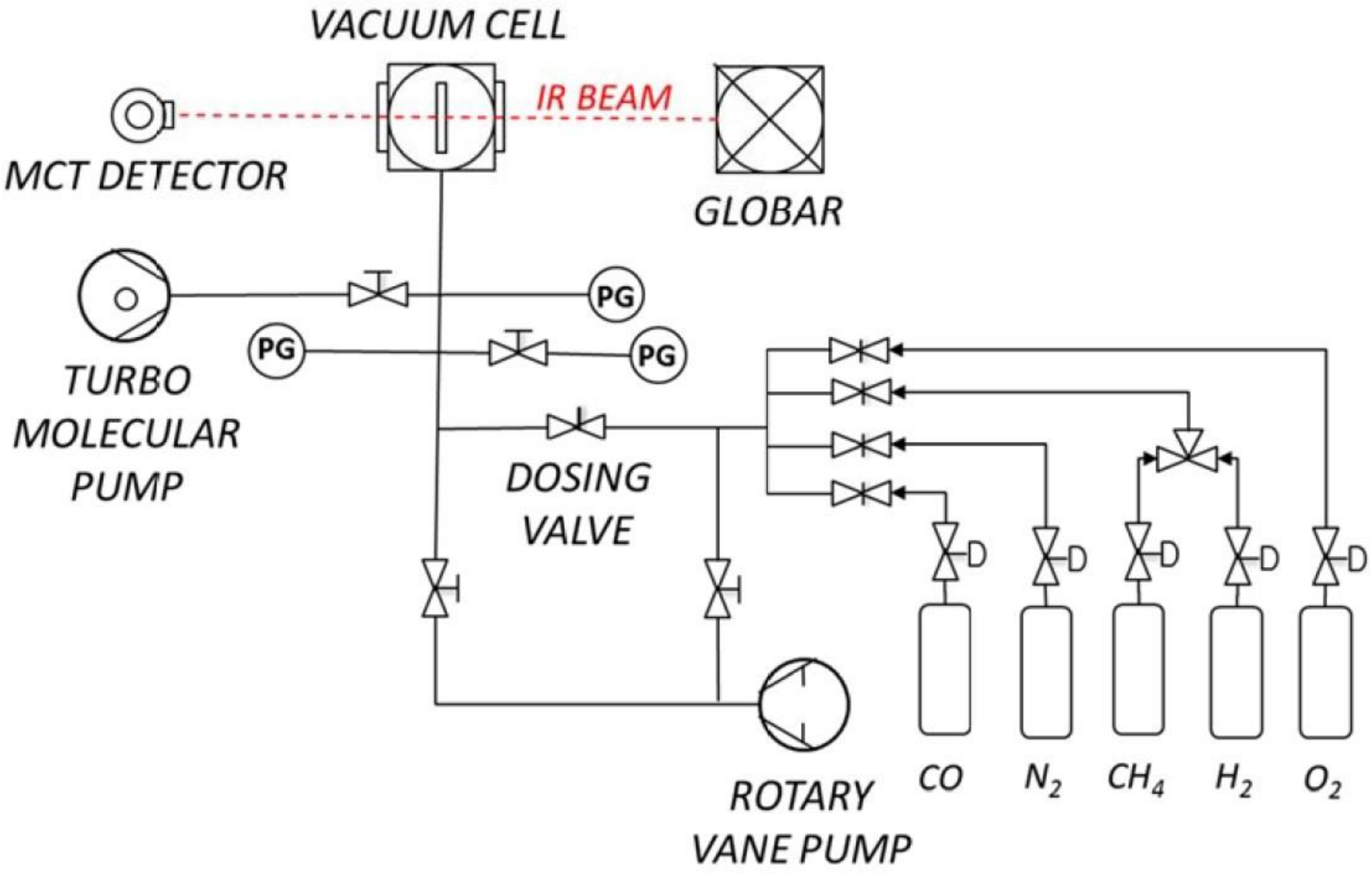

*Figure 11 Top view of the FT- IR experimental setup. Taken from* [132].

### 3.2.2 FOURIER-TRANSFORM INFRARED SPECTROSCOPY PRETREATMENT AND CO DOSING PROCEDURE

For the oxidative pretreatment, the vacuum chamber was filled with 100 mbar $O_2$, and the rest was filled with 800 mbar inert gas ($N_2$). For reductive pretreatment, the chamber was filled with 100 mbar $H_2$, and the rest with 800 mbar inert gas ($N_2$). For “calcination”,
900 mbar inter gas ($N_2$) filled the vacuum chamber.

The prescribed temperature was 573 K with a heating ramp of 5K/min starting from 300K (room temperature). At 573 K the camber was filled with the used atmosphere (oxidative, reductive or “calcination” atmosphere) and kept at 573 K for approximately one hour. After cooling the temperature down to room temperature, the CO adsorption cycle started. This Pretreatment procedure was done in order to make sure that 50% of the ligands are removed from the surface.

For the CO adsorption process the first cycle is shown in Figure 12. In the first step (=$Pre_1$ in Figure 12) the temperature increased to 425 K in oxidative, reductive or “calcination” atmosphere. In the second step, the highest temperature was kept for 10 minutes (= $Pre_2$ in Figure 12).

As next, in $Pre_3$, during the cooling phase the certain atmosphere (oxidative, reductive or “calcination” atmosphere) was removed from the vacuum chamber and the cooling phase in HV to room temperature (= $Pre_4$) was initiated.

Subsequently, the sample was cooled to room temperature, after which 5 mbar of CO were dosed stepwise into the vacuum chamber (=$Pre_5$). This atmosphere was maintained for 15 minutes.

In $Pre_6$, after the CO dosing phase, a 15-minute evacuation phase commenced.

Above and in Figure 12 the first cycle was explained in detail. The difference to the following cycles are the difference in $Pre_1$ and $Pre_2$, where the highest temperature varies by plus 50 K.

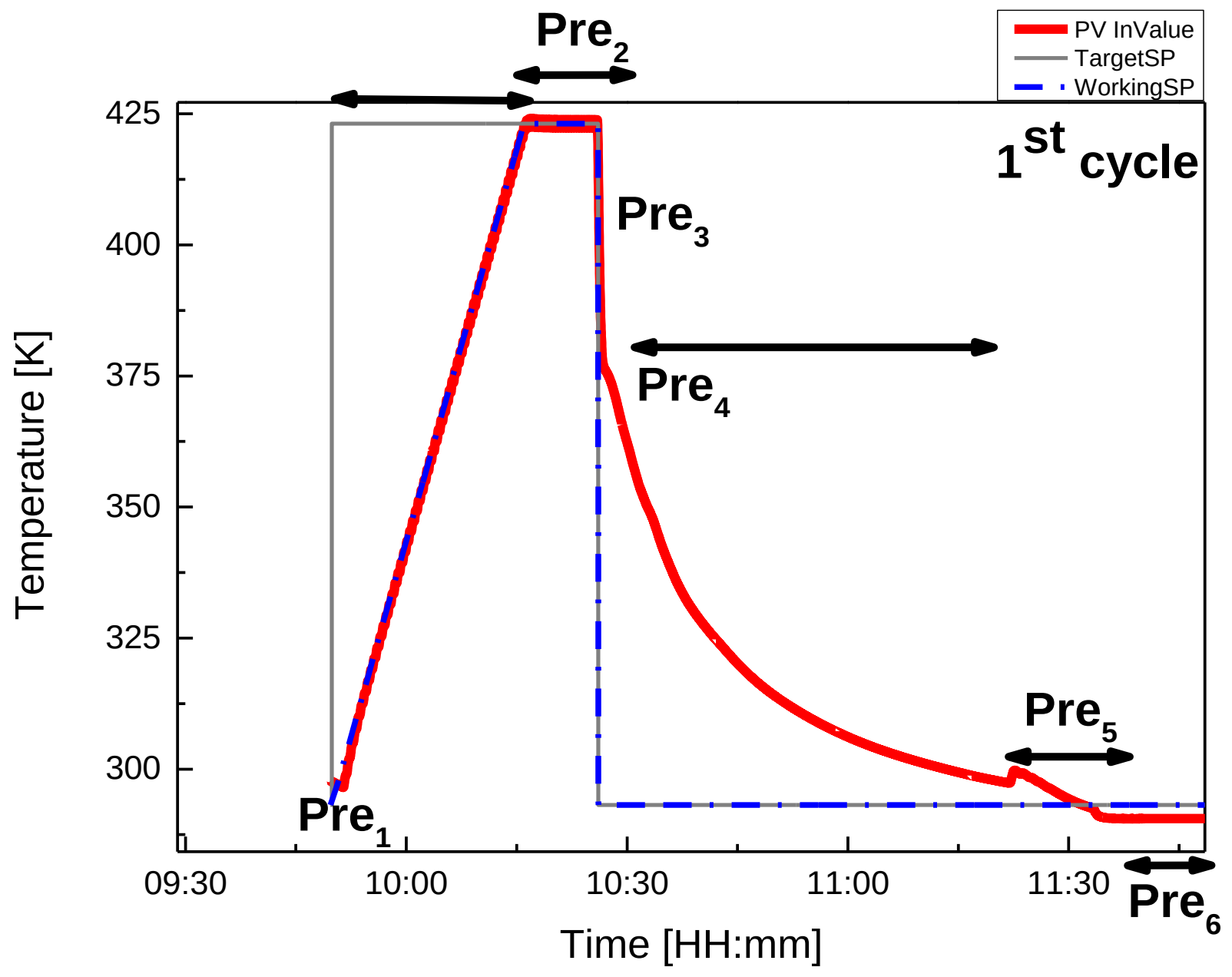

*Figure 12 This graph contains information about the heating cycle of the FT-IR CO adsorption tests. The actual oven curve (red line, PV InValue) shows a peak before CO is dosed stepwise into the sample chamber.*

### 3.2.3 SAMPLE PREPARATION

The samples for the Fourier - Transformation Infrared Spectroscopy were first pulverized inside the mortar with a pestle, which was made of soapstone and re-filled into a press.

The samples were pressed into self-supporting discs with a diameter of 4 mm and a thickness of few µm (< 0.4 mm). The weight of this discs are lower than 0.08 $mg/mm^2$.
The discs were placed into the wafer holder. As mentioned already above, the infrared apparatus was arranged with a vacuum cell such that samples could be implanted after pressing them into small discs.

### 3.2.4 DATA ANALYSIS

In-situ FT – IR spectra were measured with an MTC detector and collected with the OPUS software version 6.5. Before starting with any measurements (such as the CO cycle or the pretreatment procedure) the background was taken with and without the sample disc. The software is measuring via subtractions the FT – IR background to the measured FT – IR spectra. Results are shown in the "Results" chapter, subsection "Fourier Transformation Infrared Spectroscopy".

## 3.3 KINETIC MEASUREMENTS

In this section, the experiment (Figure 13 b) is presented and explained with a special focus on the reactor. The simplest form [133] of a catalytic reactor consists of a cylindrical tube packed with the catalysts in pellet or powder form and surrounded by a cooling medium. The feed gas mixed with the used reactant gases (CO and $O_2$) enters the cylindrical tube from the top.

This type of reactor is recommended for studying gradientless catalytic kinetics in which almost uniform temperature and pressure conditions are maintained by design and operational procedures.

Kinetic measurements of 2 wt. % $Au_{38}(SR)_{24}/CeO_2$ gold nanoclusters and $CeO_2$ support material were measured by Thorsten Boehme, Vienna University of Technology.

### 3.3.1 EXPERIMENTAL SETUP

Catalytic activity evaluations were carried out in a continuous flow fixed-bed quartz tube reactor (Figure 13) with an inner diameter of 4 mm, the tube area A of 12.57 $mm^2$ and a gas hourly space velocity (GHSV) of 15 000 mL/(gh) under ambient pressure.
In Figure 14, a detailed description of the fixed-bed reactor is given.

The fixed-bed reactor was connected to the feed gas supply, which maintained continuous gas flow through the reactor. The reactor was equipped with a type K thermocouple. The thermocouple was placed in the quartz tube above the sample. The aim was to keep this position as stable as possible in order to have comparable reproducible temperature values.

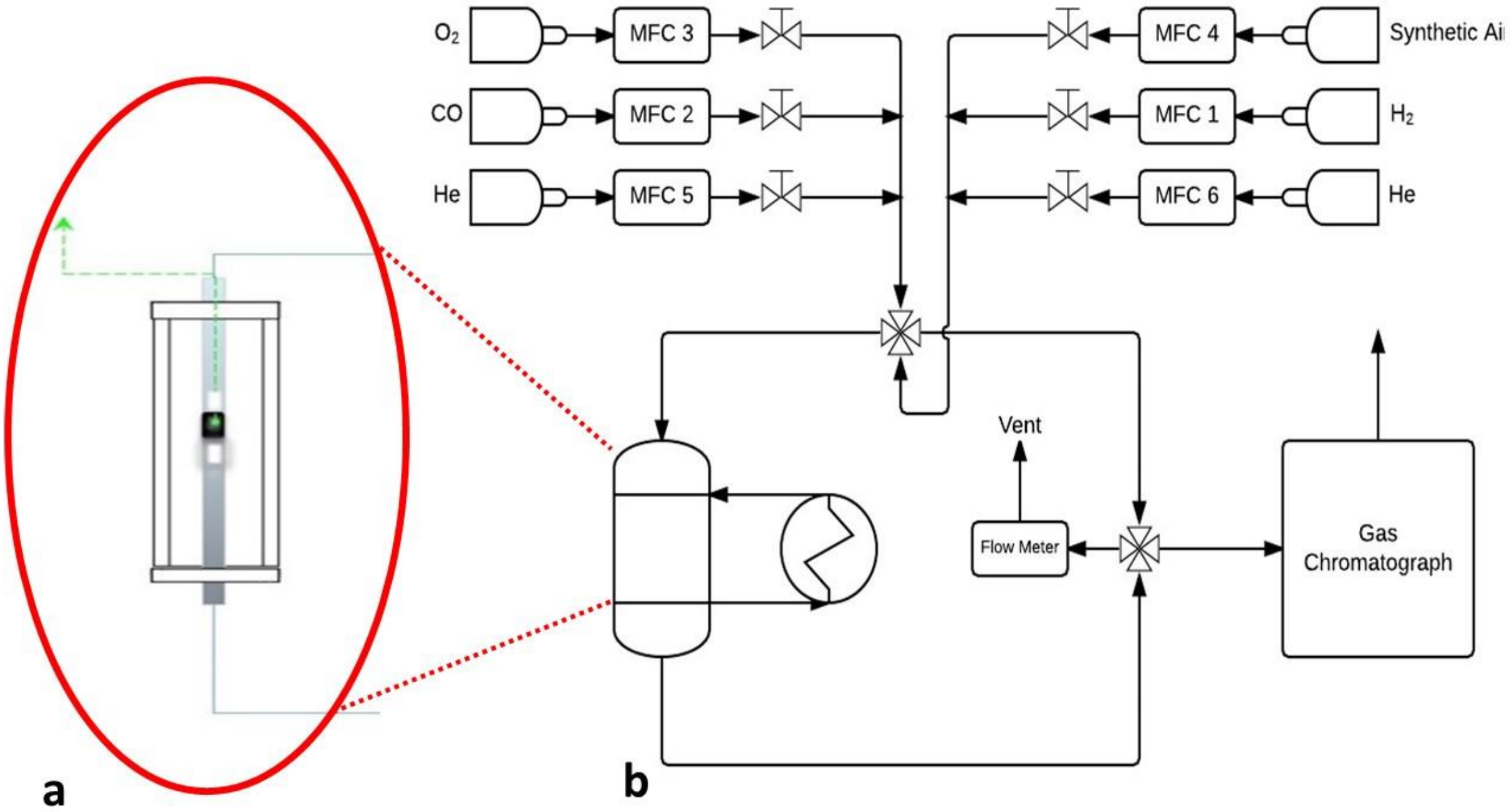

*Figure 13 A sketch of the experimental setup for the catalytic measurements in the fixed-bed flow reactor. (a) shows a detailed view of the fixed flow bed reactor and (b) gives an overview of the entire experimental setup.*

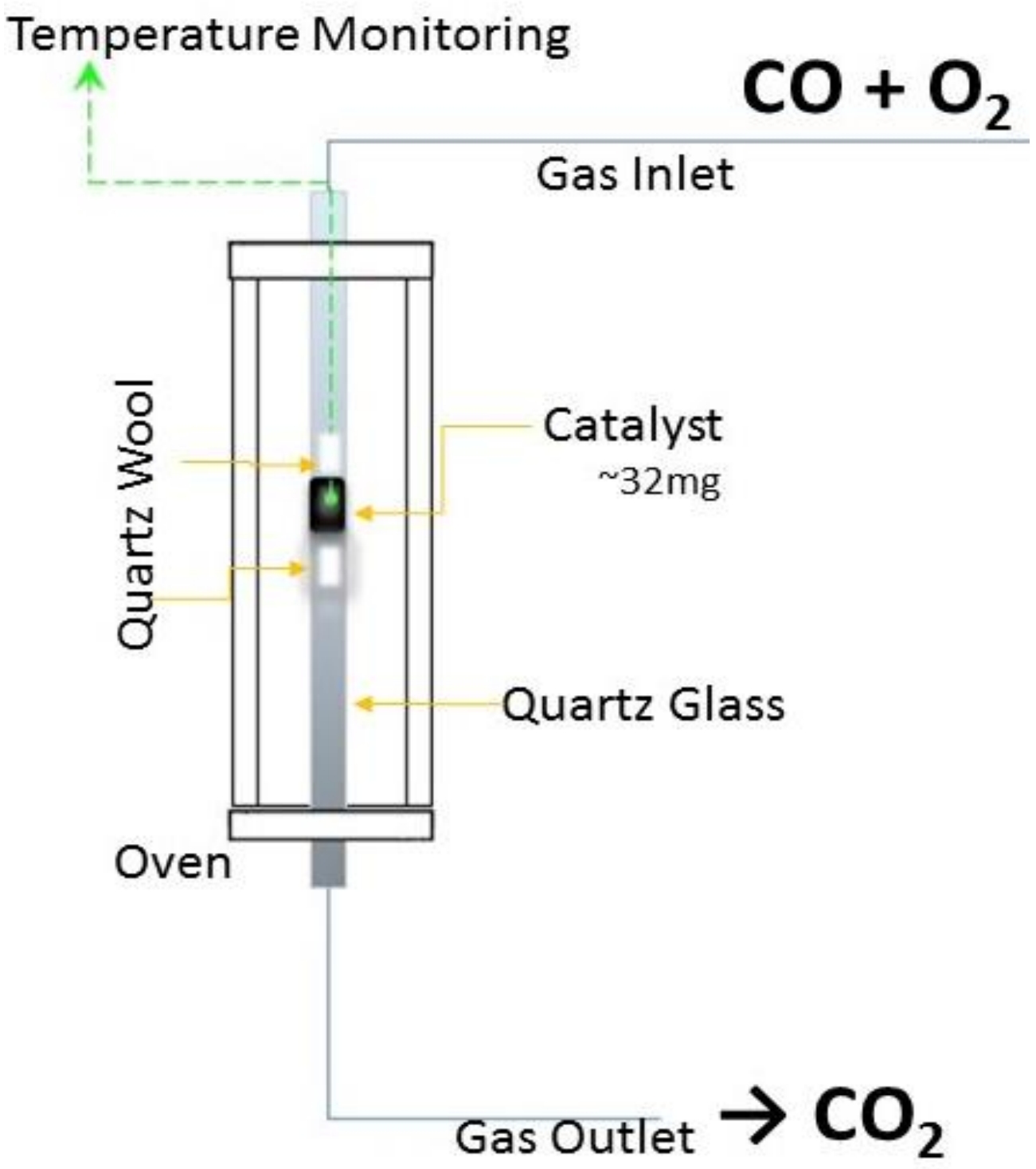

*Figure 14 Zoom – in figure of the fixed-bed flow reactor as schematic drawing with details.*

### 3.3.2 KINETIC MEASUREMENT PRETREATMENT AND CO OXIDATION PROCEDURE

The activity of the catalyst was measured after applying oxidative (synthetic air), inert (He) and reductive ($H_2$) pretreatments respectively. The pretreatment temperature was set to 473 K, since this should be the temperature at which the ligands are partially removed from the gold nanocluster surface.

The heating ramp for the pretreatment procedure was of 10 $Kmin^{-1}$, starting from room temperature to 473 K with a dwell time of 30 minutes. The conditions during pretreatment for each of the investigated catalysts are listed in Table 3.

*Table 3 Pretreatment atmospheres used for the kinetic measurements samples.*

| Sample | Pre. Atmosphere |
|---|---|
| $CeO_2$ | Inert & Oxidative |
| $Au/CeO_2$ | Oxidative, Reductive & no pretreatment |
| 1 wt. % $Au_{38}/CeO_2$ | |
| 2 wt. % $Au_{40}/CeO_2$ | |
| $Au/ Al_2O_3$ | |
| 1 wt. % $Au_{38}/Al_2O_3$ | |
| 2 wt. % $Au_{40}/Al_2O_3$ | |

A total flow of 50 mL/min was chosen for all pretreatment and reaction gas mixtures. CO oxidation was performed at room temperature. The heating ramp for the CO oxidation was 5 $Kmin^{-1}$.

The CO oxidation measurements were performed with various temperature ramp rates as shown in Figure 15. The starting temperature was room temperature (300 K). Then, the reactor (see Figure 14) was fed with the reaction mixture with the reaction mixture of 10% $O_2$ and 5% CO in He (CO : $O_2$ = 1:2) .The heating temperature steps were done essentially with changing $CO:CO_2$ ratios (see Figure 15).

The changes of activity were followed by an online-gas chromatograph (GC) until no CO conversion was observed.

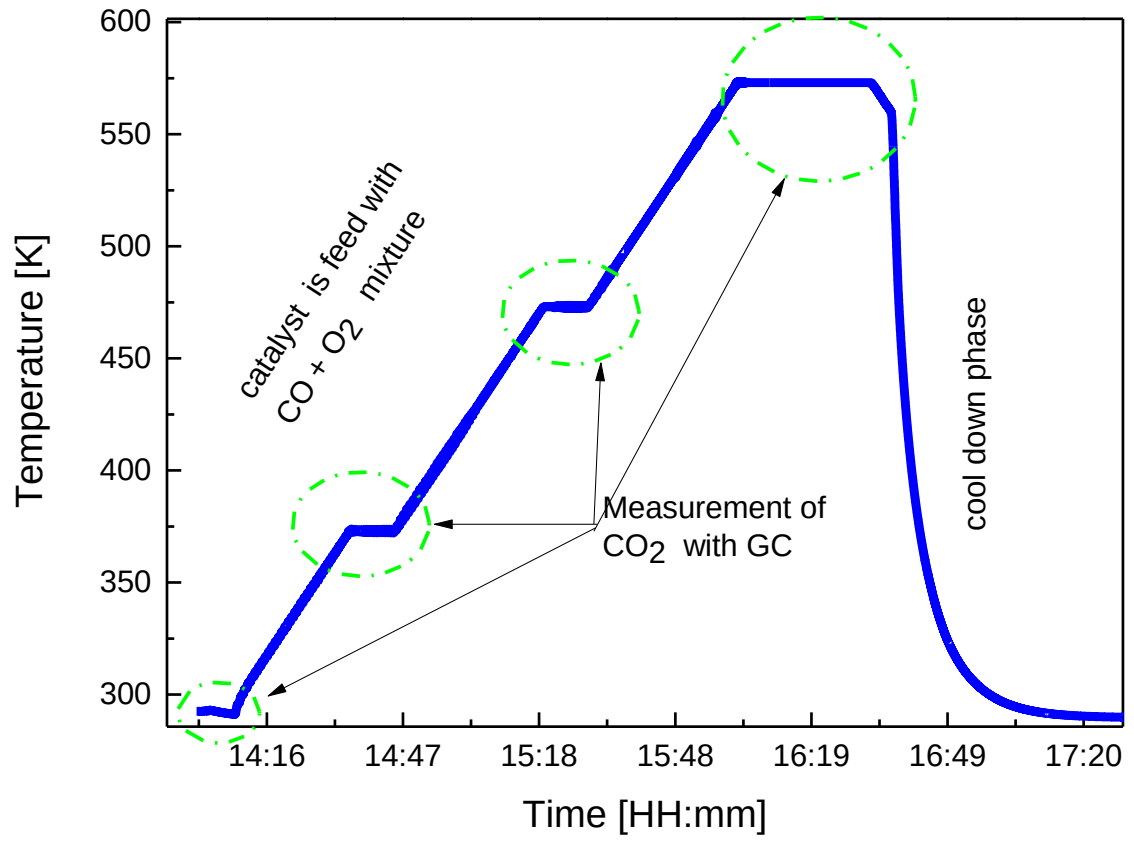

*Figure 15 An example heating ramp during the measurement of the kinetics tests, starting from 300 K to 575 K, stepwise. During the kinetic tests, steps were done at temperatures at which significant $CO_2$ conversions were measured.*

### 3.3.3 SAMPLE PREPARATION

For CO oxidation, 32 mg powder of 2 wt. % $Au_{38}(SR)_{24}/Al_2O_3$ and 2 wt. % $Au_{38}(SR)_{24}/CeO_2$ catalyst was used and 50 mg powder of 1 wt. % $Au_{40}(SR)_{24}/Al_2O_3$ and 1 wt. % $Au_{40}(SR)_{24}/CeO_2$ were used for the kinetic tests. The different amounts of catalysts were chosen differently due to the availability of the tested material.

The sample were filled inside the cylindrical tube reactor in powder form and kept in position within the quartz wool (see Figure 14).

### 3.3.4 DATA ANALYSIS

The measured CO to $CO_2$ ratios, consisting of three GC injections at each temperature step (15 minutes holding time at each temperature, see Figure 15 green-dotted encircled spots) and a heating ramp of 5 $Kmin^{-1}$.

#### a. DEFINTION OF REACTION RATE

The reaction rate r [134] in [ $mol\ g^{-1}\ min^{-1}$] is given as

$$r = \frac{P \times x_{CO} \times f_r}{R\ T} \times \frac{conv.}{m_{Cal}} \tag{2}$$

where P is the total pressure in the reactor, $x_{CO}$ gives the $CO_2$ fraction in the feed gas, $f_r$ is the total flow rate given in ml $min^{-1}$ and R is the gas constant defined as 8.314 J $K^{-1}$ $mol^{-1}$. T is the absolute temperature in K. The second ratio includes conv., which denotes the conversion value measured with the FID detector as the peak area. The peak area is proportional to the gas quantity with the same proportional factor for $CO_2$. The denominator in the break $m_{Cal}$ is the mass of the catalyst in the reactor in grams.

#### b. DEFINTION OF ACTIVATION ENERGY

The activation energy [135], [136] of catalysts was determined for all catalysts prepared and both for the oxidized and reduced samples. It is calculated starting from the Arrhenius equation

$$r = A\ exp\left(-{}^{E_a}/_{R\ T}\right) \tag{3}$$

$$ln(r) = ln(A) - \left(\frac{E_a}{R\,T}\right) \quad (4)$$

In equation (3) and (4) A is the pre-exponential factor, which is a constant for each chemical reaction that defines the rate due to the frequency of collisions in the correct orientation. The unit of A is identical to the rate constant and depends on the reaction order. In Literature this factor is known as the frequency factor or attempt frequency of the reaction. The apparent activation energy [J $mol^{-1}$] is $E_a$, R is the gas constant defined as 8.314 J $K^{-1}$ $mol^{-1}$ and T is the absolute temperature in Kelvin.

According to equation (3), ln(r) is plotted against $\frac{1}{RT}$ with the slope giving the activation energy. The error of the activation energies was determined to be about 7%, a comparatively high value.

### c. DEFINITION OF TURN-OVER-FREQUENCY

For the Turn-Over-Frequency (TOF in $sec^{-1}$) the computation was taken from [48]. TOF is defined as

$$TOF = \frac{M_{CO}}{M_{Au}} \quad (5)$$

$M_{CO}$ is the number of converted CO molecules in one second and $M_{Au}$ corresponds to the number of exposed Au sites. $M_{CO}$ is defined as

$$M_{CO} = Conversion \cdot Total\ Flow\ Rate \cdot CO\ Concentration \quad (6)$$

The conversion value corresponds to the CO conversion in percent, as next the total flow rate in mL/min and the CO concentration in the feed gas in percent.

The number of exposed gold sites $M_{Au}$ is written as

$$M_{Au} = Amount\ of\ the\ sample \cdot Au\ wt\% \cdot Au\ exposure \quad (7)$$

Here, the amount of the sample is given in mg, Au weight corresponds to the Au weight in percent and Au exposure corresponds to ~50% of gold atoms on the surface from its core – shell structure.

## 3.4 CATALYSTS AND CATALYSTS SUPPORT MATERIAL PREPARATION

The samples used in the current work are shown in Table 4, together with descriptions of their purpose for the current research and the motivation behind their use.

### 3.4.1 COMMERICALLY AVILABLE CATALYSTS

Samples of $CeO_2$ [137] and $Al_2O_3$ [138] are commercially available from Sigma Aldrich. For more details see [130] and [137]. The catalysts were prepared at the University of Geneva from members of Prof. Buergi's Group and kindly provided by Dr. N. Barrabes-Rabanal.

### 3.4.2 SYNTHESIZED CATALYSTS

1 wt.% Au/$CeO_2$ and 1 wt.% Au/$Al_2O_3$ were prepared as reference catalysts at the Vienna University of Technology by Thorsten Boehme. The measurements were kindly provided Thorsten Boehme. The synthesis recipe follows the one provided in [139].
For the gold nanoparticles supported on alumina [139] were prepared by the deposition-precipitation method (DP) with two-time adjustment of pH before and after addition of support to gold precursor. The $HAuCl_4$ ($4.2 \times 10^{-3}$M) solution was heated to 353 K and the pH was adjusted by dropwise addition of 0.5 M NaOH. The suspension was thermostated at 353 K and stirred vigorously for 2 hours.
To remove residuals, chloride and unreacted gold species, the precipitates were washed few times. It was dried over night at 373 K and calcined at 723 K for four hours.

$Au_{40}(SR)_{24}$ and $Au_{38}(SR)_{24}$ were synthesized by Prof. Buergi's group and kindly provided by Dr. N. Barrabes-Rabanal. The synthesis process was done according to the synthesis recipe published by S. Knoppe et al. [90]. The authors employed the size exclusion chromatography on a semipreparative scale after the synthesis procedure. The isolated fractions were characterized with spectroscopic tools such as ultraviolet visible spectroscopy. A picture summarizing the procedure is shown in Figure 16.

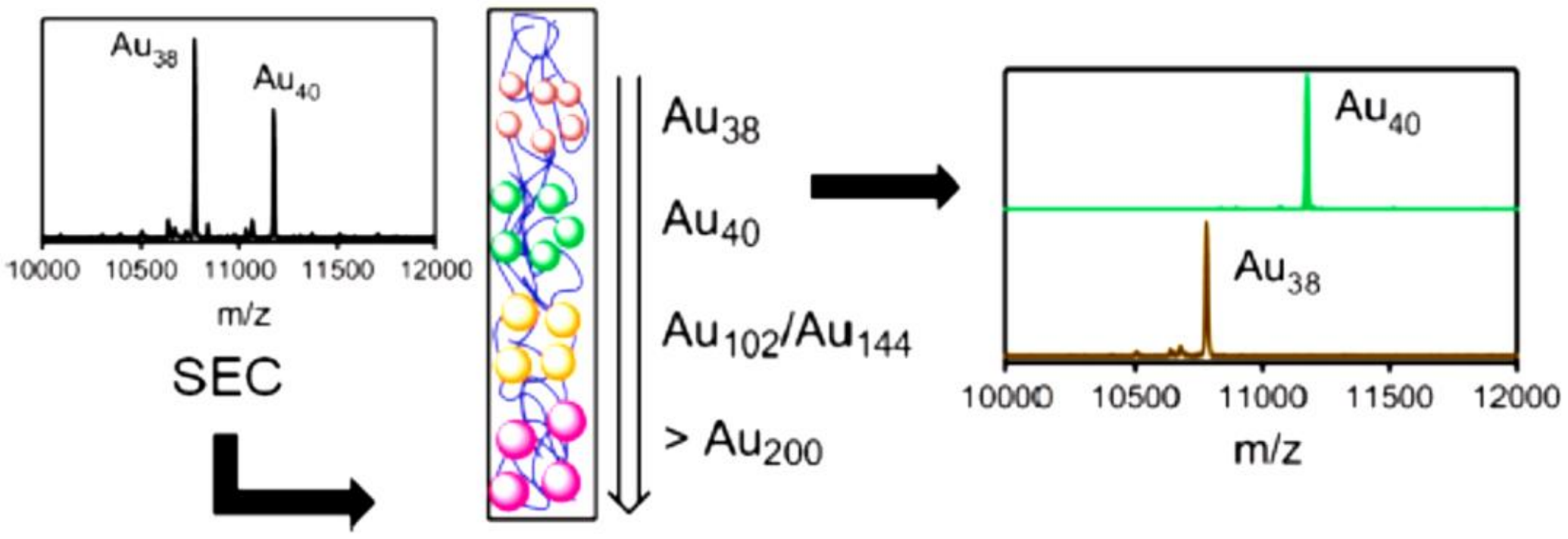

*Figure 16* *Adapted from* [90] *shows briefly how these gold nanoclusters were extracted and measured in order to fractionize the correct gold nanoclusters.*

Using the impregnation method, the gold nanoclusters were implanted into $CeO_2$ and into commercially available $Al_2O_3$. The percentage of gold nanoclusters on the two different support materials are 1 wt.% $Au_{40}(SR)_{24}/Al_2O_3$, 1 wt.% $Au_{40}(SR)_{24}/CeO_2$, 2 wt.% $Au_{38}(SR)_{24}/CeO_2$ and 2 wt.% $Au_{38}(SR)_{24}/Al_2O_3$.

Gold nanocluster catalysts were stored at low temperatures, in a dark surrounding and have to protected from sun light. Gold nanoclusters are highly photoluminescent [124] and is known for the redox behavior of gold [140].

A temperature increase can lead to undesirable sinter processes of the precursor with the ligands, support material or gold atoms and this would lower the catalytic activity of the gold nanocluster catalysts. To avoid this, the gold nanoclusters were not calcined.

*Table 4 In this table, all catalysts are listed with the experiments in which they are used and their application in reference or comparative measurements.*

<table>
<tr><th></th><th></th><th colspan="2">Source</th><th colspan="3">Experiments</th><th colspan="2">Application</th></tr>
<tr><th></th><th></th><th>Commercial</th><th>Synthesized</th><th>TGA</th><th>FT-IR</th><th>Kinetic</th><th>Reference Measurements</th><th>Comparative Measurements</th></tr>
<tr><td>1</td><td>$CeO_2$</td><td>✓</td><td></td><td>✓</td><td>✓</td><td>✓</td><td>✓</td><td></td></tr>
<tr><td>2</td><td>$Al_2O_3$</td><td>✓</td><td></td><td></td><td></td><td></td><td>✓</td><td></td></tr>
<tr><td>3</td><td>1 wt.% $Au/CeO_2$</td><td></td><td rowspan="8">✓</td><td></td><td></td><td>✓</td><td>✓</td><td></td></tr>
<tr><td>4</td><td>1 wt.% $Au/Al_2O_3$</td><td></td><td></td><td></td><td>✓</td><td>✓</td><td></td></tr>
<tr><td>5</td><td>$Au_{40}(SR)_{24}$</td><td></td><td>✓</td><td></td><td>✓</td><td></td><td>✓</td></tr>
<tr><td>6</td><td>1 wt % $Au_{40}(SR)_{24}/Al_2O_3$</td><td></td><td>✓</td><td>✓</td><td>✓</td><td></td><td>✓</td></tr>
<tr><td>7</td><td>1 wt % $Au_{40}(SR)_{24}/CeO_2$</td><td></td><td>✓</td><td>✓</td><td>✓</td><td></td><td>✓</td></tr>
<tr><td>8</td><td>$Au_{38}(SR)_{24}$</td><td></td><td>✓</td><td></td><td>✓</td><td></td><td>✓</td></tr>
<tr><td>9</td><td>2 wt % $Au_{38}(SR)_{24}/Al_2O_3$</td><td></td><td></td><td></td><td>✓</td><td>✓</td><td>✓</td></tr>
<tr><td>10</td><td>2 wt % $Au_{38}(SR)_{24}/CeO_2$</td><td></td><td>✓</td><td>✓</td><td>✓</td><td></td><td>✓</td></tr>
</table>

# 3.Chapter: RESULTS

In this chapter, the results are presented and described in detail. Starting point are the results for the Thermogravimetric Analysis, Fourier – Transformed Infrared Spectroscopy and Kinetic Tests.

## 4.1 THERMORGRAVITMETRIC ANALYSIS RESULTS

The TGA measurements were performed on a STA 449 F3 Jupiter (NETZSCH) thermal analyzer under different atmospheres ($N_2$ and $O_2$) from 298 K to 823 K at a ramp of 5 K/min.

### 4.1.1 THERMOGRAVIMETRIC ANALYSIS OF BARE $Au_{40}(SR)_{24}$ GOLD NANOCLUSTERS IN OXIDATIVE AND INERT ATMOSPHERE

Figure 17 shows the thermogravimetric results of bare $Au_{40}(SR)_{24}$ gold nanoclusters in two different atmospheres, nitrogen (Figure 17, a) and oxygen (Figure 17, b). In general, TGA measurements show steps within the weight loss curve, depending on the decomposition of the ligands. Here, we observe a three-step curve for $Au_{40}(SR)_{24}$ treated in nitrogen (Figure 17, a) and a four-step curve for the $Au_{40}(SR)_{24}$ gold nanocluster in oxygen (Figure 17, b).

According to the TGA measurements, the relative weight loss of $Au_{40}(SR)_{24}$ gold nanoclusters is around 36% (see Figure 17). The blue line corresponds to $Au_{40}(SR)_{24}$ gold nanocluster catalysts in an oxygen atmosphere.

The calculations for unsupported gold nanoclusters are given below: The TGA tested object can be written as $Au_{40}(SR)_{24} \sim Au_{40}(SC_2H_4Ph)_{24} \sim Au_{40}(SC_2H_4(C_6H_5))_{24}$. This way of writing breaks the gold nanocluster into four basic elements. For 40 gold atoms, the total weight is calculated as $Au_{40}$ = 7879 amu.

The remaining atoms are S, C and H. The respective atom weights are shown in Table 5.

*Table 5 Theoretical calculation for bare gold nanoclusters for Thermogravimetric Analysis*

| Element | Mass [amu] | Total element | Total mass [amu] |
|---|---|---|---|
| Au | 170 | $Au_{40}$ | 7879 |
| S | 32 | | |
| C | 12 | $C_2$ | 24 |
| | | $C_6$ | 72 |
| H | 1 | $H_4$ | 4 |
| | | $H_5$ | 5 |
| | | $(SR)_{24}$ | 3294 |
| | | $Au_{40}$ | 7879 |
| | | $Au_{40}(SR)_{24}$ | 11172 |

For the organic thiolate staple monomer (RS – Au – SR) and dimer (RS – Au – SR – Au – SR) motif (see Figure 2), the weight sums up to 3294 amu. The exact calculation is shown below:

$$SR_{24} = (SC_2\,H_4\,Ph)_{24} = (SC_2\,H_4\,(C_6\,H_5))_{24}$$
$$= (32 + 24 + 4 + (72 + 5)) \text{ x } 24$$
$$= 137 \text{ x } 24 = 3294 \text{ amu}$$

Therefore 30 %, of the total weight of one gold nanocluster, corresponds to the weight of the organic thiolate staple monomer (RS – Au – SR) and dimer (RS – Au – SR – Au – SR) motif. While 70 % of the total weight of one gold nanocluster corresponds to the 40 gold atoms within this structure.

The complete removal of the organic species in oxygen is 30% according to theoretical calculations. This compares well with the measured value of 36%.

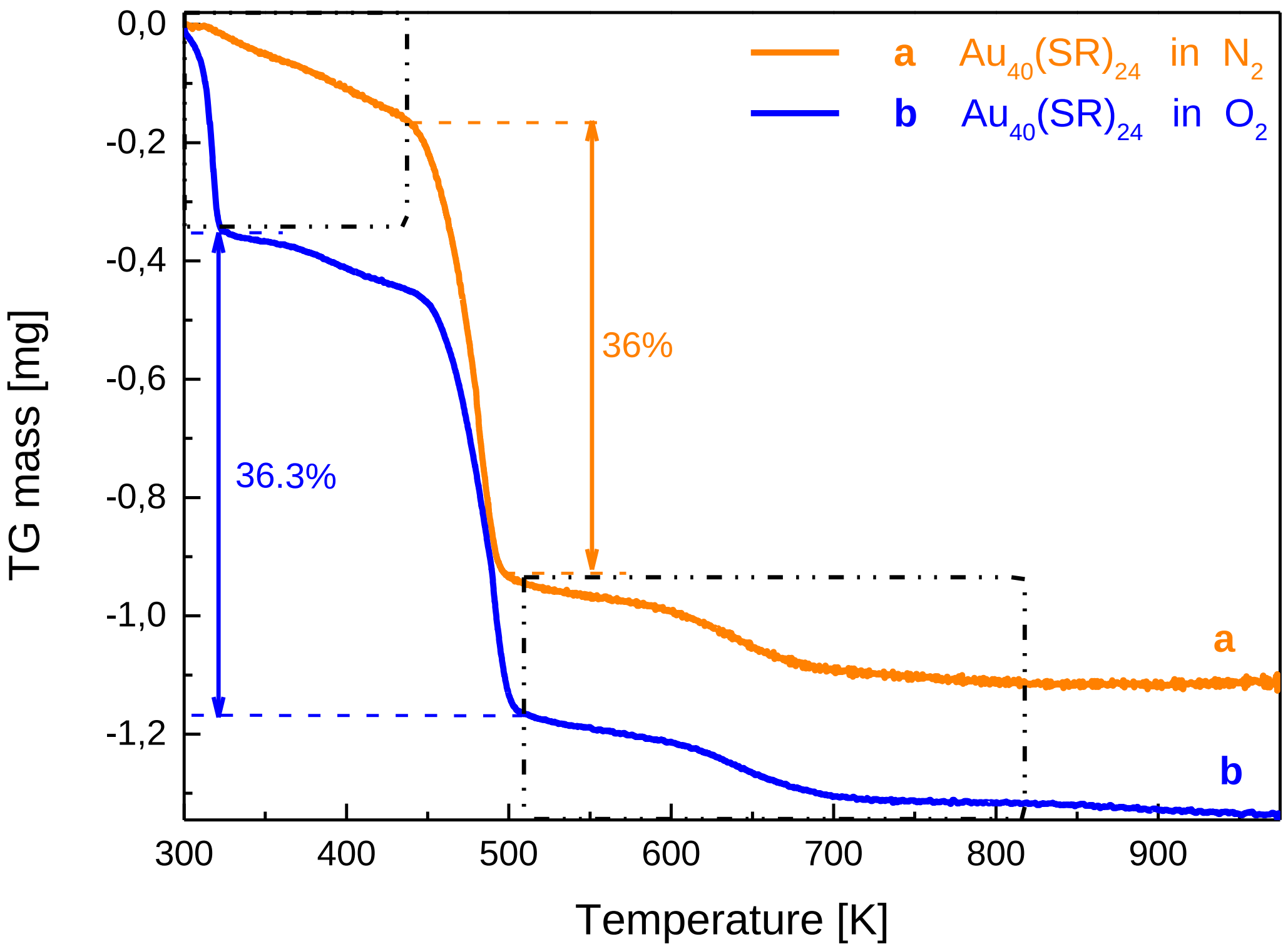

*Figure 17 $Au_{40}(SR)_{24}$ bare gold nanoclusters dissolved in dichloromethane (weight of dissolved gold nanoclusters ~30 µL = 30 mg) weight loss is measured with the help of the thermogravimetric analysis method in oxygen (b, blue line) and inert (a, orange line) atmosphere.*

As described in the chapter "Materials and Methods" (subsection "3.3.2 KINETIC MEASUREMENT PRETREATMENT AND CO OXIDATION PROCEDURE"), the starting temperature is room temperature (300 K) which is increased to 970 K.

At 970 K, a dwell time of 10 min was programmed and set. The stability of the clusters is an indication that the particles are stable even after oxidative and inert atmosphere (see Figure 19).

1

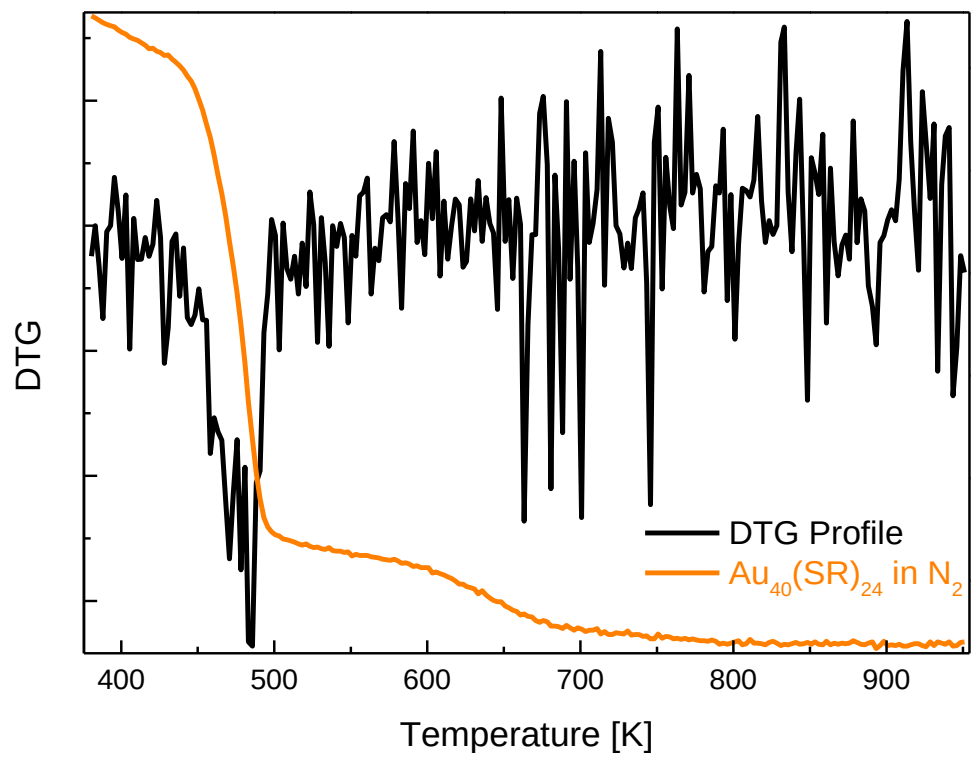

2

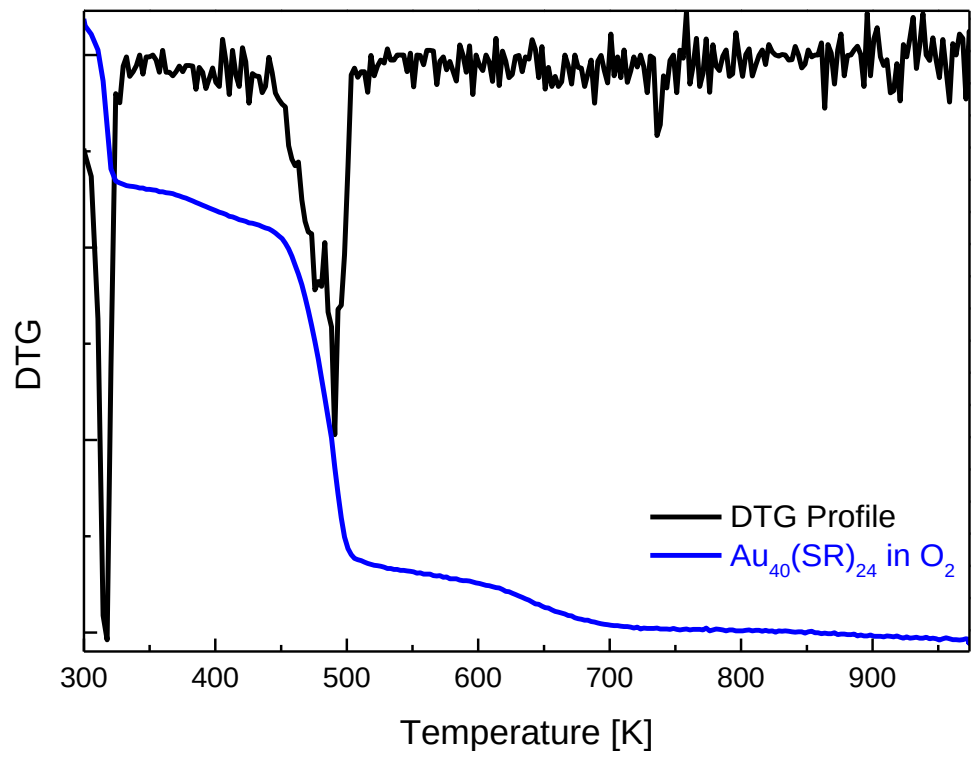

*Figure 18 In (1), the DTG Profile for $Au_{40}(SR)_{24}$ gold nanoclusters in inert atmosphere is presented. (2) shows the DTG Profile in oxygen atmosphere.*

For more insight into the thermal behavior of the gold nanocluster (here for the $Au_{40}(SR)_{24}$), Derivative-thermo Gravimetric curves are plotted in Figure 18.

According to the first derivative of the TGA measurement (DTG curve), for the gold nanocluster in $N_2$ atmosphere (see Figure 18 (a)), a negative spike occurred at 485 K. The steep drop corresponds to the gold nanocluster $Au_{40}(SR)_{24}$ mass of 36% (see Figure 18a). Another peak appeared at 663 K, 680 K, 688 K, 700 K and 745 K. Not all peaks are visible on the TGA curve.

For $Au_{40}(SR)_{24}$ in $O_2$, the DTG curve has a steep negative spike at 317 K, which corresponds to the quickly stripped moisture and decomposing molecules of the dispensed solution, dichloromethane.

A steep and negative spike is measured at 490 K, the same is again visible in Figure 18(a). At 688 K and 738 K, an additional change was measured.

For both measurements at 688 K and 738 K (the TGA measurement in $O_2$) as well as at 745 K (the TGA measurement in $N_2$ atmosphere), more changes in the weight loss were observed.

The theoretical value for this relative weight loss in oxygen (Figure 18b) is around 30 %. For gold nanoclusters $Au_{40}(SR)_{24}$, the measured value is approximately 36%, which agrees with the calculated numbers in Table 6.

In the first part of the TGA profile (see Figure 27 (1)), a steep step of approximately 8% was measured, starting already at 313 K. This is related to the presence of dichloromethane and water as well as fast desorbing ligand fractions.

From 323 K to approximately 433 K, a relative smaller weight change was measured, prior to a second steep mass drop of almost 26.3% until 518 K. This step is, as attested in literature, related to the stepwise removal of the looser organic ligands [68], [141].

The final step is marked by a weight loss of approximately 5%, which equates the complete removal of the thiolate ligands. Beyond 673 K, no weight change was measured. The constant value indicates that the gold clusters do not undergo a significant change in mass. This TGA profile is related to the $Au_{38}(SR)_{24}$ clusters with one more additional step [68], [130], [141].

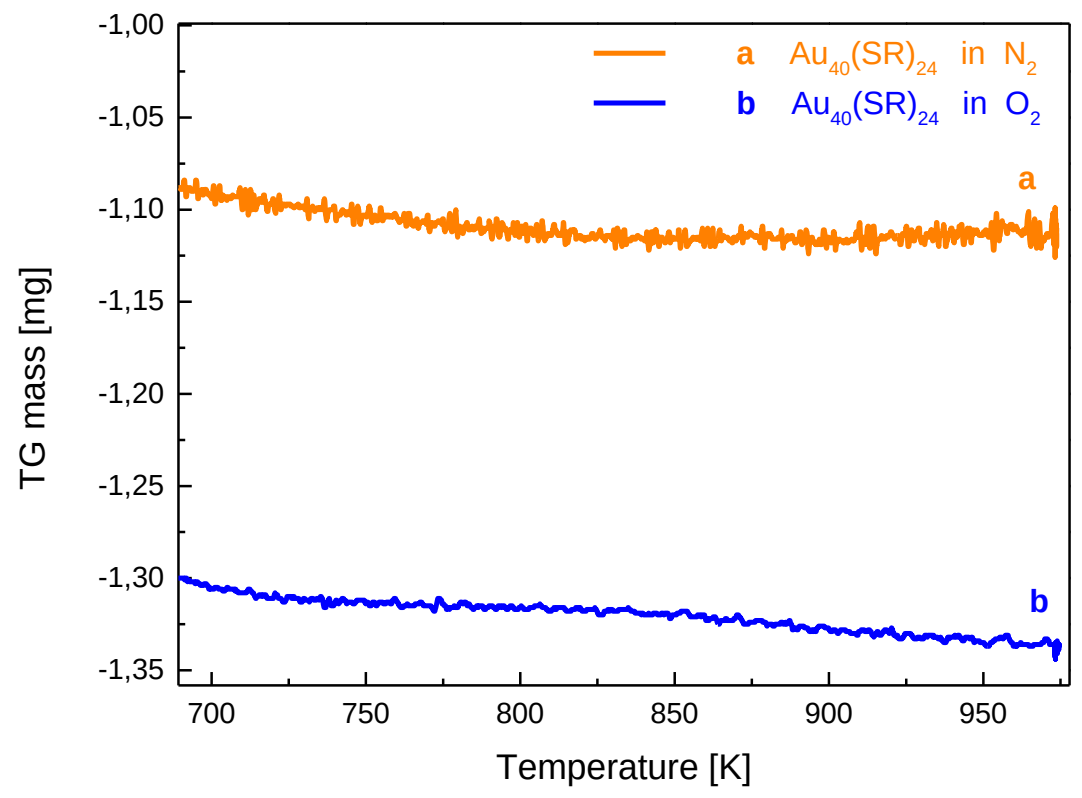

*Figure 19 Thermal stability of gold nanoclusters within the TGA measurement is shown. During temperature tuning of 270 K, from 700 K to 970 K in $O_2$ (b) and $N_2$ (a) atmosphere, the thermal curve stays constant.*

Thermal stability of the mass was tested with the TGA also (Figure 19). Figure 19

For this test, the bare gold nanocluster $Au_{40}(SR)_{24}$ in $O_2$ (Figure 19 (b)) and $N_2$ (Figure 19 (a)) atmosphere were kept at the high-end temperature for 10 minutes.

Similar tests were performed by R. Jin [107] and his colleagues to examine the thermal stability of $Au_{25}(SR)_{18}$ gold nanoclusters. These measurements confirm that the gold nanoclusters are robust and thermally stable above the ligand loss temperature.

### 4.1.2 THERMOGRAVIMETRIC ANALYSIS OF CERIA

In Figure 20 the relative weight change of the support $CeO_2$ is shown.

The relative weight change for the blank ($CeO_2$) in $O_2$ (Figure 20, b) is around 19% and in $N_2$ atmosphere the relative weight change is almost 12% (see Figure 20, a).

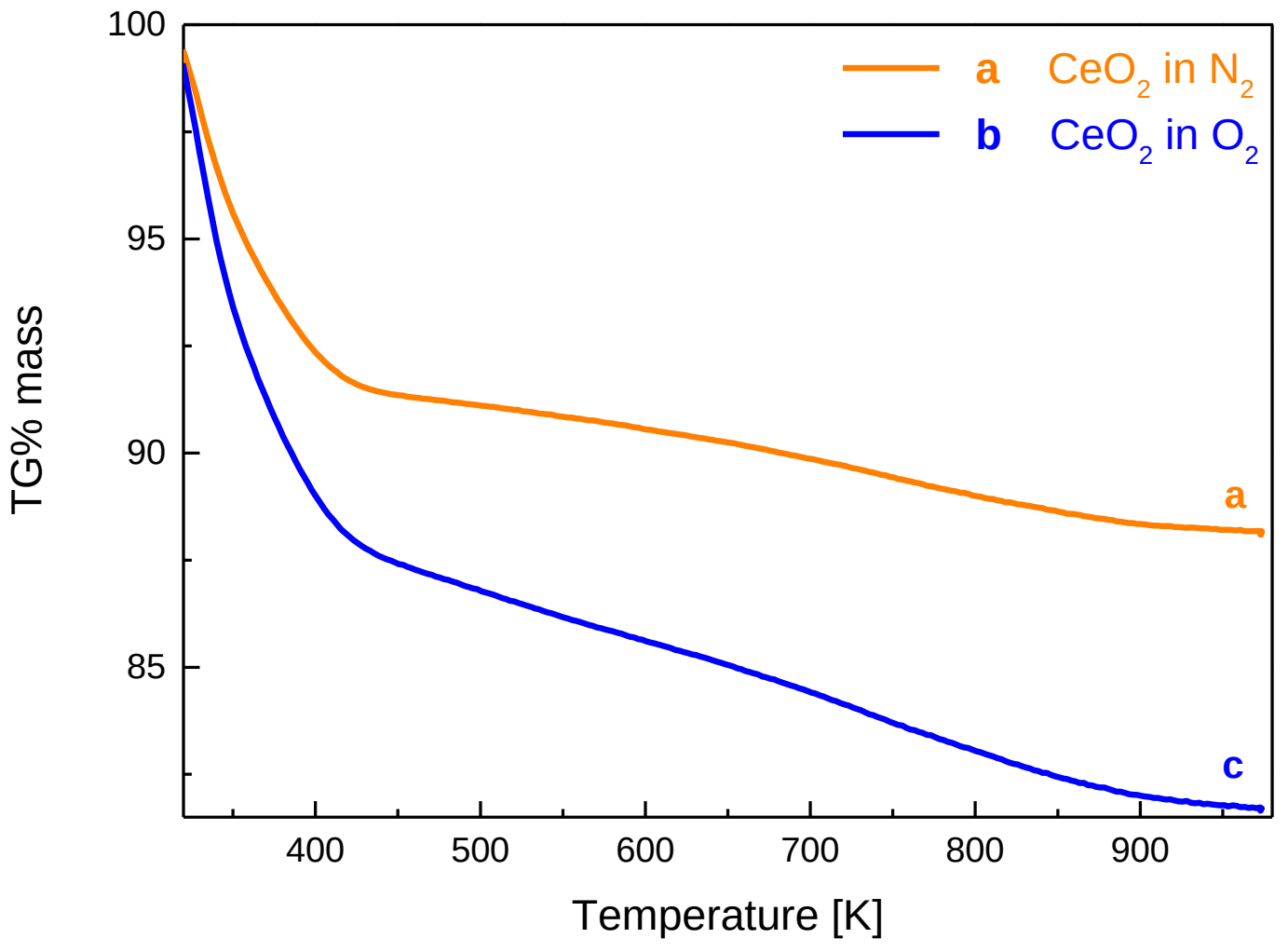

*Figure 20 The weight loss of the support material $CeO_2$ has been tested in inert (a, orange line) and oxidative (b, blue line) atmosphere.*

In general, the weight loss due to the evaporation of structural water molecules is ~17%. $Ce(OH)_4$ [142] is a hydrous oxide, represented by $CeO_2 \bullet 2\ H_2O$, which dehydrates progressively. Therefore, the decomposition of the precursor could be in form of dehydration process of the hydrated $CeO_2$. It is suggested that the difference in weight loss observed could be either [142] precipitate consisting of a partially hydrated from of ceria (i.e. $CeO_2 \bullet 2\ H_2O$), for which a ~12% weight loss on decomposition corresponds to x = 1.35 or [143] the precipitate consisted of a mixture of phases like $CeO_2 \bullet 2\ H_2O + CeO_2$.

### 4.1.3 THERMOGRAVIMETIRIC ANALYSIS OF $Au_{40}(SR)_{24}$ AND $Au_{38}(SR)_{24}$ SUPPORTED ON $CeO_2$

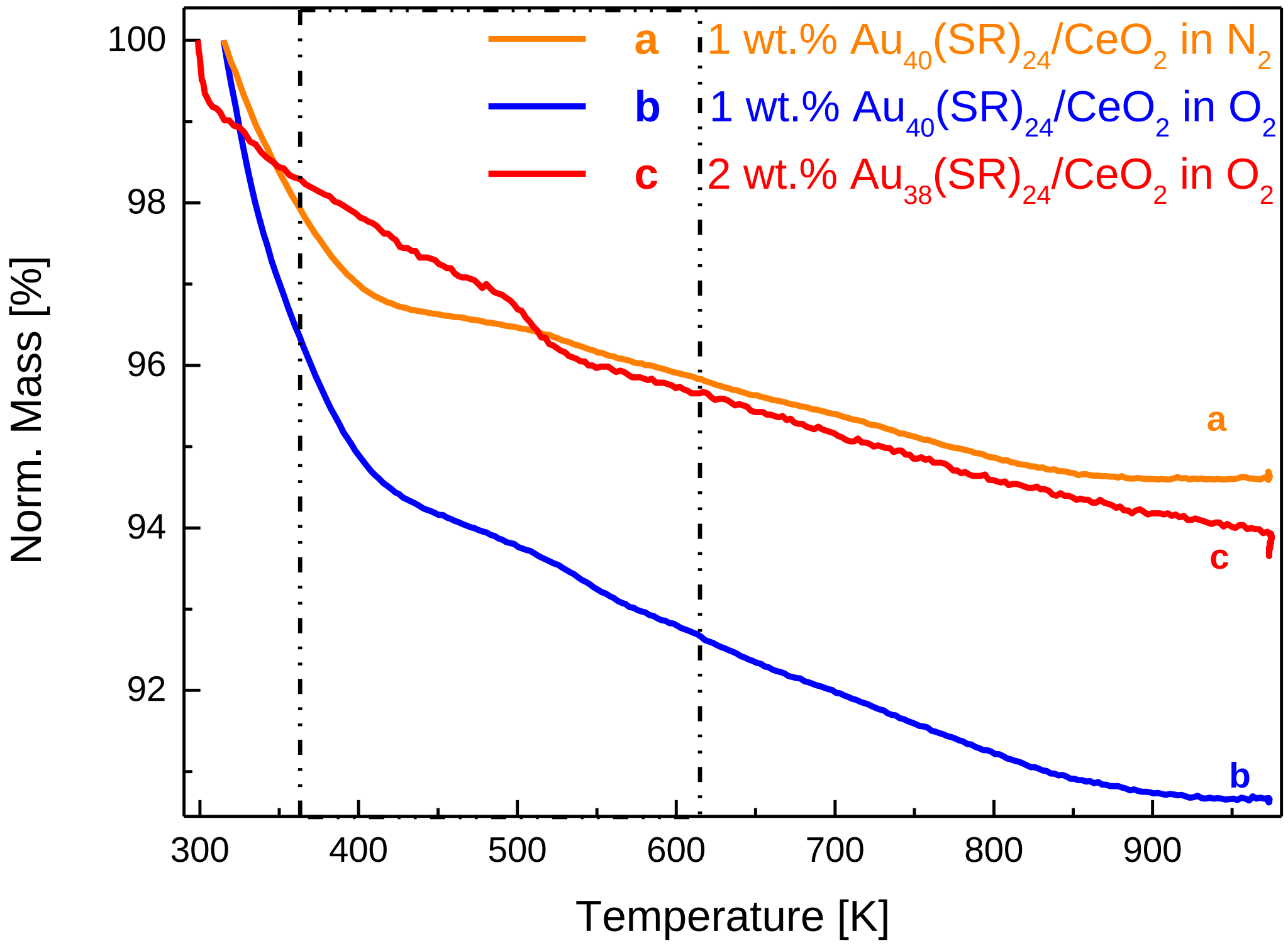

*Figure 21* *$Au_{40}(SR)_{24}$ gold nanoclusters supported on $CeO_2$ were tested in inert (a, orange line) and in oxidative atmosphere (b, blue line). These values are compared to $Au_{38}(SR)_{24}/CeO_2$ gold nanoclusters [1] tested in oxygen (c, red line).*

Thermogravimetric analysis measurements are shown in Figure 21 for 2wt. % $Au_{38}(SR)_{24}/CeO_2$ treated in oxygen (Figure 21 (c)), for 1 wt. % $Au_{40}(SR)_{24}/CeO_2$ treated in oxygen (Figure 21 (b)) and for 1 wt. % $Au_{40}(SR)_{24}/CeO_2$ treated in nitrogen (Figure 21 (a)). For a detailed account of the experimental setup, data treatment and experimental conditions, please refer to the chapter "Methods and Materials", subsection "TGA measurements".

The profile of gold nanoclusters supported by $CeO_2$ (Figure 21) evidently confirms thiol ligand removal. The support on the reducible transition metal has a total mass loss of ~6% at 940 K. More interesting is the step in between, which starts to develop approximately at 428 K with a mass loss of ~3%. This area is indicated within the square.

[1] $Au_{38}(SR)_{24}/CeO_2$ measurement was performed by Thorsten Boehme, Vienna University of Technology.

At about 490 K, the percentage drops to 96%. Then, at almost 532 K, the mass change is ~4%. The relatively steep and short drop of the thermogravimetric mass change (Figure 21 (b)), area within the black dotted square) corresponds to 30% thiol ligand removal of 2 wt. % of gold clusters on the surface.

The TGA measurement in nitrogen shows a weight loss for $Au_{40}(SR)_{24}$ supported by $CeO_2$ (Figure 21 (a)) of approximately 5%. The TGA measurement in nitrogen yielded almost 10%.

For $Au_{40}(SR)_{24}$ nanoclusters supported by $CeO_2$ (Figure 21 (a) and (b)) in both atmospheres ($O_2$, $N_2$), one step in the same temperature range can be identified. This area is bounded by the square around this temperature range.

Depending on the atmosphere, the step is consistent with 1 wt. % of $Au_{38}(SR)_{24}$ (Figure 21 (c)) being slightly higher in oxygen (~1.3%) and lower in nitrogen (0.6%).

The higher mass loss could be related to the interaction of oxygen, the purge gas, with the thiolate ligands, in which the thiolate removal is different.

Compared to the $Au_{40}(SR)_{24}$/ $CeO_2$ nanocluster, $Au_{38}(SR)_{24}$ TGA value of weight loss is ~6%, which also shows the additional $Au_{40}(SR)_{24}$ step due to the thiolate ligand removal.

### 4.1.4 THERMOGRAVIMETRIC ANALYSIS OF ANALYSIS OF $Au_{40}(SR)_{24}$ AND $Au_{38}(SR)_{24}$ SUPPORTED ON $Al_2O_3$

For the gold nanocluster $Au_{40}(SR)_{24}$ supported by $Al_2O_3$ (shown in Figure 22 (a)), a weight loss of approximately 10% was measured in $N_2$ atmosphere.

In oxygen (shown in Figure 22 (b)), the mass loss was higher, namely slightly above 17%, and compared to the results in $N_2$ atmosphere, the difference is approximately 7%. This indicates that more organic material is burned off in oxygen atmosphere.

For $Au_{38}(SR)_{24}$ supported by $Al_2O_3$ Figure 22 the TGA profile (c) shows a mass loss of ~17%. The TGA profile for $Au_{38}(SR)_{24}$ was steeper compared to $Au_{40}(SR)_{24}$ supported by $Al_2O_3$, but the start and end values as well as the mass loss are almost the same.

The only obvious change is the weight loss of the supported gold nanoclusters ($Au_{40}(SR)_{24}/Al_2O_3$ and $Au_{38}(SR)_{24}/Al_2O_3$) but without any step or change, which would indicate that these catalysts are dispends with the gold nanoclusters.

The desorption process should diverge, however, when the protective thiol ligands around the gold nanocluster interact in $O_2$ with the different support materials.

The sharp decrease in weight to from room temperature to 450 K links to the elimination of moisture. $Al_2O_3$ TGA profile can be divided into three regions [144]. From room temperature to 483 K the TGA profile indicates the release of free bound water. Between 503 K to 753 K a more pronounced change is observed. This change is related to the release of crystallization water [145].

Changes from 773 K on, these weight losses correspond to a phase change to pseudo – boehimte -γ – $Al_2O_3$ transitions, which occurs through partial dihydroxylation [144].

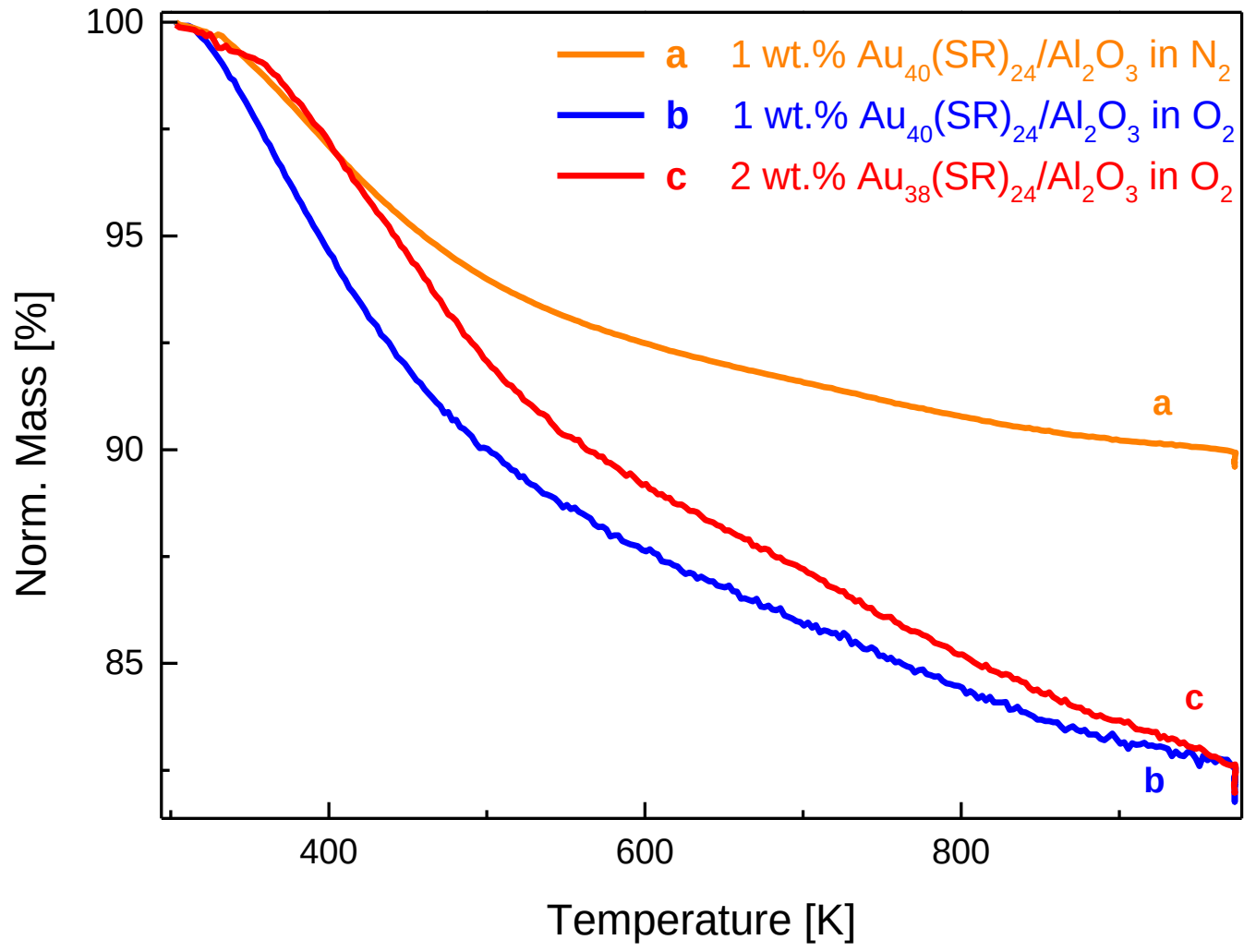

*Figure 22 $Au_{40}(SR)_{24}$ gold nanoclusters supported on $Al_2O_3$ were tested in inert (a, orange line) and in oxidative (b, blue line) atmosphere. These values are compared to $Au_{38}(SR)_{24}$/ $Al_2O_3$ gold nanoclusters[2] tested in oxygen (c, red line).*

## 4.2 FOURIER – TRANSFORM INFRARED SPECTROSCOPY RESULTS

In this subsection, the performed Fourier-Transform Infrared (FT-IR) spectroscopy measurements are shown and described.

A detailed description of the FT-IR system and the pretreatment procedure is given in the chapter “Materials and Methods” (subsection “FT-IR spectrometer”) together with a sketch of the system.

The thin line in the figures corresponds to the spectra measured after keeping the catalysts in a CO atmosphere for 10 minutes, and the thick line corresponds to a test with the CO atmosphere evacuated (at approximately $10^{-7}$ mbar high vacuum) from the sample chamber. The spectra were taken 10 minutes after the evacuation had started.

[2] $Au_{38}(SR)_{24}/Al_2O_3$ measurement was performed by TB.

### 4.2.1 FOURIER – TRANSFORM INFRARED SPECTROSCOPY OF THE SUPPORT MATERIAL $CeO_2$

The FT-IR spectra shown in this section include the results of the support material $CeO_2$.
The first set of spectra shows the CO adsorption spectra for $CeO_2$ pretreated in an oxidative atmosphere (Figure 23) and in an inert atmosphere (Figure 26).

The FT-IR spectra of the $CeO_2$ support material shows the IR peak at 2171 $cm^{-1}$ , which corresponds to CO on $Ce^{4+}$ [146], [147].
From the literature, an IR peak at 2133 $cm^{-1}$ corresponds to CO on $Ce^{3+}$ sites [148]. In the current work, the peak was found at 2121 $cm^{-1}$, as shown in Figure 23.

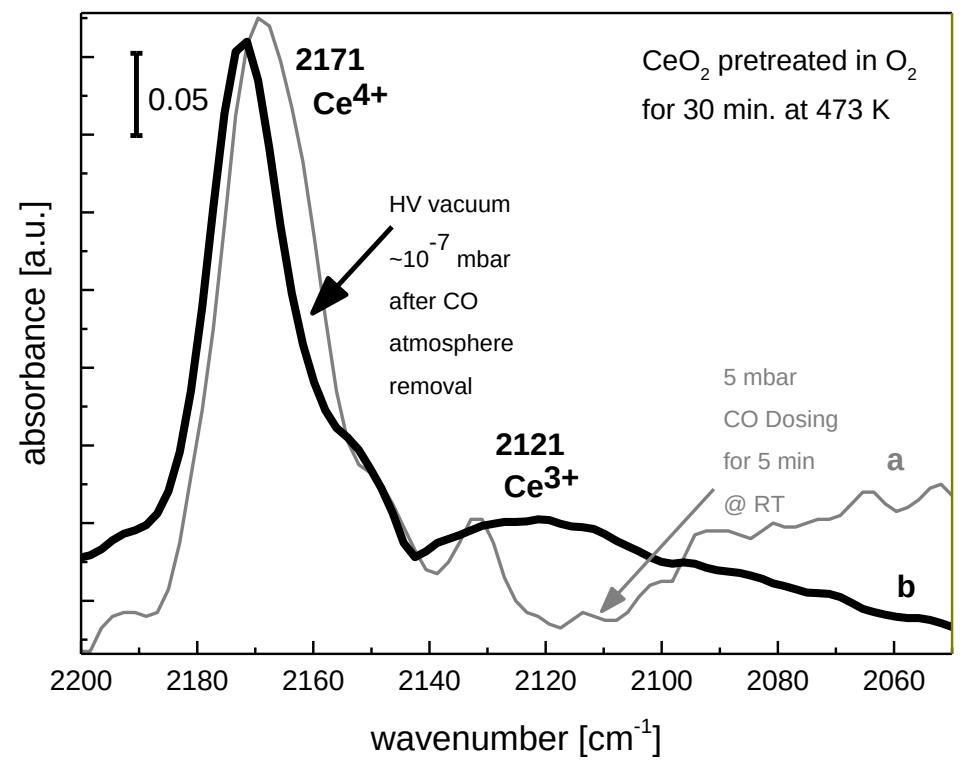

*Figure 23 $CeO_2$ is pretreated in oxygen atmosphere and dosed with 5mbar CO at 473 K. a (grey line) shows the adsorption spectrum after 10 minutes in CO atmosphere. b (black line) shows the adsorption spectrum after removal of the CO atmosphere in HV vacuum.*

This subsection comments upon the OH and carbonate regions of the FT-IR spectra for the support material $CeO_2$ under oxidative atmosphere (Figure 24 and Figure 25). Figure 25 denotes the carbonate region (the area from 1700 $cm^{-1}$ to 1300 $cm^{-1}$ is enlarged), and Figure 25 contains the OH region from 3800 $cm^{-1}$ to 3400 $cm^{-1}$.

Figure 24 shows the FT-IR data for the support material $CeO_2$ pretreated in 100 mbar oxygen at 473 K. The sample was exposed to a 5 mbar CO atmosphere for 10 minutes. The peaks were found at 1548 $cm^{-1}$, 1461 $cm^{-1}$, 1400 $cm^{-1}$ and 1303 $cm^{-1}$. These peaks appear, when $Ce^{3+}$ cations or oxygen vacancies are formed during oxidation of CO to carbonate species.

The broad peak at 1548 $cm^{-1}$ appears when high intensities are measured for C – O stretching modes of formates. For the sake of simplicity, formates are covering the region from 1500 $cm^{-1}$ to 1600 $cm^{-1}$, as the following numbers are identified as formats in Table_FT-IR (Appendix 8.1): 1548 $cm^{-1}$, 1461 $cm^{-1}$ and 1400 $cm^{-1}$. These peaks are clearly visible and dominate the IR spectrum.

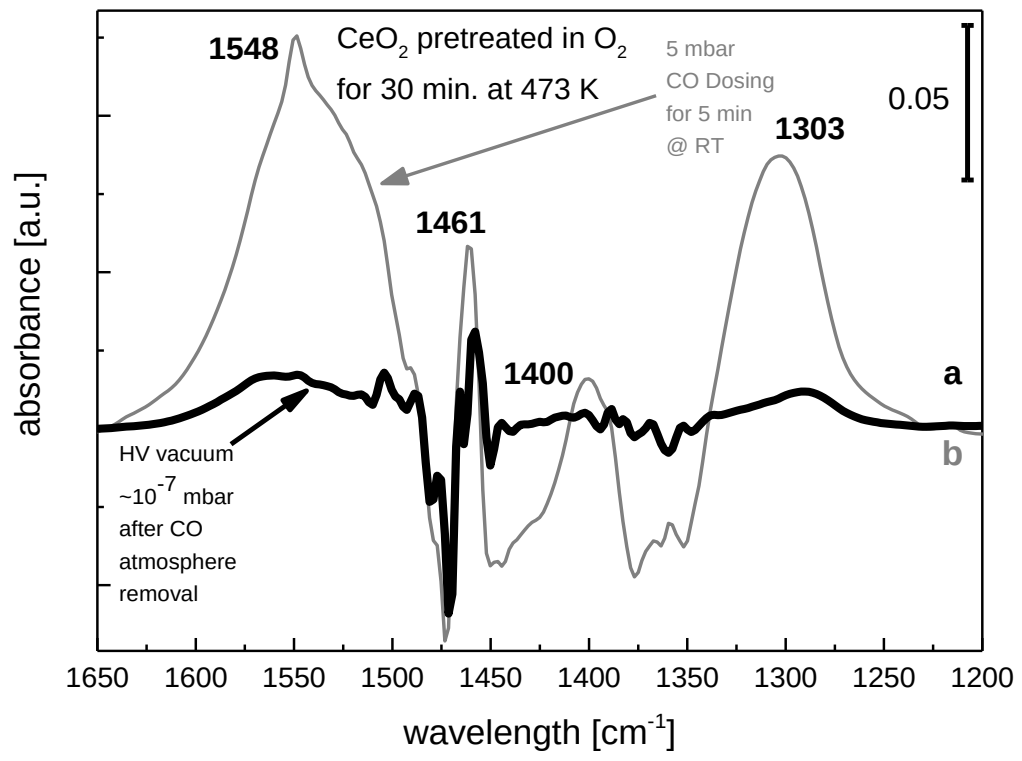

*Figure 24 Carbonates (1650 $cm^{-1}$ – 1200 $cm^{-1}$) region of $CeO_2$ at 473 K are shown. a (grey line) shows the adsorption the adsorption spectrum after 10 minutes in CO atmosphere. b (black line) shows the adsorption spectrum after removal of the CO atmosphere in HV vacuum.*

The different areas are clearly identifiable as the group of tridentate carbonate formed at the $CeO_2$ particle edges (1500 $cm^{-1}$–1451 $cm^{-1}$), bidentate carbonate (1530 $cm^{-1}$–1620 $cm^{-1}$), C-O-C bridged polycarbonates region is from 1400 $cm^{-1}$ to 1350 $cm^{-1}$ and monodentate carbonate from 1300 $cm^{-1}$ to 1370 $cm^{-1}$. In comparison, bands are visible for the monodentate and bidentate carbonates from a sample chamber filled with a 5 mbar CO atmosphere, as shown in Figure 24. A significant change was measured for a broad peak at 1400 $cm^{-1}$. This corresponds exactly to the wavenumber, where hydrogen carbonates appear.

The OH region for support material $CeO_2$ is shown in Figure 25 (a) and (b).

The bridged OH groups appear in the wavelength region lower than 3700 $cm^{-1}$. The first two peaks at 3712 $cm^{-1}$ and 3658 $cm^{-1}$ were detected after oxidative treatment. They are an indicator for OH species on the catalyst's surface.

The intense bands for terminal OH bands also appear in the wavelength region lower than 3700 $cm^{-1}$. There, the bridged broad bands of O- H stretching are observed in the wavelength region below 3600 cm$^{-1}$. This is an indication that the oxidative pretreatment processes lead to absorption of significant amount of water.

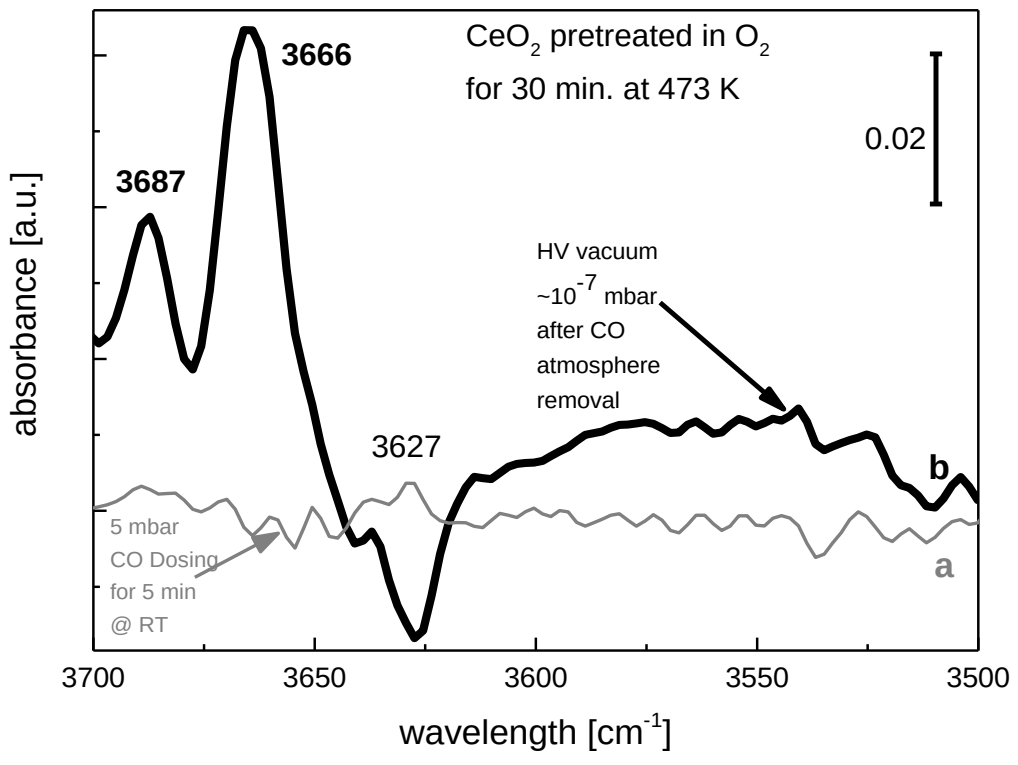

*Figure 25 OH – Stretching (3700 cm$^{-1}$ – 3500 cm$^{-1}$) regions of $CeO_2$ at 473 K are shown. a (grey line) shows the adsorption spectrum after 10 minutes in CO atmosphere. b (black line) shows the adsorption spectrum after removal of the CO atmosphere in HV vacuum.*

For the IR spectra of the support material $CeO_2$ pretreated in Nitrogen in Figure 26, a peak is found at 2157 cm$^{-1}$, which corresponds to CO on $Ce^{4+}$ cations.

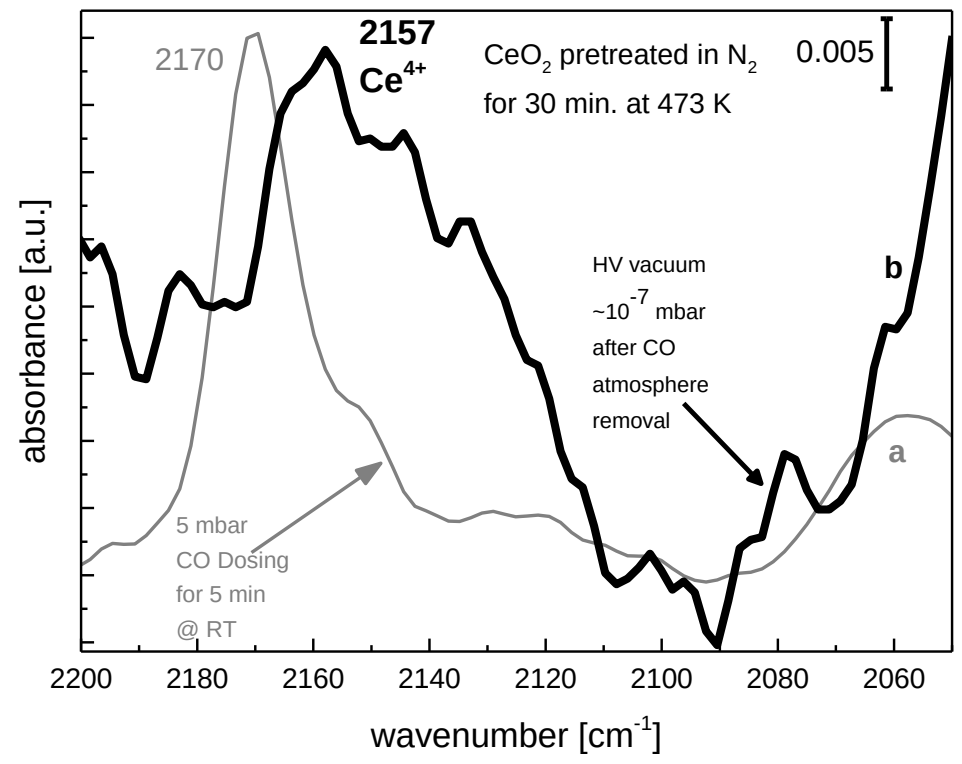

*Figure 26 $CeO_2$ is pretreated in inert atmosphere and dosed with 5mbar CO at 473 K. a (grey line) shows the adsorption spectrum after 10 minutes in CO atmosphere. b (grey line) shows the adsorption spectrum after removal of the CO atmosphere in HV vacuum.*

This subsection comments upon the OH and carbonate regions of the FT-IR spectra of the support material $CeO_2$ in inert atmosphere (Figure 27 and Figure 28). Figure 27 denotes the carbonate region and Figure 28 contains the OH region, the area from 3800 cm$^{-1}$ to 3400 cm$^{-1}$. Figure 27 shows the Carbonate region from 1650 cm$^{-1}$ to 1200 cm$^{-1}$ for $CeO_2$ catalyst pretreated at 473 K in a 900 mbar inert atmosphere.

In this atmosphere, the dominating OH groups are from the carbonate band.

1484 $cm^{-1}$ indicates a residual carbonate structure while hydrogen carbonates are at 1401 $cm^{-1}$.

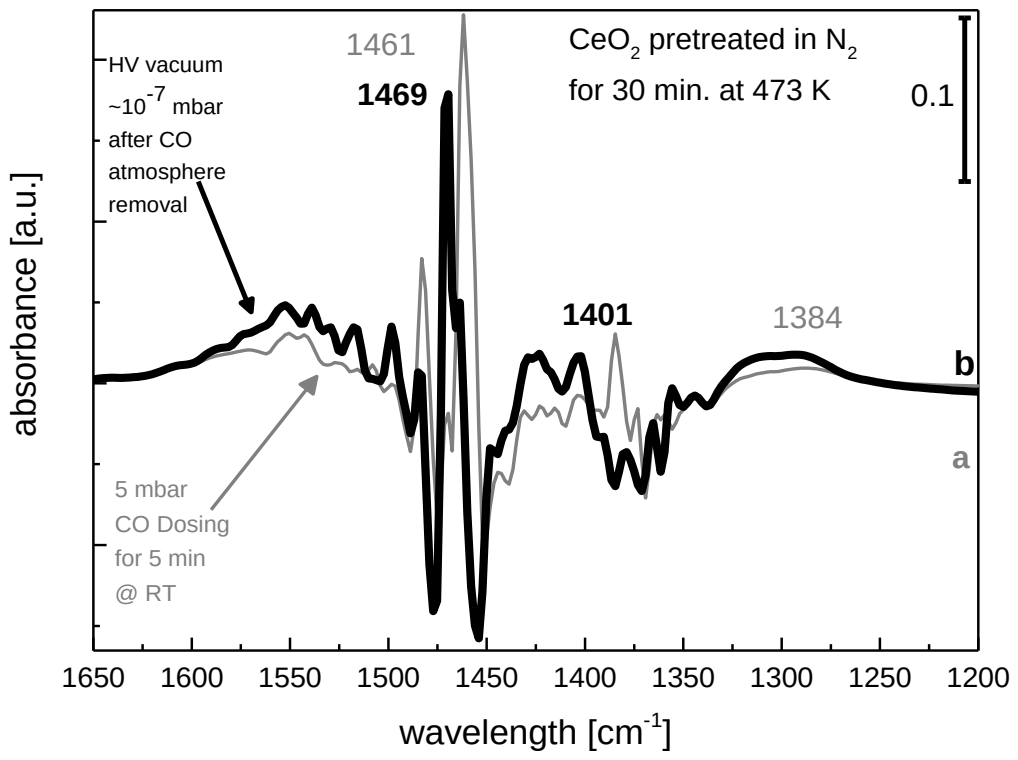

*Figure 27 Carbonates (1650 $cm^{-1}$ – 1200 $cm^{-1}$) regions of $CeO_2$ are shown. a (grey line) shows the adsorption spectrum after 10 minutes in CO atmosphere. b (black line) shows the adsorption spectrum after removal of the CO atmosphere in HV vacuum.*

The FT-IR spectrum trend is similar to the results in Figure 25 and Figure 24, as the peaks are not very intense and clear, but still the assignment to the same wavenumbers are justifiable. After smoothing the curve, the trend becomes more obvious as hydrogen carbonates, C – H stretching and hydroxyl containing species are found on the catalyst's surface.

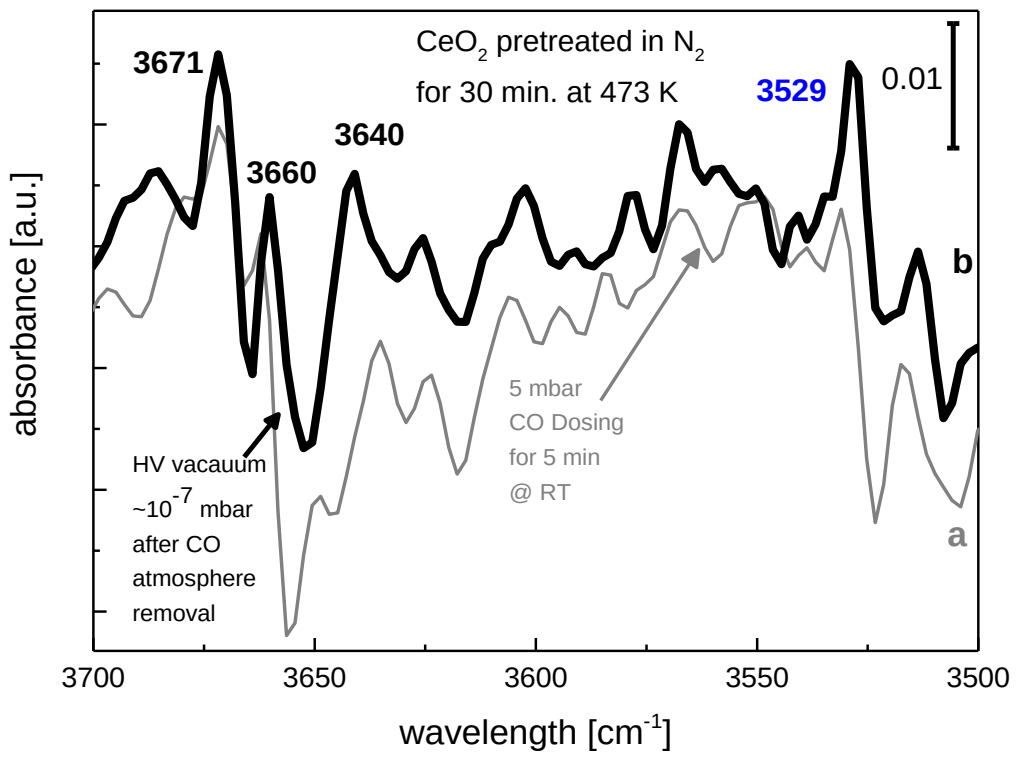

*Figure 28 OH stretching (3700 $cm^{-1}$ – 3500 $cm^{-1}$) regions of $CeO_2$ are shown. a (grey line) shows the adsorption spectrum after10 minutes in CO atmosphere. b (black line) shows the adsorption spectrum after removal of the CO atmosphere in HV vacuum*

### 4.2.2 FOURIER – TRANSFORMATION SPECTROSCOPY OF GOLD NANOCLUSTERS $Au_{40}(SR)_{24}/CeO_2$ AFTER OXYGEN PRETREATMENT

Figure 29 show the three main peaks. The first peak is at 2152 $cm^{-1}$, which corresponds to CO molecules bounded on $Au^{\delta+}$ ($0 < \delta < 1$) sites. At lower temperatures, CO bound on $Ce^{3+}$ is measured at 2123 $cm^{-1}$.

After pretreating the gold nanocluster in oxygen at lower temperatures from 373 K to 525 K, with the maximum of the peak measured at 525 K, CO adsorbed on $Au^{\delta-}$ ($0 < \delta < 1$) sites is found at 2073 $cm^{-1}$.

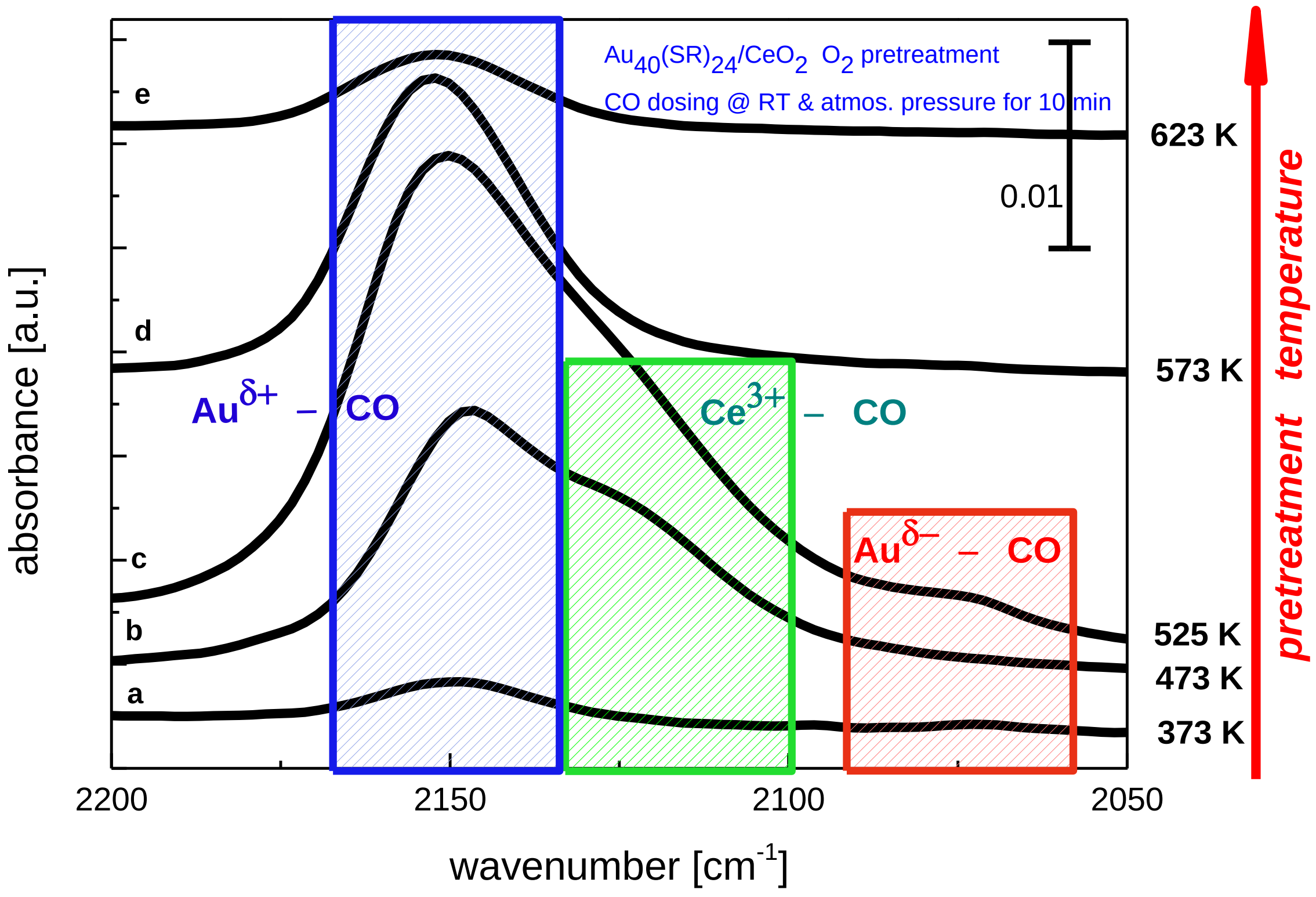

*Figure 29 $Au_{40}(SR)_{24}/CeO_2$ CO adsorption spectra, after oxygen pretreatment, are shown. The red arrow on the right side shows the increasing pretreatment procedure, after dosing the catalyst with 5 mbar CO for 10 minutes at room temperature and atmospheric pressure. At lower temperatures (spectrum a and b), the $Ce^{3+}$, $Au^{-}$and $Au^{\delta+}$- sites increase. The adsorption spectrum, at 525 K and higher shows Au $^{\delta+}$sites (spectrum c, d and e).*

### 4.2.3 OH AND CARBONATES REGION FOR $Au_{40}(SR)_{24}/CeO_2$ PRETREATED IN OXYGEN

In Figure 30, spectra are taken after every temperature starting from 373 K and increasing up to 623 K. The gold nanocluster catalyst was pretreated in 100 mbar oxygen.

The appearance of species of carbonates on the gold nanocluster catalyst $Au_{40}(SR)_{24}/CeO_2$ is confirmed from peaks at 1655 $cm^{-1}$ to 1200 $cm^{-1}$, as shown in Figure 30.
At the lowest temperature, bidentate (1562 $cm^{-1}$), monodentate (1523 $cm^{-1}$, 1462 $cm^{-1}$, 1361 $cm^{-1}$ and 1309 $cm^{-1}$) and hydrogen carbonate (1393 $cm^{-1}$) can be seen.
When increasing the temperature by about 50 K, the monodentate shows carbonates growth, and the hydrogen carbonate peak at 1393 $cm^{-1}$ gets more pronounced, while new bands at 1614 $cm^{-1}$ and 1216 $cm^{-1}$ start to grow slowly. At 423 K, the monodentate and bidentate carbonate bands grow to wide peaks, whereas the hydrogen carbonates at 1614 $cm^{-1}$, 1216 $cm^{-1}$ and 1393 $cm^{-1}$ turn into sharp peaks. The same is observed for the 1361 $cm^{-1}$ and 1467 $cm^{-1}$ monodentate carbonate peaks.
As featured in the FT-IR spectrum at 473 K, a peak at 1467 $cm^{-1}$ is observed, and from 1650 $cm^{-1}$ to 1550 $cm^{-1}$, a wide peak in the hydrogen and bidentate carbonate regions is found. From 1300 $cm^{-1}$ to 1370 $cm^{-1}$, a small rise is identifiable in the monodentate region.
At 525 K, where the thiolate ligands are completely removed, in the segment from 1650 $cm^{-1}$ to 1517 $cm^{-1}$, a rising peak differentiates this spectrum from all other FT-IR spectra. This area includes the hydrogen, bidentate and monodentate carbonates, with a shoulder at 1523 $cm^{-1}$. At higher temperatures, such as 573 K and 623 K, this wide peak splits into three peaks, as evidenced by the hydrogen (1617 $cm^{-1}$), bidentate (1565 $cm^{-1}$) and monodentate (1523 $cm^{-1}$) carbonates. Around 1477 $cm^{-1}$, another peak rises and remains as the temperature increases. At 1392 $cm^{-1}$, another peak grows higher, and new peaks form at 1361 $cm^{-1}$ and 1347 $cm^{-1}$ in the monodentate carbonate region (1300 $cm^{-1}$ to 1370 $cm^{-1}$). The broad peak at 1309 $cm^{-1}$ shifts to 1288 $cm^{-1}$ at higher temperatures.

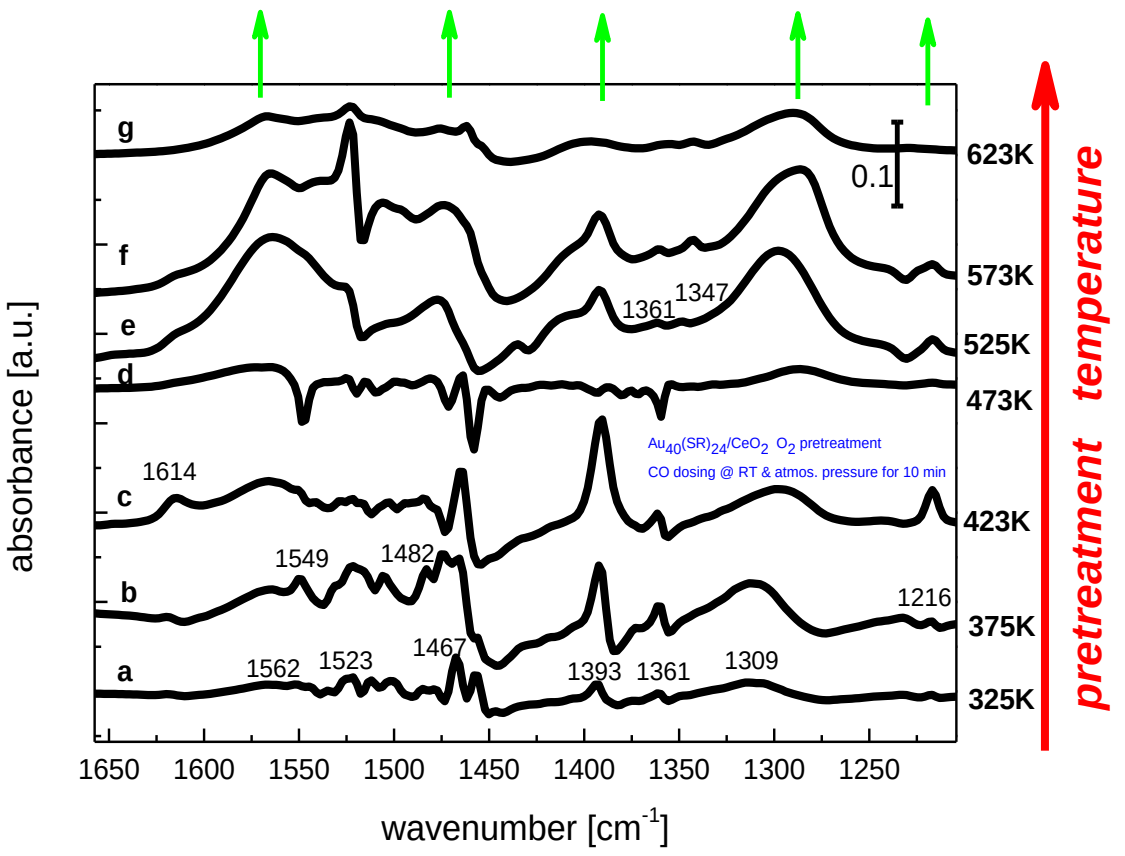

*Figure 30* *$Au_{40}(SR)_{24}/CeO_2$ CO adsorption spectra in the carbonates region (1650 cm$^{-1}$–1200 cm$^{-1}$), after oxygen pretreatment, are shown. The red arrow on the right side shows the increasing pretreatment procedure, after dosing the catalyst with 5 mbar CO for 10 minutes at room temperature and atmospheric pressure.*

Figure 31 displays the OH region. The FT -IR spectrum below 3600 cm$^{-1}$ exhibits less interaction among intense OH bands. This is an indication that water molecules are not interacting via the hydrogen bond. On the other hand, the bands above 3600 cm$^{-1}$ (>3700 cm$^{-1}$) are visible and more intense throughout the increasing temperature at each measurement.

The bridged (<3700 cm$^{-1}$) OH groups are more pronounced for temperatures up to 423 K and have an intense peak at 3716 cm$^{-1}$. At lower and higher temperatures, the peaks at 3741 cm$^{-1}$ and 3737 cm$^{-1}$ disappear and re-appear respectively. This is a sign that molecularly and dissociatively adsorbed water (Figure 31) is present in the form of hydroxyl molecules under reaction conditions, as no water was added. The OH groups represent the major surface species at temperatures up to 423 K, the temperature at which the thiolates are partially removed from the $Au_{40}(SR)_{24}/CeO_2$ gold nanocluster catalysts. From 473 K to 523 K, except at the intense peak at 3671 cm$^{-1}$, the OH groups disappear and form again in the area from 573 K to 623 K at the gold nanocluster surface.

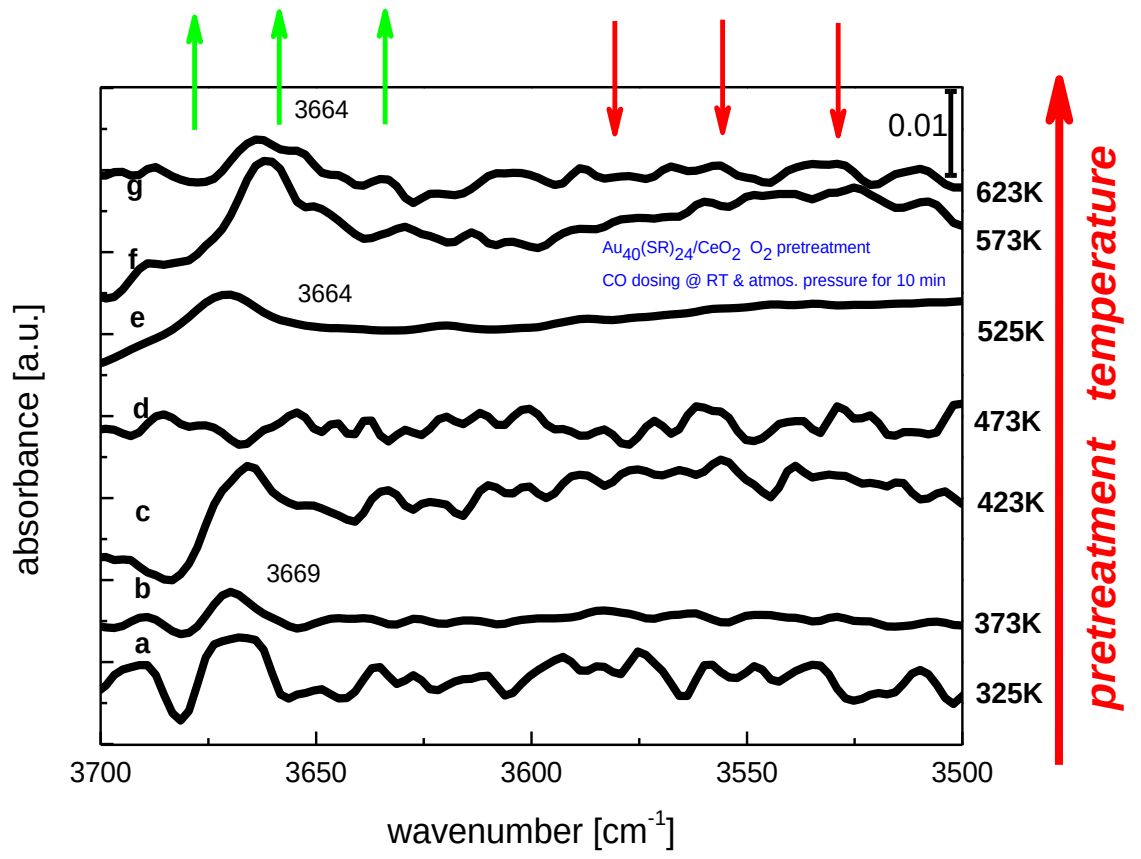

*Figure 31* *$Au_{40}(SR)_{24}/CeO_2$ CO adsorption spectra in the OH region (3700 $cm^{-1}$ – 3500 $cm^{-1}$), after oxygen pretreatment, is shown. The red arrow on the right side shows the increasing pretreatment procedure, after dosing the catalyst with 5 mbar CO for 10 minutes at room temperature and atmospheric pressure.*

### 4.2.4 FOURIER – TRANSFORMATION SPECTROSCOPY OF GOLD NANOCLUSTERS $Au_{40}(SR)_{24}/CeO_2$ IN NITROGEN

Figure 32 shows the FT-IR spectra of the $Au_{40}(SR)_{24}$ gold nanocluster catalysts pretreated in an inert atmosphere ($N_2$). The differences between the six spectra correspond to the various pretreatment temperatures from room temperature to 673 K. In Figure 32, only the most significantly changing spectra are displayed from 423 K to 673 K to illustrate this effect. The highest peak at 2173 $cm^{-1}$ corresponds to CO adsorbed on $Ce^{4+}$ sites.

The literature value is in the region from 2191 $cm^{-1}$ to 2188 $cm^{-1}$. The bands at 2150 $cm^{-1}$, 2130 $cm^{-1}$ and at higher temperatures at 2142 $cm^{-1}$ correspond to CO molecules adsorbed on $Au^{\delta+}$ sites, which according to the literature is in the region between 2152 $cm^{-1}$ and 2127 $cm^{-1}$. The CO molecules are adsorbed on $Au^{\delta-}$ sites after inert treatment, because the peak at 2078 $cm^{-1}$ increases with rising temperatures.

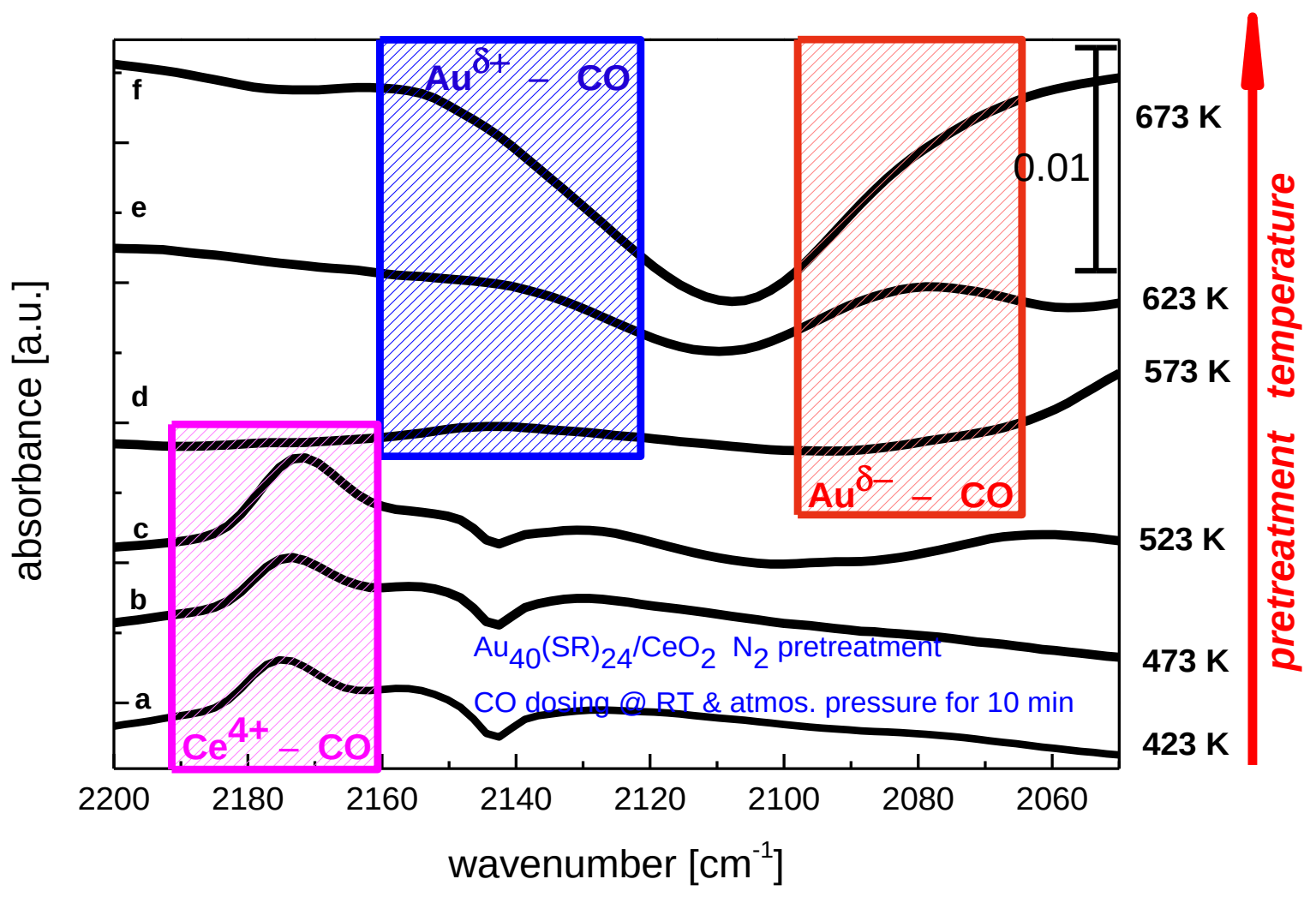

*Figure 32 $Au_{40}(SR)_{24}/CeO_2$ CO adsorption spectra, after nitrogen pretreatment, is shown. The red arrow on the right side shows the increasing pretreatment procedure, after dosing the catalyst with 5 mbar CO for 10 minutes at room temperature and atmospheric pressure. At lower temperatures (spectrum a, b and c), the $Au^-$ sites are active. The adsorption spectrum, at 573 K and higher, shows broad peaks of $Au^{\delta+}$ and $Au^-$ sites (spectrum d, e and f).*

### 4.2.5 OH AND CARBONATES REGION FOR $Au_{40}(SR)_{24}/CeO_2$ PRETREATED IN NITROGEN

Figure 33 contains the carbonates region (1750 $cm^{-1}$ to 1200 $cm^{-1}$) for $Au_{40}(SR)_{24}/CeO_2$ pretreated in a 900 mbar inert atmosphere. At the lowest temperature, 373 K, peaks are found at 1677 $cm^{-1}$, which corresponds to the bridged bidentate carbonates, and at 1614 $cm^{-1}$ hydrogen carbonate is detected on the surface. Throughout the temperature increase from 373 K to 673 K in 50 K steps, in the bidentate region (1530 $cm^{-1}$–1620 $cm^{-1}$), the peak growth differs in that the maximum jumps from 1550 $cm^{-1}$ to 1521 $cm^{-1}$ at 423 K, grows at 1550 $cm^{-1}$ and shifts to 1544 $cm^{-1}$.

Finally, this peak maximum occurs at 1565 $cm^{-1}$ while growing again by a wide margin without a certain maximum into the bidentate region.

Uncoordinated carbonate ions are found in the segment between 1470 $cm^{-1}$ and 1420 $cm^{-1}$. The shift in their peaks depends on the temperature. Significant changes are measured for the monodentate carbonate region (1530 $cm^{-1}$ – 1470 $cm^{-1}$ and 1300 $cm^{-1}$ – 1370 $cm^{-1}$). Of interest is the growth of the hydrogen carbonate peak at 1216 $cm^{-1}$, which exists at 373 K and disappears during the temperature rise but emerges again at 573 K and 673 K.

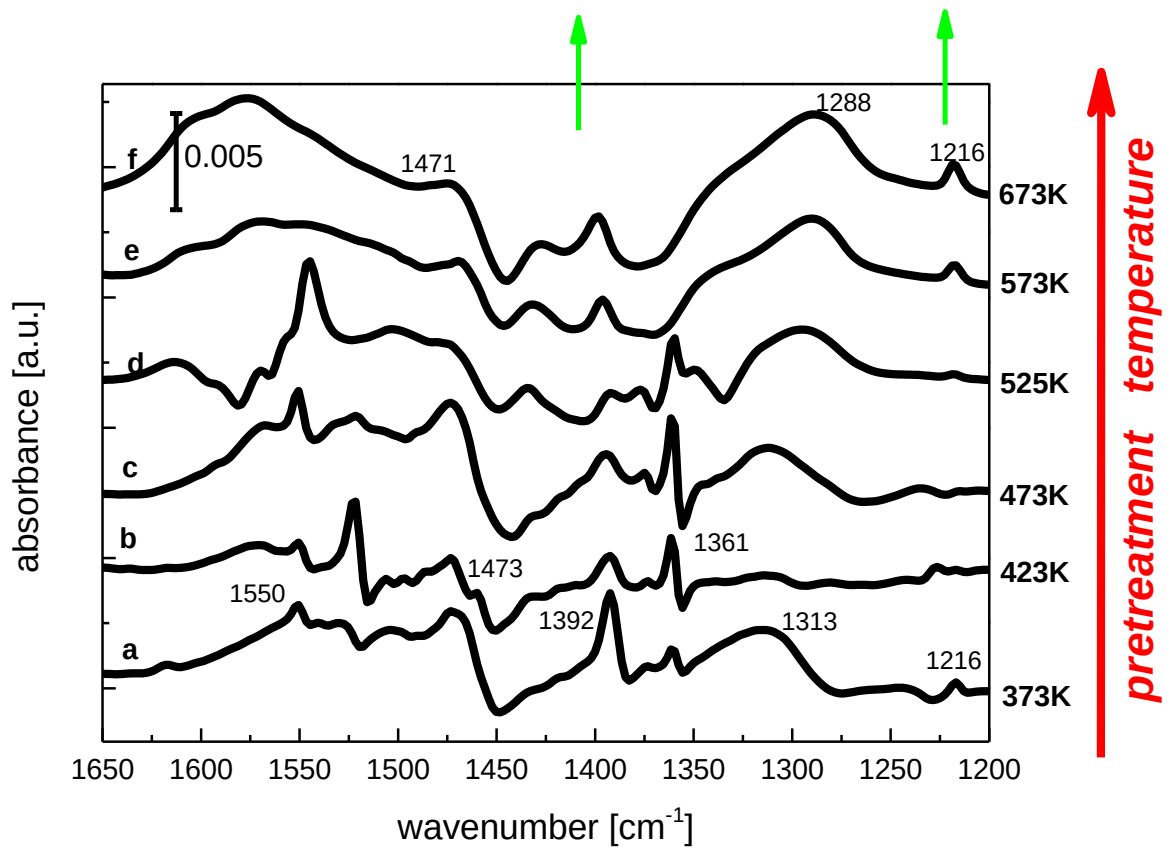

*Figure 33 Au$_{40}$(SR)$_{24}$/CeO$_2$ CO adsorption spectra in the carbonates region (1650 cm$^{-1}$–1200 cm$^{-1}$), after nitrogen pretreatment, is shown. The red arrow on the right side shows the increasing pretreatment procedure, after dosing the catalyst with 5 mbar CO for 10 minutes at room temperature and atmospheric pressure.*

In Figure 34 , the OH region is plotted from 3750 cm$^{-1}$ to 3500 cm$^{-1}$. The curve trend, depending on the temperature change, implies the existence of a characteristic terminal (<3700 cm$^{-1}$) and bridged (>3700 cm$^{-1}$) OH regions after an inert pretreatment procedure. At lower temperatures, the peak assignment is easier, since the visibility of the peaks is rather obvious. As temperature increases to 473 K, these defined peaks unbend and turn into a large peak with a minor shift of the peak maximum.

For the wavelength region >3700 cm$^{-1}$, the one single peak at 3665 cm$^{-1}$ shifts and grows to a broad band peak. After heating the gold nanocluster catalysts to 473 K, the peak vanishes and becomes more undefined (Figure 34).

Below 3600 cm$^{-1}$, a broad peak is again observed. The characteristic temperature at which the peak disappears starts already from 525 K. For the region from 3760 cm$^{-1}$ to 3450 cm$^{-1}$, the OH species are shown in Figure 34 for Au$_{40}$(SR)$_{24}$/CeO$_2$ pretreated in a 900 mbar inert atmosphere.

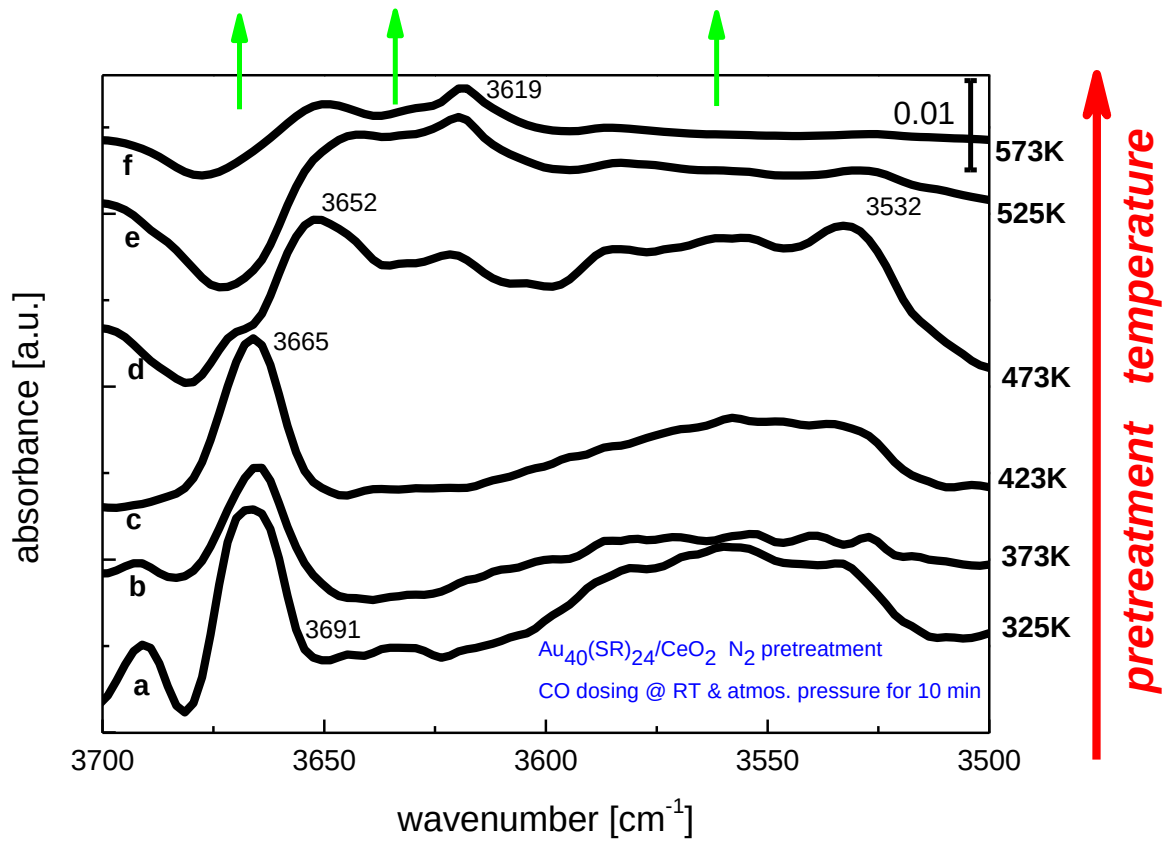

*Figure 34 Au$_{40}$(SR)$_{24}$/CeO$_2$ CO adsorption spectra in the OH region (3700 cm$^{-1}$–3500 cm$^{-1}$), after nitrogen pretreatment, is shown. The red arrow on the right side shows the increasing pretreatment procedure, after dosing the catalyst with 5 mbar CO for 10 minutes at room temperature and atmospheric pressure.*

### 4.2.6 COMPARISON BETWEEN GOLD NANOCLUSTERS Au$_{40}$(SR)$_{24}$/CeO$_2$ PRETREATED IN NITROGEN AND SUPPORT MATERIAL CeO$_2$ PRETREATED IN NITROGEN

Comparison between gold nanoclusters Au$_{40}$(SR)$_{24}$ supported on CeO$_2$ with the FT – IR results of Ceria, shows that Ce$^{3+}$ sites are active in both FT – IR spectra. The Ce$^{3+}$ shoulder at 473 K after oxidative pretreatment results from the thermal dehydration of the support material. This coincides with the TGA results perfectly as described there in detail.

The intensity for the Ce$^{3+}$ band related to CO absorbed on the support simultaneously increases with increasing pretreatment temperature and with the increasing intensity of Au$^{\delta-}$.

On the one hand, cationic gold adsorption sites such as Au$^{\delta-}$ are exposed at the catalysts surface interacting with Ce$^{3+}$, after partially This allows an electronic transfer from the support material to the gold nanoclusters.

On the other hand, Au$^{\delta+}$ CO adsorption active sites are measured after oxidative pretreatment at 473 K, which results from the continuous removal of thiolate ligands. The can be understood, considering that thiolate ligands are electron acceptors that reduce the surface gold atoms electron deficient.

From Literature [149], it is known, that CO molecules adsorbed on transition – or precious – metal sites react with oxygen supplied by the metal – ceria interface.

In general, water dissociation takes place on an oxygen – vacancy site of ceria to form hydrogen and atomic oxygen, which reoxidizes ceria.

Formation of carbonate and carboxylate species are an indication of CO oxidation by surface species and of concomitant ceria reduction.

Li with co-workers [98] and Tabakova [149] measured this at already 90 K. Li's work observed this behavior on ceria at RT and suggested that the co – ordinately unsaturated surface sites may play a key role in the CO oxidation.

In Literature [149], know is that nanosized gold particles cause modification of the surface properties of ceria.

Ceria is reduced ($Ce^{3+}$ ) and uncoordinated sites near the very small gold clusters with
d ≤ 1 mn are produced.

Oxygen, which is adsorbed on the gold clusters and some of the surface oxygen atoms of ceria reacts, giving rise to the formation of water and oxygen vacancies and/or $Ce^{3+}$ defects on ceria. The presence of these defects allows an electron transfer from the support to the gold transfer from the support to the gold particles and to the gold clusters.

A depletion of the band at 3663 $cm^{-1}$ can be connected with the production of formate species at the borderline with $Ce^{3+}$ by increasing the temperature.

CO co-adsorption with oxygen species on gold step sites, close to the support and Ce3+ where oxygen can be with CO to from $CO_2$.

Mono- and doubly bridging OH groups on $Ce^{3+}$ [18, 26]. At lower frequencies, a weak bond related to the electronic transition of $Ce^{3+}$ located at subsurface defective lattice sites, grows up to 2127 $cm^{-1}$.

### 4.2.7 FOURIER – TRANSFORMATION SPECTROSCOPY OF GOLD NANOCLUSTERS $Au_{40}(SR)_{24}/Al_2O_3$ IN OXYGEN AND NITROGEN

For $Au_{40}(SR)_{24}/Al_2O_3$, no CO adsorption is observed (Figure 35, after oxygen pretreatment and Figure 36, after inert pretreatment), either after oxygen pretreatment or the inert pretreatment procedure. The reason for this lies in the support material, which is not accessible for the CO molecules to adsorb at the catalyst's surface.

The broad peaks in Figure 36 and Figure 35 are CO peaks in the gas phase. The peaks in Figure 35 (b) and Figure 36 (f) are very broad. Here, CO in gas phase is measured.

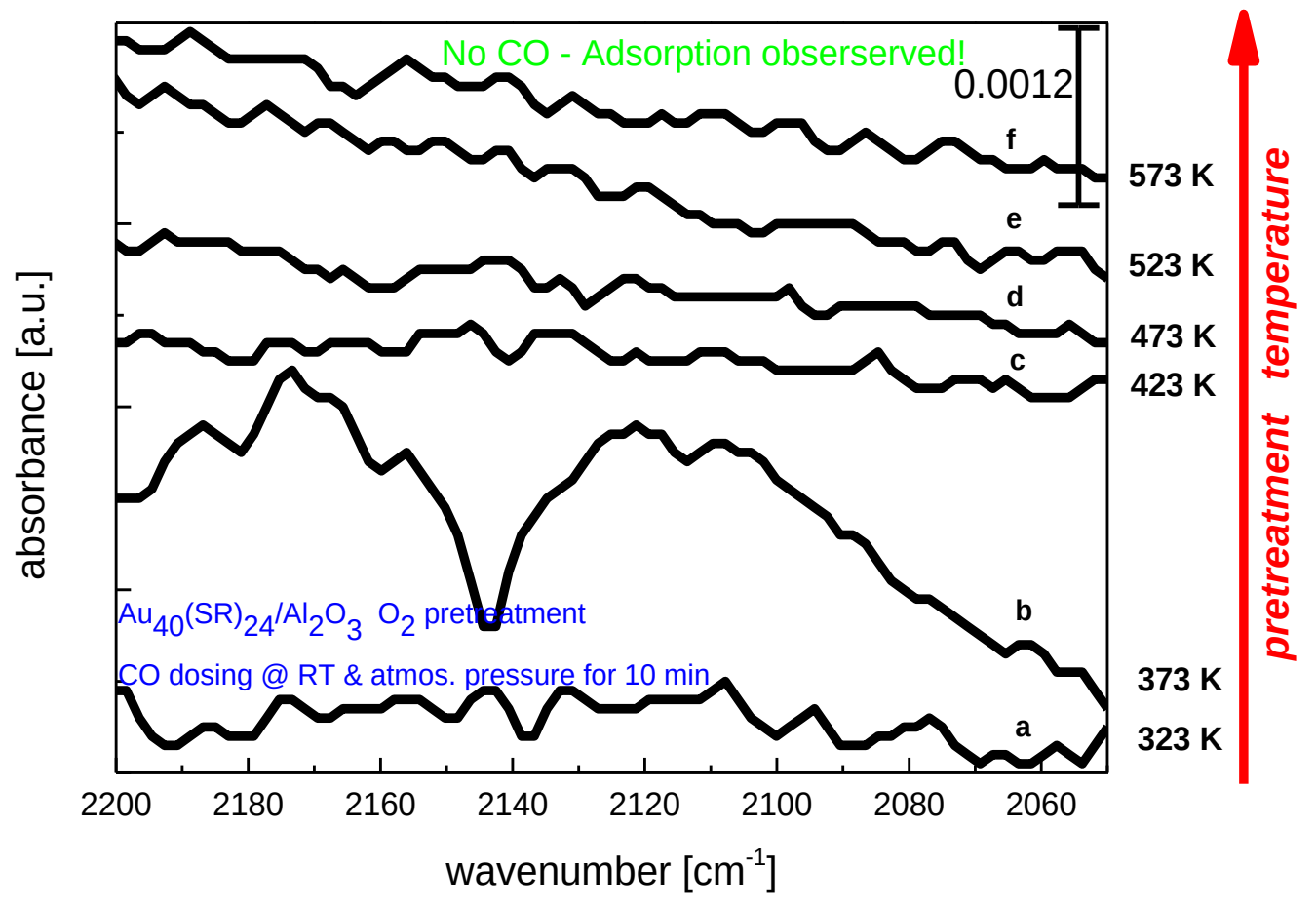

*Figure 35 Au$_{40}$(SR)$_{24}$/Al$_2$O$_3$ CO adsorption spectra, after oxidative pretreatment, is shown. The red arrow on the right side shows the increasing pretreatment procedure, after dosing the catalyst with 5 mbar CO for 10 minutes at room temperature and atmospheric pressure. No CO adsorption is observed.*

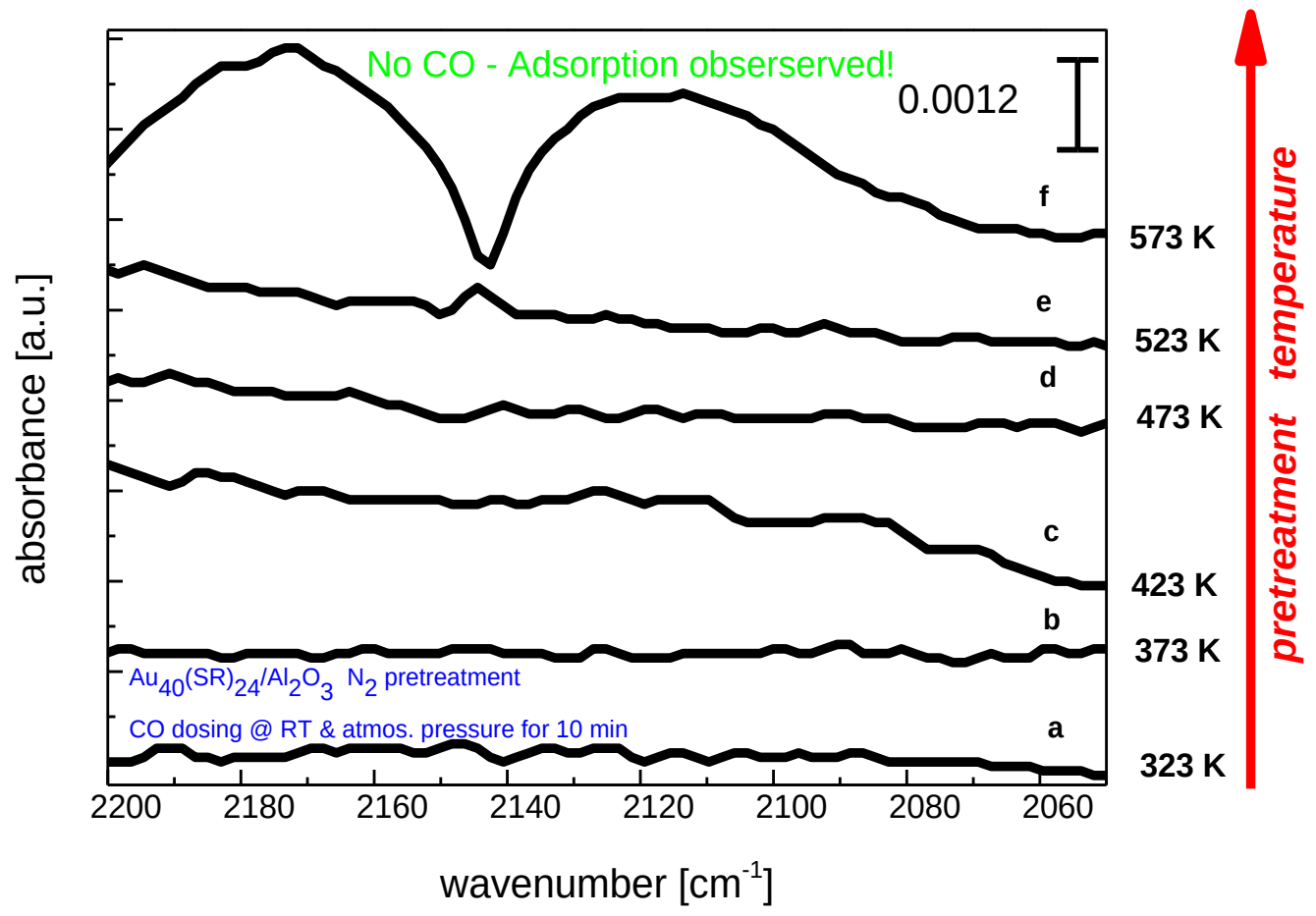

*Figure 36 Au$_{40}$(SR)$_{24}$/Al$_2$O$_3$ CO adsorption spectra, after inert pretreatment, is shown. The red arrow on the right side shows the increasing pretreatment procedure, after dosing the catalyst with 5 mbar CO for 10 minutes at room temperature and atmospheric pressure. No CO adsorption is observed.*

### 4.2.8 FOURIER – TRANSFORMATION SPECTROSCOPY OF GOLD NANOCLUSTERS $Au_{38}(SR)_{24}/CeO_2$ IN OXYGEN

This subsection contains the FT-IR CO adsorption spectra for $Au_{38}(SR)_{24}/CeO_2$ in oxidative atmosphere. The detailed pretreatment procedure is described in chapter "Materials and Methods" (subsection "FT- IR Pretreatment Procedure").

The CO adsorption results for $Au_{38}(SR)_{24}/CeO_2$ gold nanocluster catalysts are shown in Figure 37.

After pretreatment at 473K in 100 mbar oxygen and doses of CO, the peak at 2128 $cm^{-1}$ grows until saturation is reached at 5 mbar. For a low temperature test in oxygen, the CO dosing was measured in the following steps: 1 mbar, 2 mbar, 5 mbar and 10 mbar.

Subsequently, the same sample was heated to 473 K (Figure 37 (a)).

The peaks after removing the CO atmosphere can be found at 2134 $cm^{-1}$ and 2157 $cm^{-1}$, corresponding to $Au^{\delta+}$ active site and $Ce^{4+}$. The literature values for $Au^{\delta+}$ range from 2152 $cm^{-1}$ to 2127 $cm^{-1}$.

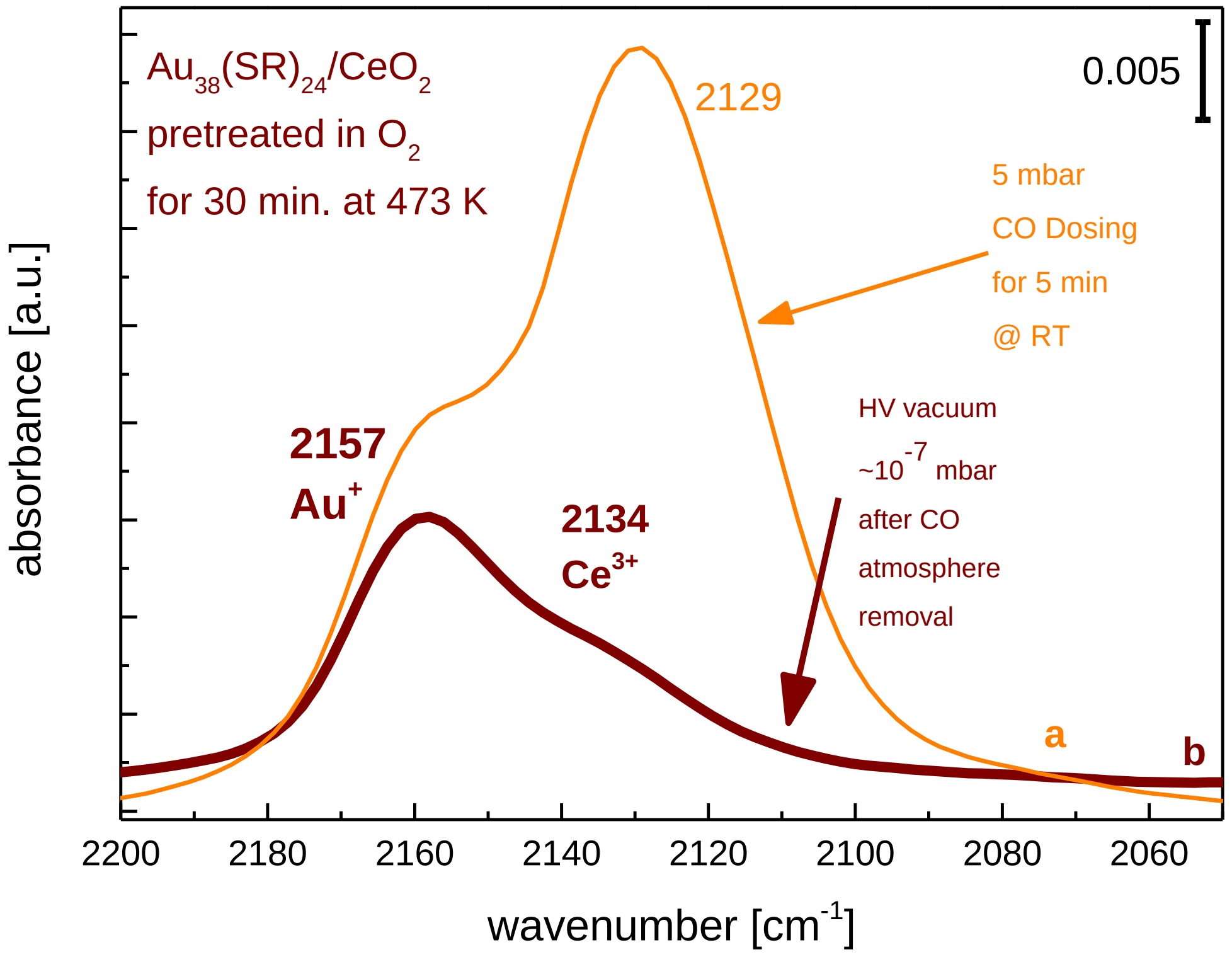

*Figure 37* *$Au_{38}(SR)_{24}/CeO_2$ is pretreated in oxygen atmosphere[3] and dosed with 5mbar CO. a (orange line) shows the adsorption spectrum after 10 minutes in CO atmosphere. b (red line) shows the adsorption spectrum after removal of the CO atmosphere in HV vacuum*

### 4.2.9 OH AND CARBONATES REGION FOR $Au_{38}(SR)_{24}/CeO_2$ PRETREATED IN OXYGEN

Under a high vacuum (~$10^{-7}$ mbar) the OH and carbonates (Figure 38 and Figure 39) show significant changes and more information about the measured system.
After pretreatment in 100 mbar oxygen at a low temperature (375K), the carbonates and OH regions (Figure 38 and Figure 39) do not exhibit any obvious change in the FT-IR spectrum. In both cases, a decrease in the FT-IR spectra is observed over the entire spectrum.

The carbonates region (Figure 38) exhibits dominating peaks at 1571 $cm^{-1}$, 1548 $cm^{-1}$ and 1301 $cm^{-1}$. These peaks are assigned to the prominent group of monodentate carbonates, which according to the literature range from 1530 $cm^{-1}$ to 1470 $cm^{-1}$.

---

[3]Figure 48 – Figure 54 were measured by Thorsten Boheme, Vienna University of Technology.

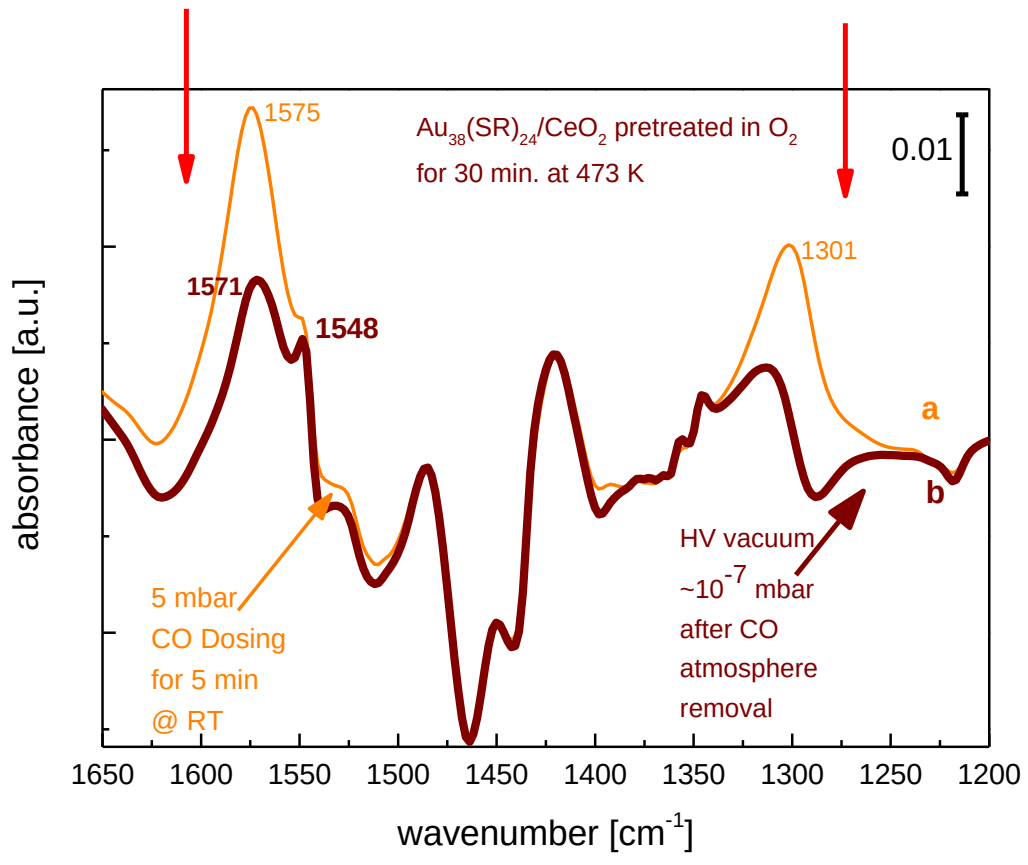

*Figure 38 The carbonates (1650 cm$^{-1}$–1200 cm$^{-1}$) region of $Au_{38}(SR)_{24}/CeO_2$ is shown. a (orange line) shows the adsorption spectrum after 10 minutes in CO atmosphere. b (red line) shows the adsorption spectrum after removal of the CO atmosphere in HV vacuum.*

In Figure 39, the OH region is shown. Bridged broadband OH stretching below 3600 cm$^{-1}$ is visible, which demonstrates that water molecules strongly interact via hydrogen bonds. The characteristic OH bands (> 3700 cm$^{-1}$) as well as terminal and bridged (<3700 cm$^{-1}$) OH bands can also be observed.

This behavior is prominent for molecularly and dissociatively adsorbed water involving hydroxyl molecules.

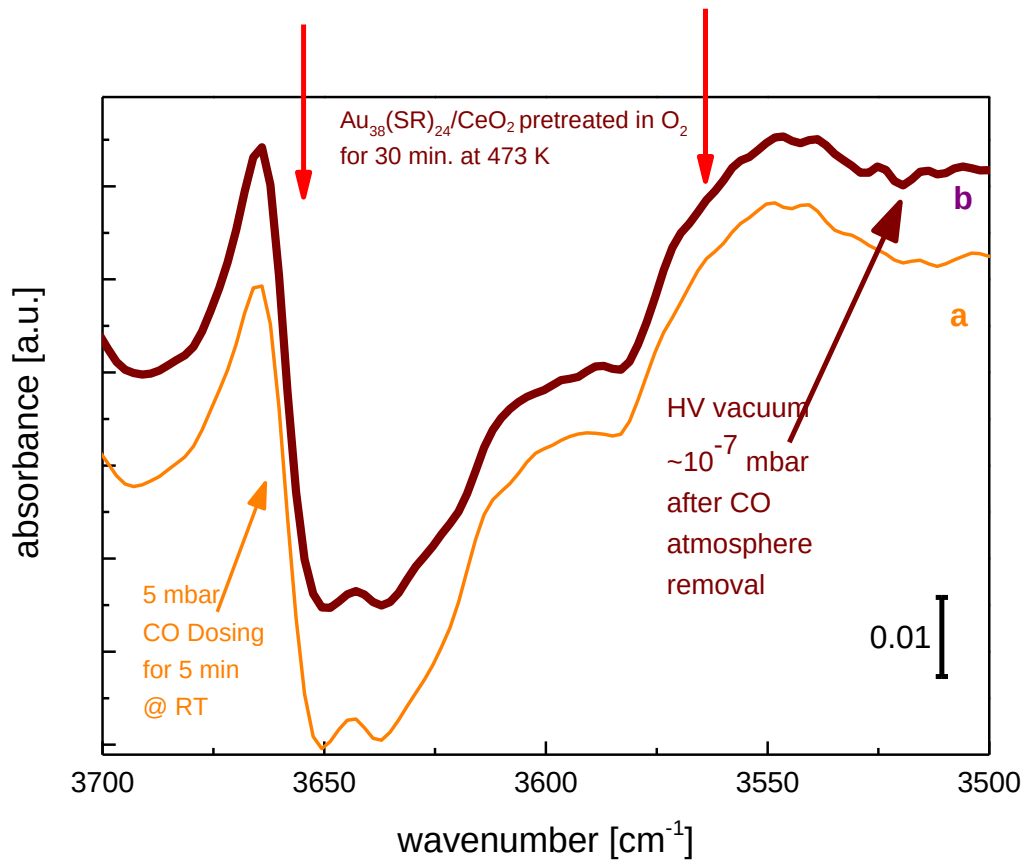

*Figure 39 The OH – stretching (3700 cm$^{-1}$–3500 cm$^{-1}$) region of $Au_{38}(SR)_{24}/CeO_2$ is shown. a (orange line) shows the adsorption spectrum after 10 minutes in CO atmosphere. b (red line) shows the adsorption spectrum after removal of the CO atmosphere in HV vacuum.*

## 4.2.10 FOURIER – TRANSFORMATION SPECTROSCOPY OF GOLD NANOCLUSTERS $Au_{38}(SR)_{24}/CeO_2$ IN NITROGEN

The gold nanocluster catalysts were tested at first after heating to 323K in a 900 mbar inert atmosphere, prior to stepwise dosing of CO into the sample chamber (1 mbar, 2 mbar, 3 mbar, 4 mbar and 5 mbar).

This yields interesting results: The evacuated spectrum at ~$10^{-7}$ mbar is displayed in Figure 40, where a dominating factor is the CO adsorption on $Ce^{3+}$ sites at 2137 $cm^{-1}$.

$Au^{\delta+}$ can be found in the wavenumber region from 2152 $cm^{-1}$ to 2127 $cm^{-1}$ according to the research of Wu et al. [48]. While this peak is barely measurable.

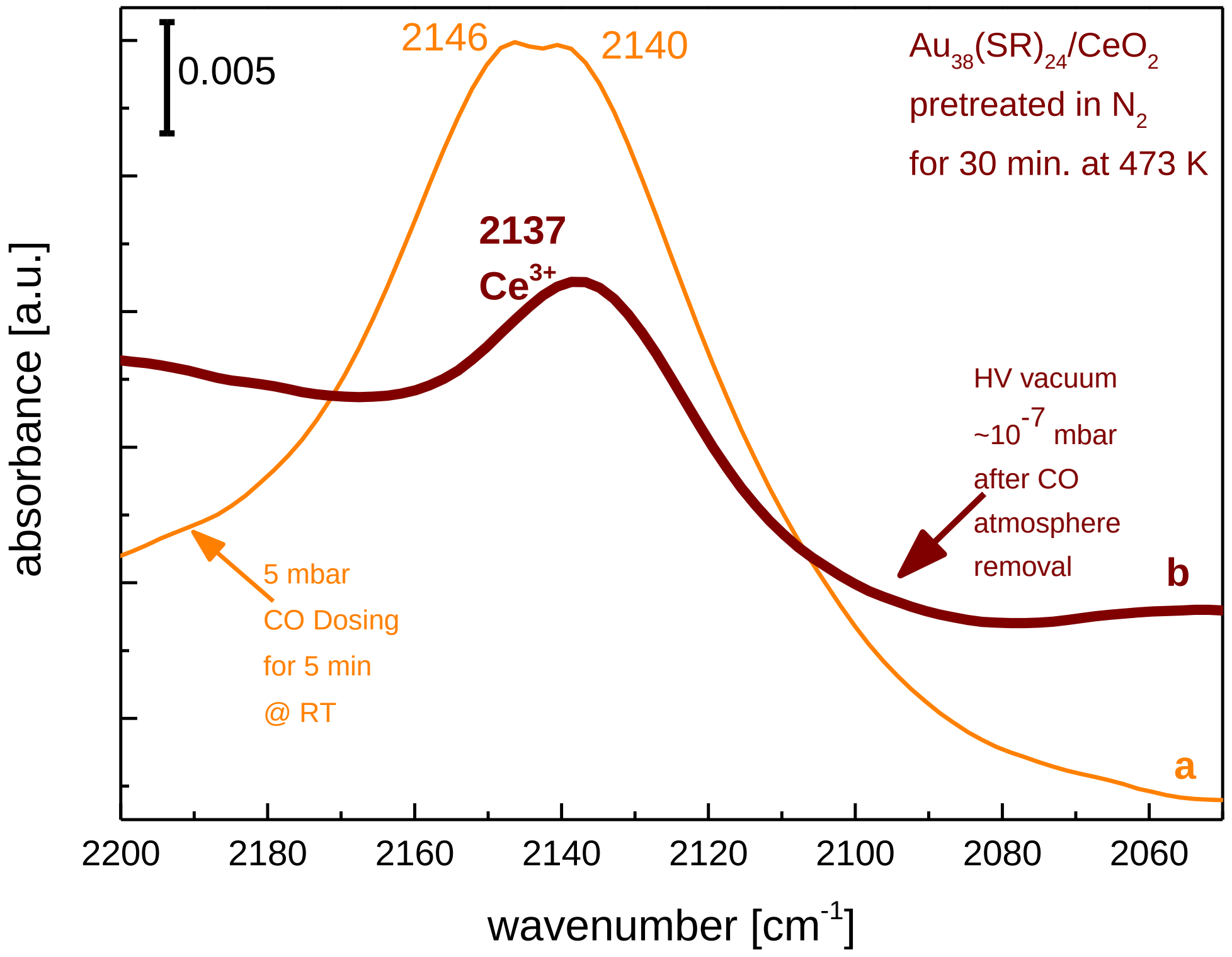

*Figure 40 $Au_{38}(SR)_{24}/CeO_2$ is pretreated in nitrogen atmosphere and dosed with 5mbar CO. a (orange line) shows the adsorption spectrum after 10 minutes in CO atmosphere. b (red line) shows the adsorption spectrum after removal of the CO atmosphere in HV vacuum.*

### 4.2.11 OH AND CARBONATES REGION FOR $Au_{38}(SR)_{24}/CeO_2$ PRETREATED IN NITROGEN

Taking the OH (Figure 42) and the carbonates (Figure 41) regions under examination reveals that for the OH region, the FT-IR spectrum after removing the CO atmosphere to a ~$10^{-7}$ mbar vacuum leads to the shift of a few peaks.

The carbonates region (Figure 41) exhibits peaks at 1523 $cm^{-1}$, 1434 $cm^{-1}$, 1396 $cm^{-1}$ and 1347 $cm^{-1}$. These peaks are assigned to the prominent group of monodentate carbonates, which according to the literature range from 1530 $cm^{-1}$ to 1470 $cm^{-1}$. With the help of the inert pretreatment procedure, a broad peak is observed throughout the range from 1700 $cm^{-1}$ to 1200 $cm^{-1}$. This indicates that water dissociates and molecularly bounds as hydroxyls.

Considering Figure 41, the carbonates in the region from 1550 $cm^{-1}$ to 1311 $cm^{-1}$ according to the literature suggest the monodentate (1530 $cm^{-1}$ to 1470 $cm^{-1}$), bi-dentate (1530 $cm^{-1}$ to 1620 $cm^{-1}$) and carboxyl groups (1400 $cm^{-1}$ to 1530 $cm^{-1}$).

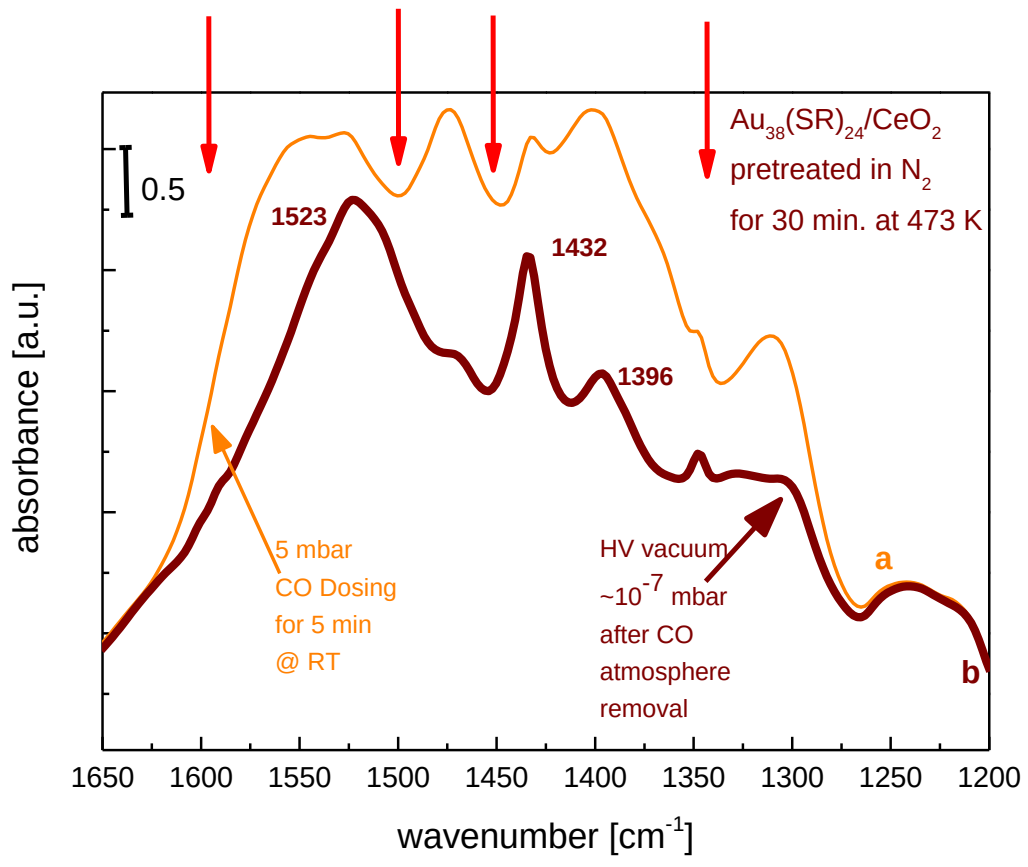

*Figure 41 The carbonates (1650 $cm^{-1}$–1200 $cm^{-1}$) region of $Au_{38}(SR)_{24}/CeO_2$ is shown. a (orange line) shows the adsorption spectrum after 10 minutes in CO atmosphere. b (red line) shows the adsorption spectrum after removal of the CO atmosphere in HV vacuum.*

In Figure 42, this idea emerges more clearly, as in the region >3700 $cm^{-1}$, which corresponds to the terminal bound OH groups, the bridged bound O-H stretching is more intense than in the region <3700 $cm^{-1}$. The OH groups represent the major surface species after activation.

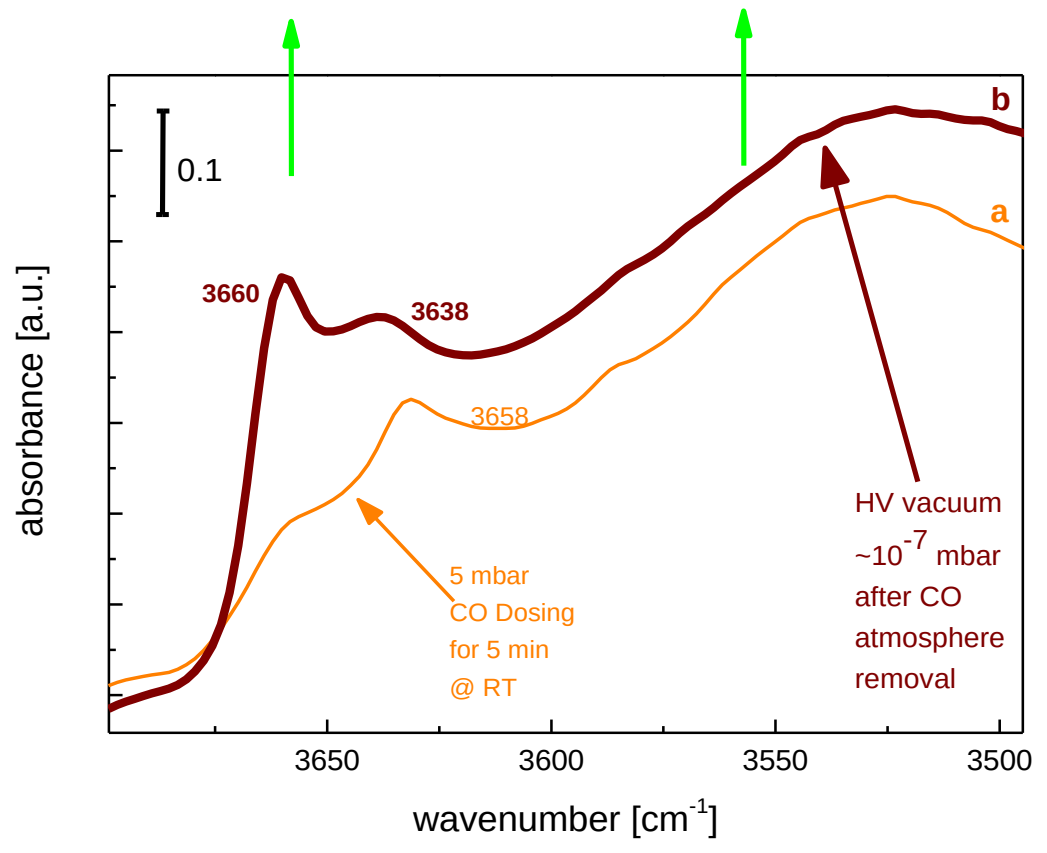

*Figure 42 The OH – stretching (3700 cm$^{-1}$–3500 cm$^{-1}$) region of $Au_{38}(SR)_{24}/CeO_2$ is shown. a (orange line) shows the the adsorption spectrum after 10 minutes in CO atmosphere. b (red line) shows the adsorption spectrum after removal of the CO atmosphere in HV vacuum.*

#### 4.2.12 COMPARISON $AU_{40}(SR)_{24}/CeO_2$ PRETREATED IN OXYGEN WITH $AU_{38}(SR)_{24}/CeO_2$ PRETREATED IN OXYGEN

Comparing the FT – IR spectra for $Au_{40}(SR)_{24}/CeO_2$ with $Au_{38}(SR)_{24}/CeO_2$ shows that $Ce^{3+}$ and $Au^{\delta+}$ (where $\delta = 1$ for $Au_{38}(SR)_{24}/CeO_2$) sites are active for CO adsorption on the catalysts surfaces.

Characteristic IR features of thiolate ligands are not prominent on room temperature treated samples, probably because they are obscured by water on the ceria surface and background noise of the experimental setup.

For $Au_{38}(SR)_{24}\backslash CeO_2$, positively charged gold sites are active and $Ce^{3+}$ sites are measured after oxygen pretreatment. At lower temperatures, after inert pretreatment procedure, $Ce^{3+}$ sites are stronger in intensity.

At lower temperatures, when these results are compared to the high temperature FT-IR tests with $Au_{40}(SR)_{24}/CeO_2$, it emerges that intensities of the active gold sites are lower. For the OH and carbonates regions, the availability of mono-, bi- and tridentate carbonates and hydrogen carbonates is higher in intensity, which indicates that water is present on the catalyst's surface. This correlates with the strong $Ce^{4+}$ and $Ce^{3+}$ peaks. In the chapter "Appendix", the IR band assignment is given in the relevant table (see Table 8 and Table 9 ).

### 4.2.13 FOURIER – TRANSFORMATION SPECTROSCOPY OF GOLD NANOCLUSTERS $Au_{40}(SR)_{24}/Al_2O_3$ IN OXYGEN AND NITROGEN

CO oxidation on the first generation of $Au_{38}(SR)_{24}/Al_2O_3$ gold nanocluster catalysts shows no CO adsorbed on the surface of the catalysts. The pretreatment procedure (in an oxygen, inert or reduced atmosphere at 473K) does not make any difference regarding CO adsorption. This argument underpins the statement that no CO adsorption is measured in Figure 43 (a) in 100 mbar oxygen, (b) 900 mbar inert, and (c) 100 mbar reduced atmospheres.

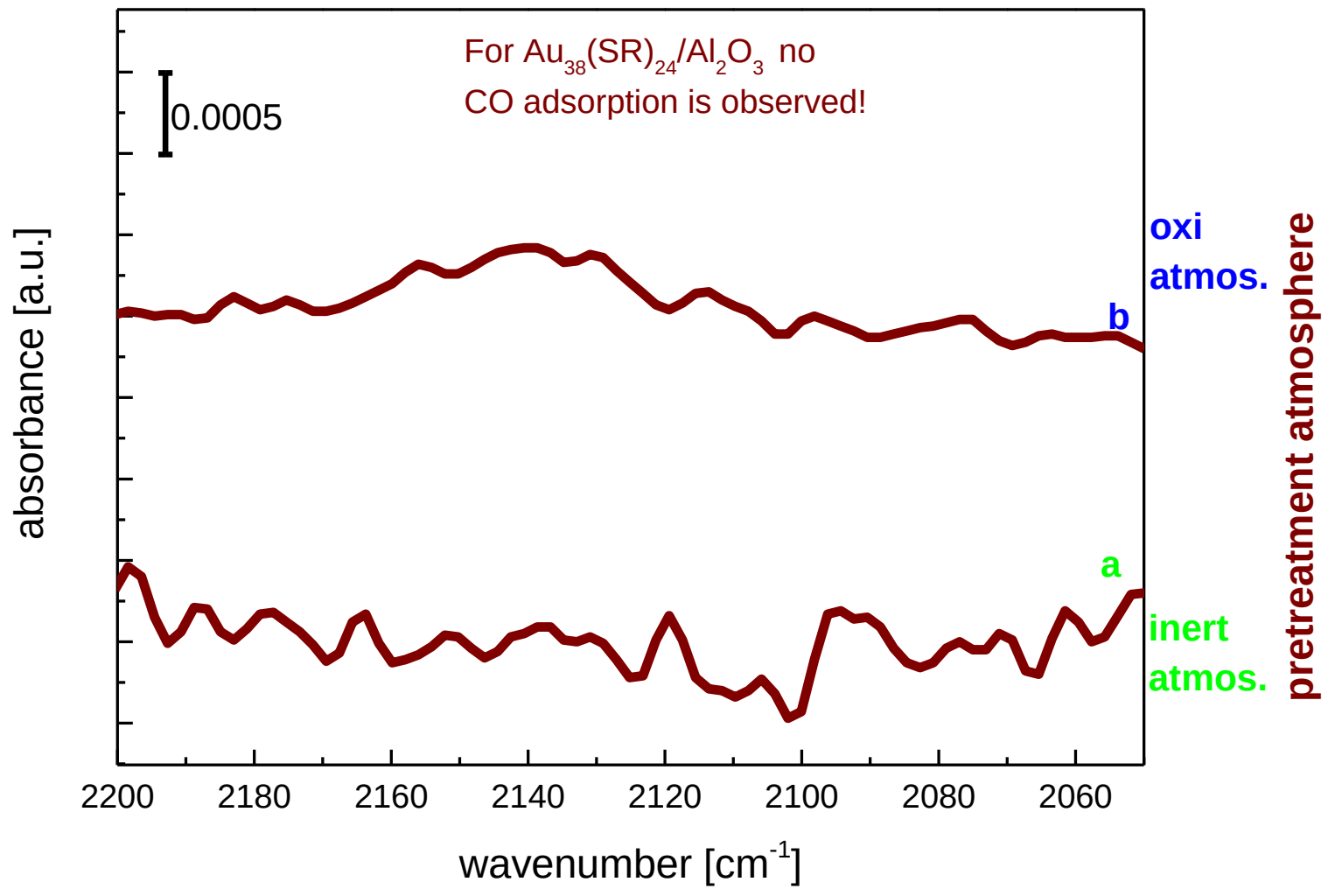

*Figure 43 $Au_{38}(SR)_{24}/Al_2O_3$ CO adsorption spectra, after oxidative (b) and inert (a) pretreatment, is shown, after dosing the catalyst with 5 mbar CO for 10 minutes at room temperature and atmospheric pressure. No CO adsorption is observed.*

Figure 43 (a) and (b) yield almost identical results; the peak remains at 2140 $cm^{-1}$ and disappears at 2148 $cm^{-1}$ and 2130 $cm^{-1}$. These peaks result from CO in the gas phase. Obviously, CO is not bound on the surface of the gold $Au_{38}(SR)_{24}/Al_2O_3$ nanocluster catalysts. After the reductive pretreatment procedure with 100 mbar hydrogen, the situation is more obvious. Peaks are found at 2175 $cm^{-1}$ and 2121 $cm^{-1}$. After removal of the CO atmosphere, they invariably disappear.

### 4.2.14 COMPARISON $Au_{40}(SR)_{24}/Al_2O_3$ WITH $Au_{38}(SR)_{24}/Al_2O_3$

For $Al_2O_3$ supported gold nanocluster catalysts, no CO adsorption was observed. This leads to the deduction that CO adsorption is weaker compared to $CeO_2$ supported gold nanoclusters, since this has been measured independent of any pretreatment atmosphere.

## 4.3 KINETIC MEASUREMENTS

CO oxidation was performed with gold nanoclusters of types 1 wt. % $Au_{40}(SR)_{24}/CeO_2$, 1 wt. % $Au_{40}(SR)_{24}/Al_2O_3$, 2 wt. % $Au_{38}(SR)_{24}/CeO_2$, 2 wt. % $Au_{38}(SR)_{24}/Al_2O_3$ and gold particles $Au/CeO_2$, $Au/Al_2O_3$.

The samples used are not commercially available and had to be prepared properly.

The aim was to pretreat the gold nanoclusters at a predetermined temperature to remove the thiolate ligands partially (by approximately 50%). TGA tests in oxygen and nitrogen determined that the correct temperature for this is 473 K.

The initial temperature was room temperature (300 K). After feeding the reactor (see Figure 14) with the reaction mixture of 10% O2 and 5% CO in He (CO: $O_2$ = 1:2), then the reaction gas was introduced into the quartz tube through the catalyst and the $CO_2$ level was measured.

### 4.3.1 KINETIC MEASUREMENTS ON THE SUPPORT MATERIAL $CeO_2$

The $CeO_2$ support has an excellent oxygen storage capacity, which in the field of catalysts marks it as one of the most important transition metal supports.

From the high storage capacity at the surface of $CeO_2$, the interesting question arises at which temperatures this effect is visible. For this reason, the $CeO_2$ support by itself was pretreated in synthetic air and measured in the range from 313 K to 373 K (corresponding to the black line in Figure 44).

From 353 K to 373 K, CO oxidation by $CeO_2$ was tested without any pretreatment (corresponding to the green line in Figure 44). In both cases, after feeding the support material $CeO_2$ with $O_2$ and CO plus some inert gas in a ratio of 2:1 with stepwise heating, no $CO_2$ was detected with a flame ionization detector.

$CO_2$ conversion was not observed during this temperature range, i.e. $CeO_2$ was not converting CO to $CO_2$. This result is positive, since our aim was to bind the CO and $O_2$ molecules directly to the gold atoms.

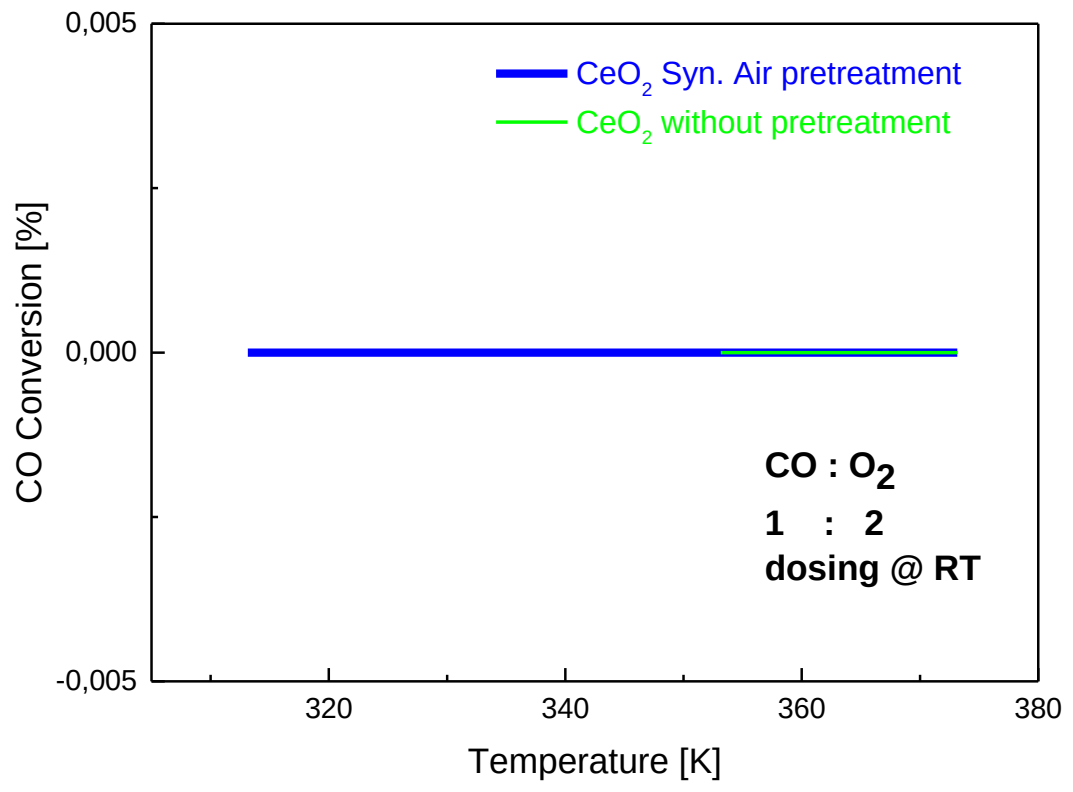

*Figure 44 The CO conversion of the support material $CeO_2$ is tested in synthetic air (blue line) and $CeO_2$ (green thin line) is tested without any pretreatment procedure[4] in the temperature range from 300 K to 370 K.*

### 4.3.2 KINETIC MEASUREMENTS ON $CeO_2$ SUPPORTED GOLD NANOCLUSTERS

The catalytic activity of the 1 wt. % $Au_{40}(SR)_{24}/CeO_2$, 2 wt. % $Au_{38}(SR)_{24}/CeO_2$ and $Au/CeO_2$, catalysts is shown in Figure 45 and details are pictured in Figure 46, Figure 47 and Figure 48. In the figures gold nanoparticles $Au/CeO_2$ are shown in red line, gold nanoclusters 1 wt. % $Au_{40}(SR)_{24}/CeO_2$ are shown in blue dotted lines and gold nanoparticle 2 wt. % $Au_{38}(SR)_{24}/CeO_2$[5] are shown by green dotted lines after reductive (square) or oxidative (circle) pretreatment and without pretreatment (star). The $Au/CeO_2$ (Figure 46) catalysts converted CO to $CO_2$ prior to 400 K, depending on the pretreatment temperature.

For synthetic air, the conversion reached ~8% at 318 K, in comparison to the $H_2$ atmosphere, for which the conversion value reached approximately 14% at 301 K. Conversion for the catalysts without any pretreatment attained ~5% at 323 K. According to these measurements, CO conversion is therefore very quick, and for the $CeO_2$ supported gold particles.

[4] The measurements in Figure 44 were performed by Thorsten Boehme, Vienna University of Technology.
[5] $Au_{38}(SR)_{24}/CeO_2$ measurement was performed by Thorsten Boheme, Vienna University of Technology.

At 373 K, gold nanocluster 1 wt. % $Au_{40}(SR)_{24}/CeO_2$ catalysts had already converted ~7% after pretreatment with synthetic air (red line with red stars in Figure 45 and details are shown in Figure 47).

After synthetic air pretreatment, the starting temperature shifted to just above 373 K, and 73% was measured at 423 K. In case of $H_2$ pretreatment at 373 K, the conversion percentage was 1.4%. The final temperature was observed in increments of 100 K. This means that for the synthetic air pretreatment, the final temperature was 373 K, after no pretreatment procedure 473 K, and for catalysts pretreated in $H_2$, the end temperature shifted to 573 K. Above 573 K, all catalysts achieved total conversion.

In tests on 1 wt. % $Au_{40}(SR)_{24}$ gold nanoclusters supported by $CeO_2$ (Figure 45), the usefulness of this new class of nanoclusters was demonstrated.

Tests on 2 wt.% $Au_{38}(SR)_{24}/CeO_2$ demonstrated, after pretreatment with oxygen, observable activity already at room temperature (details are shown in Figure 48).

At 313 K, the conversion level was 7%. Total conversion was completing at 353 K, as 100% $CO_2$ was detected after combining the catalyst with a mixture of $O_2$ and CO.

The results of $Au_{38}(SR)_{24}/CeO_2$ catalysts, after exposure to a reducing atmosphere of $H_2$, was approximately 2% at 353 K and for non-pretreated $Au_{38}(SR)_{24}/CeO_2$ catalysts at 353 K the $CO_2$ level was measured at approximately 1.2%.

Total conversion after the different pretreatment procedures was attained at 473 K.

After reductive pretreatment, total conversion was reached with a more variable slope compared to the catalysts without any pretreatment. This is visible even at the beginning, since the conversion ratio of CO: $CO_2$ is higher for the gold nanocluster catalysts pretreated in a reduced atmosphere.

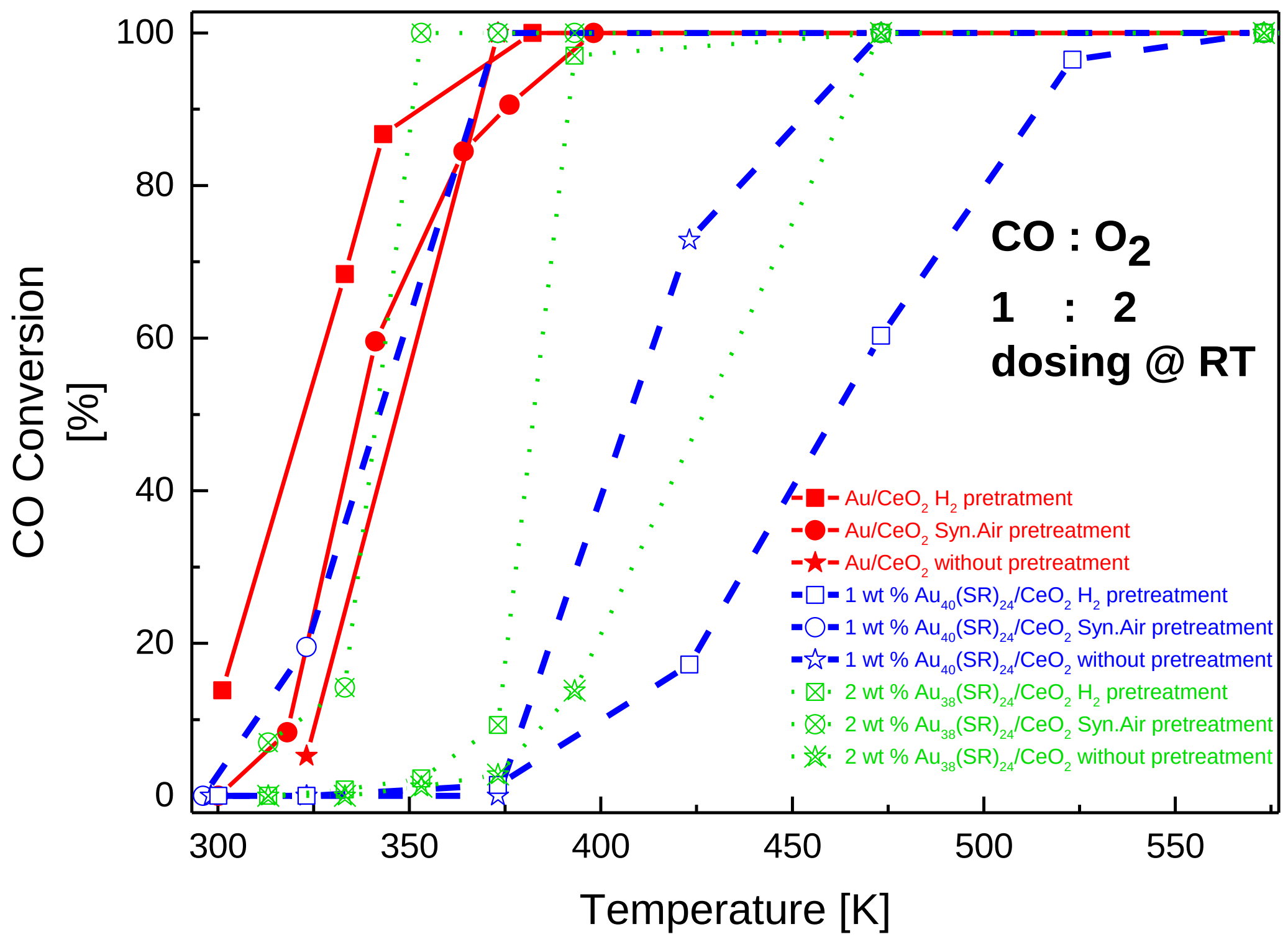

*Figure 45 This figure summarizes all catalysts Au (red line), $Au_{40}$ (blue dotted) and $Au_{38}$[6] (green dotted) supported on $CeO_2$ after reductive (square) or oxidative (circle) pretreatment and without pretreatment (star). Details are shown in Figure 46, Figure 47 and Figure 48.*

[6] $Au_{38}(SR)_{24}/CeO_2$ measurement was performed by Thorsten Boheme, Vienna University of Technology.

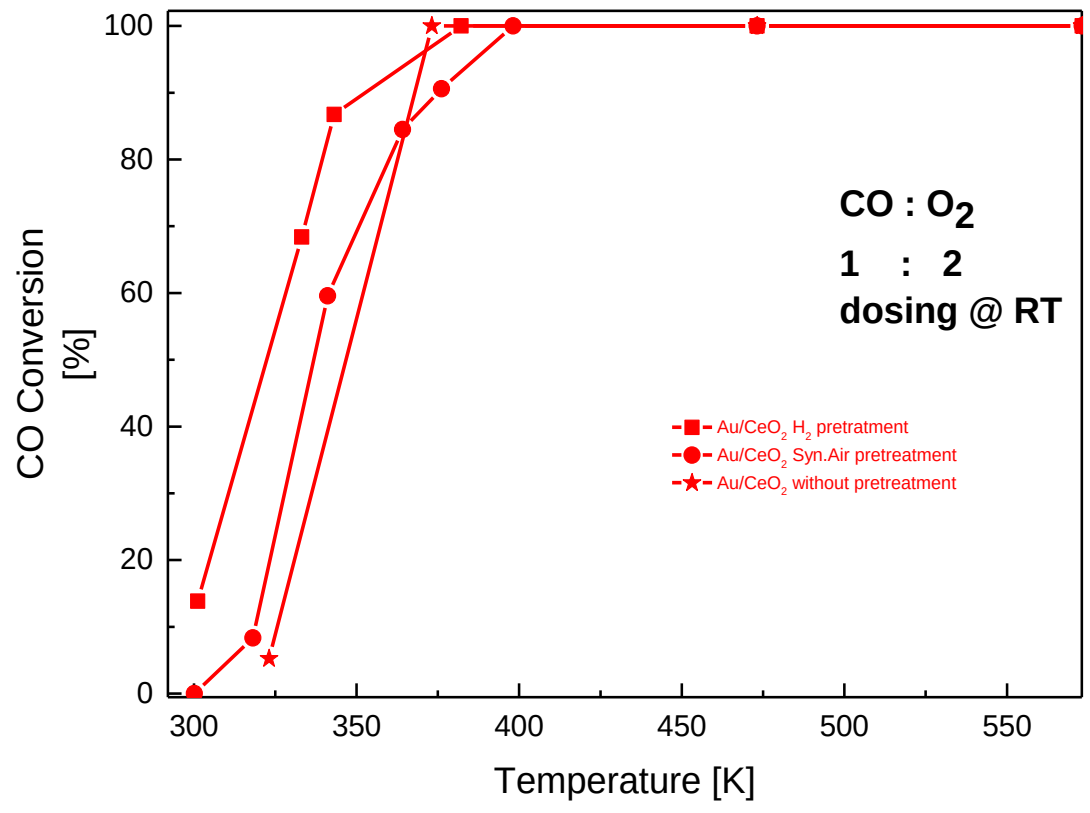

*Figure 46 This figure shows the kinetic measurement of Au/ $CeO_2$ after reductive (red square), oxidative (red circle) pretreatment and without pretreatment (red star).*

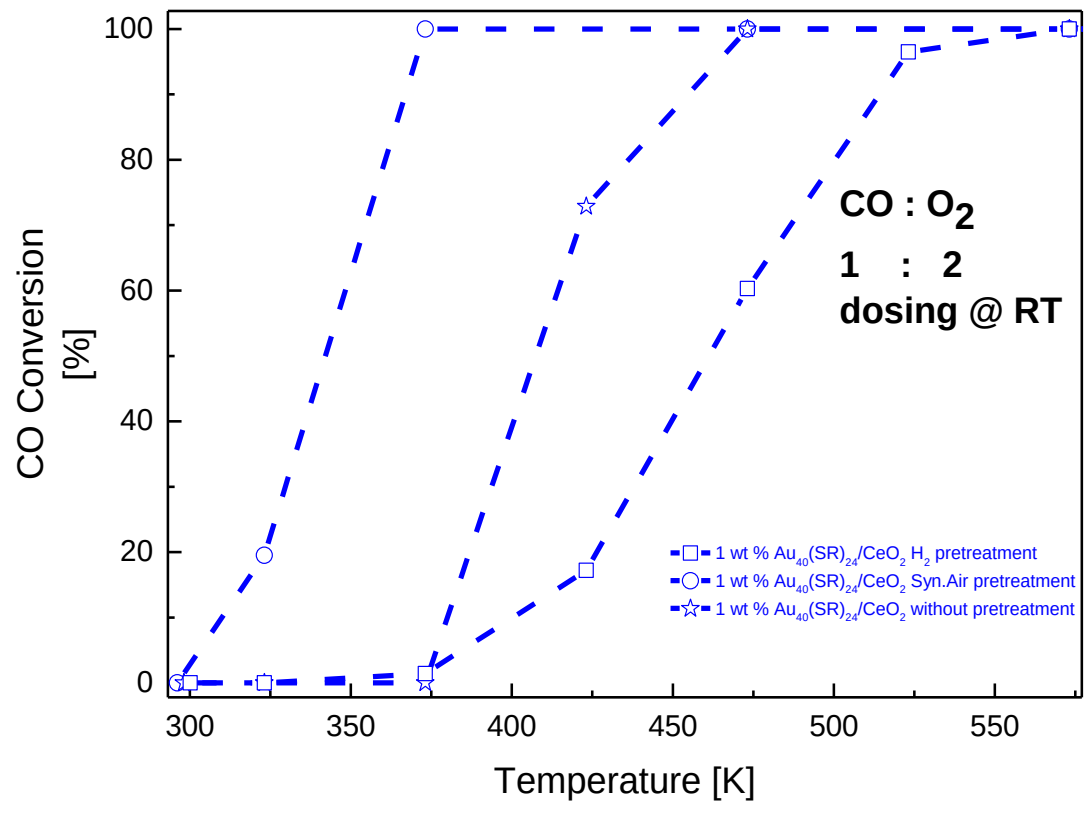

*Figure 47 This figure shows the kinetic measurement of 1 wt. % $Au_{40}(SR)_{24}/CeO_2$ after reductive (blue square), oxidative (blue circle) pretreatment and without pretreatment (blue star).*

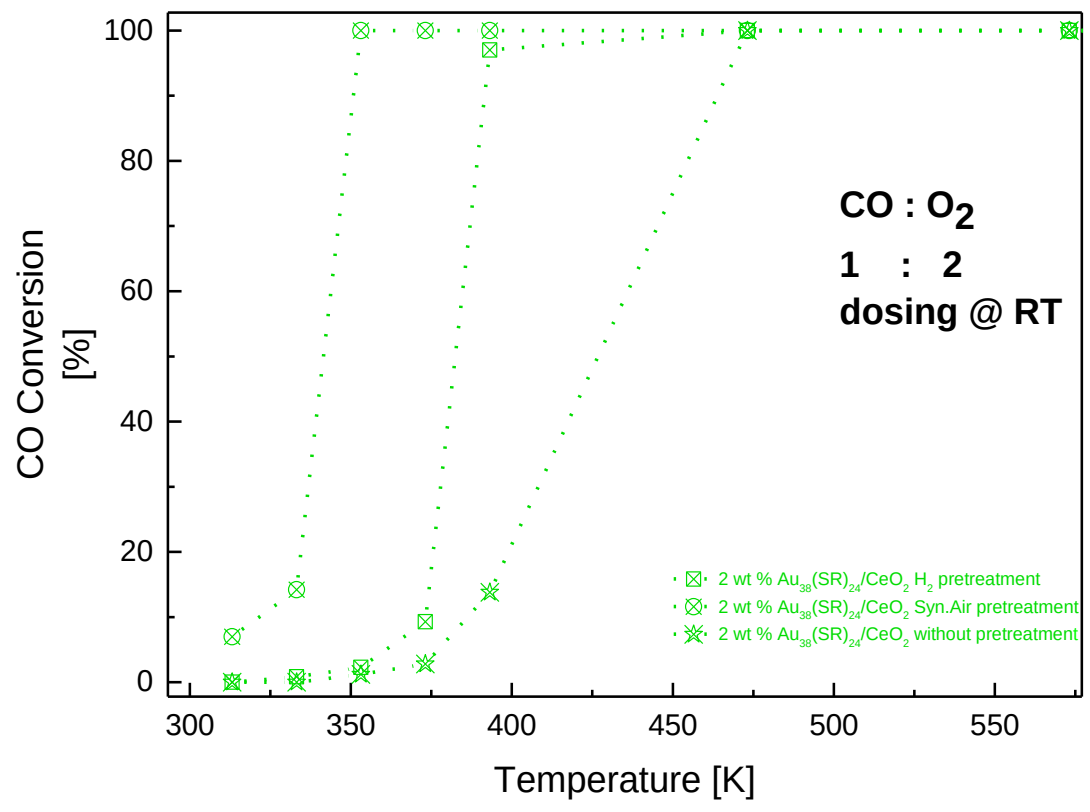

*Figure 48[7] This figure shows the kinetic measurement of 2 wt. % $Au_{38}(SR)_{24}/CeO_2$ after reductive (green square), oxidative (green circle) pretreatment and without pretreatment (green star).*

[7] $Au_{38}(SR)_{24}/CeO_2$ measurement was performed by Thorsten Boheme, Vienna University of Technology.

### 4.3.3 KINETIC MEASUREMENTS GOLD NANOCLUSTERS SUPPORTED ON ALUMINIUM OXIDE

The gold nanocluster catalysts with $Al_2O_3$ support (see Figure 49), only after $H_2$ pretreatment, enabled a conversion level of 3% at 303 K. For this catalyst, the measured plateau from 373 K to 473 K was interesting.

In this region, the lowest conversion measured was 59%, while the highest was 63%. This was only observed for this catalyst and with this pretreatment. Complete conversion was measured at 573 K, which is common for all gold particle catalysts with $Al_2O_3$ support.

After pretreatment at 473 K with synthetic air, CO conversion starts before 323 K, where a value of 3% was measured, whereas the catalyst became more active after 373 K (CO conversion of 7%).

After the pretreatment with $H_2$ and the cycle without pretreatment, total conversion finished at 375 K.

The second class of catalysts, 1 wt. % $Au_{40}(SR)_{24}/Al_2O_3$, were interesting to measure since the starting temperature was, depending on the pretreatment procedure (Figure 49, details are shown in Figure 51), approximately 600 K.

For synthetic air and no pretreatment procedure, the starting temperature was 603 K. At this temperature, with synthetic air, pretreated gold nanocluster conversion reached 7% and without pretreatment procedure, it reached 4%.

Gold particles (Figure 50) supported by $Al_2O_3$ exhibit a slow curving slope, indicating a smaller rate of change compared to the first catalyst class as discussed above. The total $CO_2$ conversion was completing at 573 K for all catalysts.

Gold nanoclusters 2 wt.% $Au_{38}(SR)_{24}/Al_2O_3$ CO conversion stays below the conversion rate for 1 wt. % $Au_{40}(SR)_{24}/Al_2O_3$ (for comparison see Figure 49). 2 wt.% $Au_{38}(SR)_{24}/Al_2O_3$ gold nanocluster catalysts are neglectable. Still, after $O_2$ pretreatment procedure higher conversion values (at 773 K ~7 % CO conversion, Figure 52) are measured compared to the pretreatment procedure with reductive and without pretreatment atmospheres.

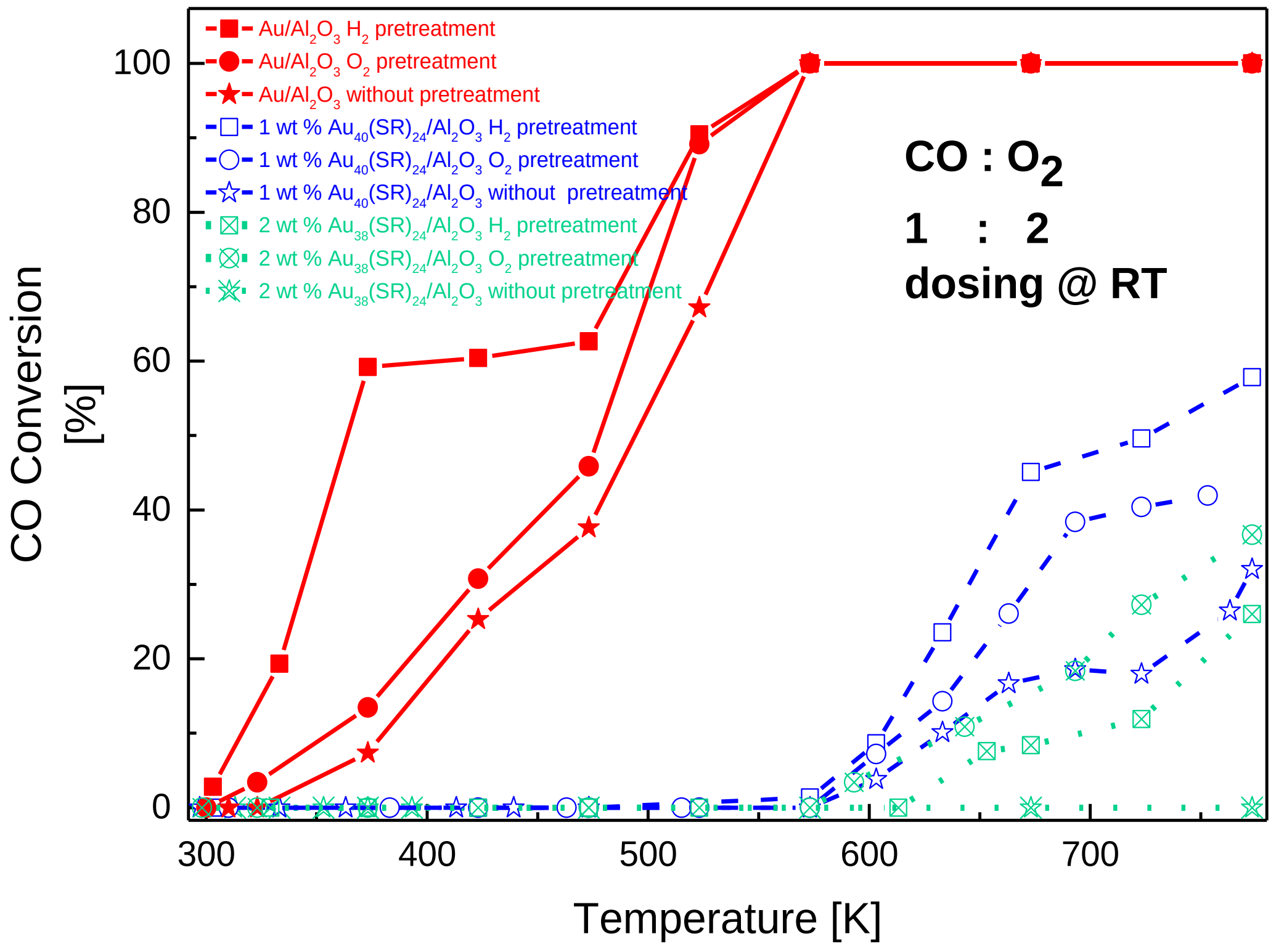

*Figure 49 This figure summarizes all catalysts Au (red line), $Au_{40}$ (blue dotted) and $Au_{38}$[8] (green dotted) supported on $CeO_2$ after reductive (square) or oxidative (circle) pretreatment and without pretreatment (star). Details are shown in Figure 50, Figure 51 and Figure 52.*

[8] $Au_{38}(SR)_{24}/CeO_2$ measurement was performed by Thorsten Boheme, Vienna University of Technology.

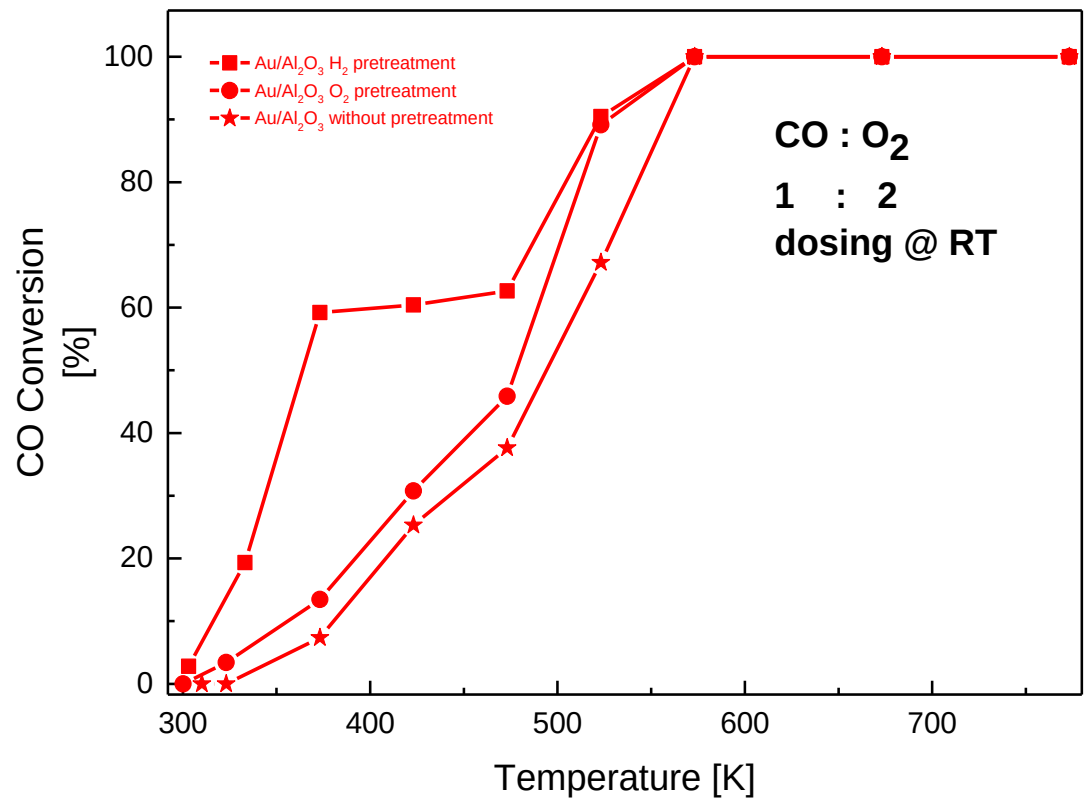

*Figure 50 This figure shows the kinetic measurement of Au/$Al_2O_3$ after reductive (red square), oxidative (red circle) pretreatment and without pretreatment (red star).*

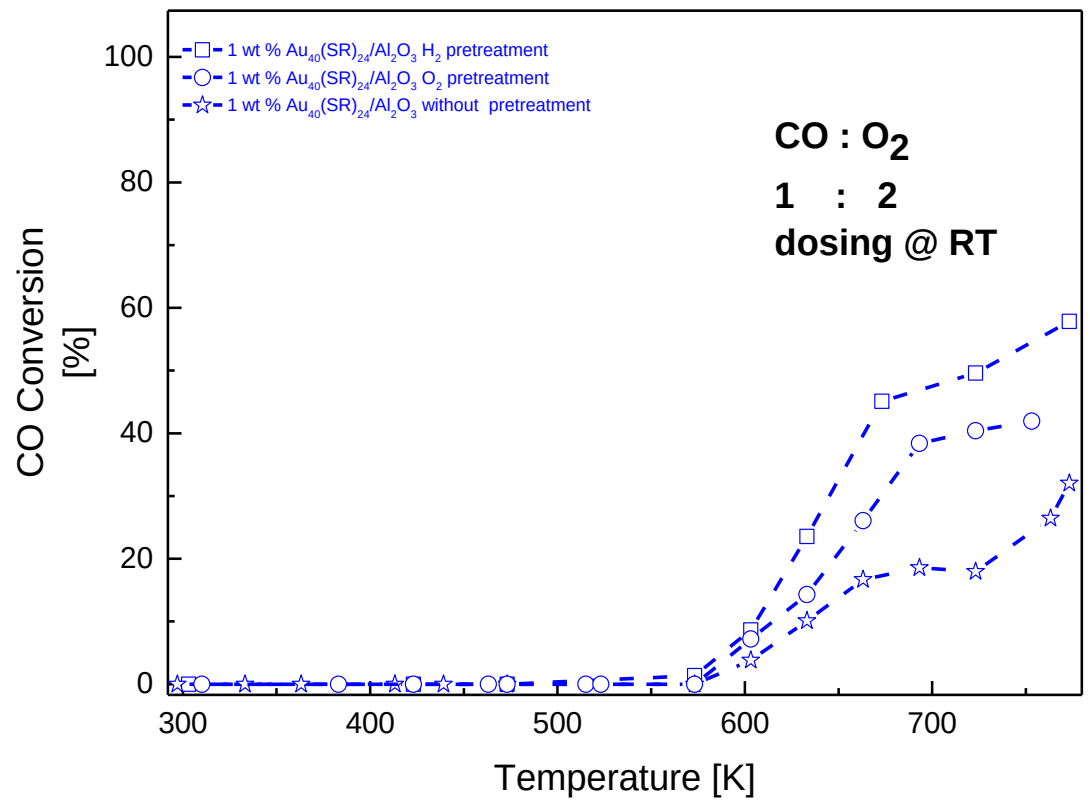

*Figure 51 This figure shows the kinetic measurement of 1 wt. % $Au_{40}(SR)_{24}/Al_2O_3$ after reductive (blue square), oxidative (blue circle) pretreatment and without pretreatment (blue star).*

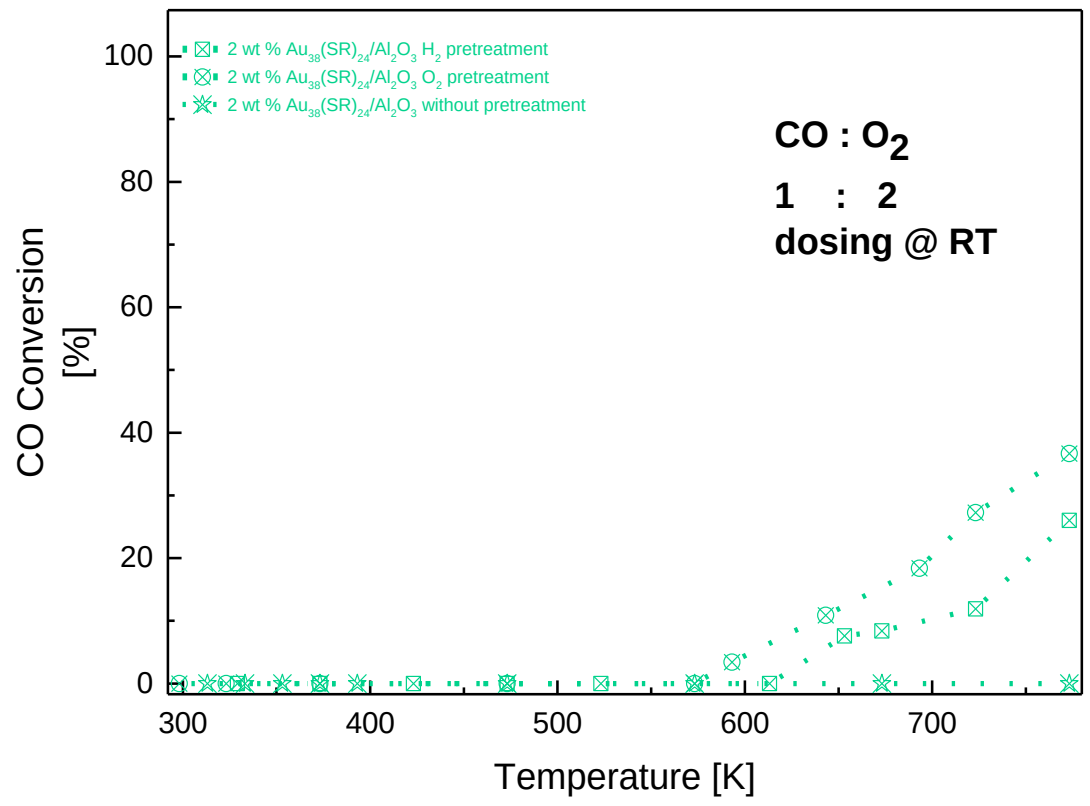

*Figure 52 This figure[9] shows the kinetic measurement of 2 wt. % $Au_{38}(SR)_{24}/Al_2O_3$ after reductive (blue square), oxidative (blue circle) pretreatment and without pretreatment (blue star).*

[9] $Au_{38}(SR)_{24}/CeO_2$ measurement was performed by Thorsten Boheme, Vienna University of Technology.

### 4.3.4 KINETIC PARAMETER

The kinetic parameters are calculated as described in Table 6 and Table 7.
In Table 6 the reaction rate and the activation energy are calculated. These values clearly show that the reaction rate and activation energy is for gold nanoclusters supported on $CeO_2$ is lower than for those gold nanocluster catalysts $Al_2O_3$. The same observation can be made for Table 7 for the TOF values.

*Table 6 Reaction Rate and Activation Energy for gold nanocluster catalysts are shown.*

| | Pretreatment Procedure | Reaction Temperature **T** [K] | Reaction Rate **r** [mol $s^{-1}$ $g^{-1}$ $10^{-5}$] | Activation Energy $\mathbf{E_a}$ [kJ $mol^{-1}$] |
|---|---|---|---|---|
| $Au_{40}(SR)_{24}/CeO_2$ | $O_2$ | 373 | 13 ± 0.3 | 25 ± 5 |
| | $H_2$ | | 1.9± 0.05 | 67 ± 5 |
| | No pre. | | 0 | 59 ± 5 |
| $Au_{38}(SR)_{24}/CeO_2$ | $O_2$ | | 17± 0.3 | 47 ± 5 |
| | $H_2$ | | 2 ± 03 | 54 ± 5 |
| | No pre. | | 0.5± 0.05 | 66± 5 |
| $Au_{40}(SR)_{24}/Al_2O_3$ | $O_2$ | 723 | 10 ± 0.5 | 14 ± 5 |
| | $H_2$ | | 12 ± 0.05 | 2 ± 5 |
| | No pre. | | 46± 0.5 | 68± 5 |
| $Au_{38}(SR)_{24}/Al_2O_3$ | $O_2$ | | 30± 0.05 | 37± 5 |
| | $H_2$ | | 70± 0.05 | 52± 5 |
| $Pd_2Au_{36}/Al_2O_3$ | $O_2$ | 703 | 83± 0.05 | 40± 5 |
| | $H_2$ | 723 | 16± 0.07 | 63± 5 |

*Table 7 TOF values for gold nanocluster catalysts.*

| Sample | Pretreatment Atmosphere | Reaction Temperature [K] | CO Conversion [%] | TOF [$sec^{-1}$] |
|---|---|---|---|---|
| $Au_{40}(SR)_{24}/CeO_2$ | $O_2$ | 341 | 50 | 0.065 ± 0.2 |
| $Au_{40}(SR)_{24}/CeO_2$ | $H_2$ | 473 | 50 | 0.065± 0.2 |
| $Au_{40}(SR)_{24}/CeO_2$ | without pre. | 407 | 50 | 0.06± 0.25 |
| $Au_{38}(SR)_{24}/CeO_2$ | $O_2$ | 341 | 50 | 0.088± 0.2 |
| $Au_{38}(SR)_{24}/CeO_2$ | $H_2$ | 382 | 50 | 0.088± 0.2 |
| $Au_{38}(SR)_{24}/CeO_2$ | without pre. | 427 | 50 | 0.088 ± 0.2 |
| $Au_{40}(SR)_{24}/Al_2O_3$ | $O_2$ | 648 | 20 | 0.052 ± 0.2 |
| $Au_{40}(SR)_{24}/Al_2O_3$ | $H_2$ | 626 | 20 | 0.052± 0.2 |
| $Au_{40}(SR)_{24}/Al_2O_3$ | without pre. | 730 | 20 | 0.052± 0.2 |
| $Au_{38}(SR)_{24}/Al_2O_3$ | $O_2$ | 750 | 20 | 0.026± 0.2 |
| $Au_{38}(SR)_{24}/Al_2O_3$ | $H_2$ | 660 | 20 | 0.026± 0.2 |
| $Au_{38}(SR)_{24}/Al_2O_3$ | without pre. | 0 | 20 | 0.026± 0.2 |
| $Pd_2Au_{36}(SR)_{24}/Al_2O_3$ | $O_2$ | 600 | 10 | 0.027± 0.2 |
| $Pd_2Au_{36}(SR)_{24}/Al_2O_3$ | $H_2$ | 776 | 10 | 0.027± 0.2 |

# 4.Chapter DISCUSSION

In the current chapter, arguments related to the field of cluster science and gold nanocluster catalysts will be debated. After that, the results will be discussed. In the following subsection, the results will be compared to the information given in the chapter “Introduction” and particularly in the subsection “Literature Review”. On this basis, conclusions are drawn.

## 5.1 GENERAL DISCUSSION

Before going into a detailed discussion of the different parts of the experiment, interesting open questions in cluster science relevant to the work presented in this thesis will be explored. Some relevant questions and topics in the cluster science community have received increasing attention in the course of the last few years. They have resulted in new questions, which are compiled from several relevant articles [125], [150]–[154] in cluster science in relation to the work presented here. The notable questions, in previous work and in this work, include:

- Does the thermal $O_2$ pretreatment process remove the surface thiolate ligands on $Au_{40}(SR)_{24}$ gold nanoclusters?
- If the external surface of $Au_{40}(SR)_{24}$ is where CO oxidation occurs, then a drastic support effect exists?
- Is the partial removal of thiolates around the gold nanoclusters necessary in order to achieve better CO adsorption?

## 5.2 RESULTS

In Figure 17, the gold nanocluster catalysts $Au_{40}(SR)_{24}/CeO_2$ pretreated in oxygen and inert atmosphere are shown. Significant are the step changes in this measurement.

Overall, three main steps are visible for the TGA measurements of $Au_{40}(SR)_{24}$ gold nanoclusters in oxygen (Figure 17 (b)).

One additional step was observed as compared to the twin gold $Au_{38}(SR)_{24}$ nanocluster catalysts, where at higher temperatures, organic elements are burned off, as published 2015 in [130].

For bare gold nanoclusters $Au_{40}(SR)_{24}$, the computed (~30%) values diverge slightly from the measured (36%) values, which is an indicator that more organic material is found in the tested sample.

The TGA measurements for ceria, are decreasing fast and are showing high losses due to dehydration processes of ceria [155].

Reviewing the $CeO_2$ TGA profile, these results seem appropriate when taking into account surface impurities such as physically adsorbed water

Figure 21 presents catalysts supported on $CeO_2$, such as $Au_{40}(SR)_{24}$ are measured in oxygen (Figure 21 (a)) and inert atmosphere (Figure 21 (b)) in addition to $Au_{38}(SR)_{24}$ measured in oxygen (Figure 21 (c)).

In all three TGA profiles, the second step from 400 K to 600 K, the burning of the carbon of the material, is visible and was measured.

The TGA measurement shows the additional $Au_{40}(SR)_{24}$ step due to the thiolate ligand removal. These results are in accordance with literature values [130].

For $Au_{40}(SR)_{24}\backslash Al_2O_3$ and $Au_{38}(SR)_{24}/Al_2O_3$ (Figure 22(c)) supported clusters, a clear TG profile corresponding to the thiol ligands was not observed neither in oxygen atmosphere (Figure 22 (a)) nor in inert atmosphere (Figure 22 (b)), although a relative mass loss of no more than a few percent was measured.

The relative mass loss depending on temperature and time for the gold nanoparticles supported by $Al_2O_3$ clearly validates that no thiolate removal is visible, compared to Figure 11 in [156] .

For FT-IR results of $CeO_2$ after oxidative pretreatment procedure at 473K (Figure 23 (a)), it is apparent that CO adsorption on $Ce^{4+}$sites is found after evacuation in HV vacuum (~$10^{-7}$mbar) at the expected wavenumber (experimental: 2171 $cm^{-1}$ and literature value: 2177 $cm^{-1}$).

In Figure 25, a broad peak was found around 3600 $cm^{-1}$, which indicates that molecular water is found on the catalyst's surface. The expected reflection of the H-O-H deformation is valid (Figure 25).

This behavior leads to the conclusion that a significant amount of adsorbed water is available on the surface after oxidative pretreatment procedures at 473 K.

In conclusion, $CeO_2$ adsorbs CO at the surface and oxidative pretreatment procedures enhance this process for active $Ce^{4+}$ sites after pretreatment at 473 K.

These measurements show that partially charged Au sites are active for CO adsorption after partial thiolate removal. Partial removal of thiolates does not affect CO adsorption in a negative way. Moreover, the partial removal of thiolates enhances the CO adsorption on the gold nanocluster.

It is concluded, after analyzing the TGA data, that partial removal of thiolates is preferable to complete removal. However, when $CeO_2$ measurement was performed at the adjusted temperature range, water formed on the catalyst's surface.

For the OH and carbonates regions, it is noteworthy that the spectra change at 473 K, after inert and oxidative pretreatment. At this temperature, thiolates are partially removed.

Of interest are the OH and the carbonates regions, as discussed above for the $CeO_2$ support. Here, molecular and dissociative water is measured for the pure $CeO_2$.

FT-IR measurements demonstrate that $CeO_2$ activates $O_2$. CO is activated on de-thiolated gold sites and oxidized by $CeO_2$ lattice oxygen [48], [130].

This argument underlines that the partial removal of thiolates is superior for suppressing the absorbance interaction of the support material $CeO_2$ with CO molecules and to enhance the vibration spectrum of CO with the active $Au^{\delta+}$ sites.

As temperatures increase, the subsequent sites become more pronounced, $Au^{+}$ sites disappear.

These IR bands are most intense after 423 K treatment, as measured for $Au_{40}(SR)_{24}$ and $Au_{38}(SR)_{24}$ after oxygen pretreatment at 423 K.

The bands gradually decrease in intensity and finally disappear at a treatment temperature of 523 K.

For $CeO_2$ supported $Au_{40}(SR)_{24}$ and $Au_{38}(SR)_{24}$, the FT-IR results prefer oxygen pretreatment at 423 K for gold nanocluster catalysts. Most active is the $Au^{\delta+}$ site.

In the cases of $Au_{40}(SR)_{24}$ and $Au_{38}(SR)_{24}$ supported on $Al_2O_3$, gold nanoclusters deliver almost no results. The support material $Al_2O_3$ could suppress the Au-CO vibration around 437 K.

Kinetic measurements for $CeO_2$ supported gold nanoclusters showed that the $Au_{40}(SR)_{24}$ supported gold nanoclusters show the most effective CO conversions compared to $Au_{38}(SR)_{24}$. While the gold nanoparticle catalyst is converting CO faster and more efficient to 100% $CO_2$ before reaching 400 K. Gold nanoparticle catalysts have the drawback, that over few hundred gold atoms are forming the nanoparticles.

For $Al_2O_3$ supported gold nanoclusters the CO conversion is slowly and less efficient at high temperatures (>400 K) the CO conversion starts.

The kinetic parameters reflect the outcome of the measurements. Obviously, the $CeO_2$ supported gold nanoclusters are favorable, since the values for the kinetic parameters such as reaction rate, activation energy and TOF are small compared to $Al_2O_3$ supported gold nanocluster catalysts.

## 5.3 CLOSING REMARKS

Desorption of the thiolate from the gold nanoclusters depends on the gas atmosphere and is a multistep process. The outcomes from the supported $Au_{40}(SR)_{24}/CeO_2$ and $Au_{38}(SR)_{24}/CeO_2$ gold nanocluster are solid. No clear conclusion can be reached based on the performed TGA measurements for $Au_{40}(SR)_{24}/Al_2O_3$ and $Au_{38}(SR)_{24}/Al_2O_3$.

Thermal pretreatment of $Au_{40}(SR)_{24}/CeO_2$ catalysts in $O_2$ significantly boosts its catalytic activity.

For FT-IR tests, the partially positively charged Au species ($Au^{\delta+}$ for $0 < \delta < 1$) are the most active sites after pretreatment with 100 mbar oxygen at 473 K on $Au_{40}(SR)_{25}/CeO_2$ and $Au_{38}(SR)_{24}/CeO_2$ gold nanocluster catalysts, while the Au sites in +1 and $-\delta$ ($0<\delta<1$) charged states contributes to CO adsortion only above room temperature.

On the other hand, thermal treatment of $Au_{40}(SR)_{24}/CeO_2$ catalysts in $N_2$ seems to be ineffective.

These results demonstrate that thermal $O_2$ pretreatment of the $CeO_2$ support is remarkably effective in enhancing the activity for CO adsorption at moderate temperatures, which is reflected in the FT – IR measurements.

The activity of the gold nanocluster catalysts supported on $Al_2O_3$ for CO oxidation proved to be less than that for the catalyst with $CeO_2$ support.

Comparing the kinetic measurements to the results with the $CeO_2$ and $Al_2O_3$ supported gold nanocluster catalyst, the overall CO conversion of $CeO_2$ supported gold nanoclusters starts at a lower temperatures (< 400 K). As can be seen from the measurements, the conversion procedure seems to be progressing slowly, with a total conversion at 573 K.

With these results, the gold particles supported by $CeO_2$ are in clear contrast to the gold particles catalysts supported by $Al_2O_3$.

Thiolate ligands do not seem to pose an apparent positive or negative effect on the reaction kinetics for CO oxidation at the accessible Au sites.

Large thiolate ligands lose their flexibility and may be rigidly bound to the shell of $Au_{40}$ nanoclusters, thus preventing the reactant from adsorbing and further reaction.

Thiolate ligands must be at least partially removed to allow $Au_{40}$ nanoclusters to be active for CO oxidation.

The results of CO oxidation catalyzed by intact $Au_{40}(SR)_{24}/CeO_2$ catalysts in which the surface thiolates are partially removed by a temperature calcination process are comparable to previous work for $Au_{38}(SR)_{24}/CeO_2$ [151], in which the surface thiolate ligands were completely removed by a temperature calcination process.

The results here imply that the ligands do not inhibit the activity of the catalysts, as assumed initially, at least in CO oxidation. The drastic CO adsorption in FT – IR effect and CO oxidation within kinetic measurements indicates that the interfacial interaction between gold clusters and the support is a key factor in answering this question.

For $Au_{40}(SR)_{24}$ and $Au_{38}(SR)_{24}$, one may argue that their properties do not differ much, and both can be adapted for the gold nanocluster catalysts measured in this work.

The properties of $CeO_2$ regarding $O_2$ activation due to its rich oxygen vacancies are well-known from previous work. It can be concluded that $Au_{40}$ and $Au_{25}$ clusters supported by $CeO_2$ behave similarly for CO oxidation, as the model proposed by Jin [84] assumes that the $Au_{40}$ structure must be a twin core of $Au_{25}$ icosahedra connected via two joint gold atoms.

Since the $Au_{40}(SR)_{24}$ structure is not known, future work on the catalytic mechanism for CO and $O_2$ activation and detailed surface reactions will certainly need new and relevant insights.

# 5.Chapter: OUTLOOK

The field of cluster catalysis is new and has been explored only recently. Much work in every direction is yet to be done. Unresolved are questions regarding the structure model, but also the basic question of whether and how these promising candidates for green chemistry and catalysis can be used efficiently.

It is beyond the scope of the current thesis to describe more than a very small portion of existing research.

What if the support material was doped with lanthanum [157]? Such supports have already been successfully used in traditional catalysis [105].

Does conversion, as it is claimed for $Au_{25}(SR)_{18}$ [48], happen for all gold nanocluster catalysts between the interface support and gold nanocluster catalysts via the MvK mechanism?

An interesting approach lies in the oxidation of ethanol. Ethanol may be oxidized to acetaldehyde. Ethanol may also be oxidized to acetic acid, depending on the reagents and conditions. Several acetaldehyde formation experiments at room temperature have been proposed [64], [158]–[160].

Research has been conducted with gold nanoparticles, but not with thiolate stabilized gold nanoclusters $Au_{40}(SR)_{24}$.

Good et al. [157] measured CO oxidation values for gold catalysts (see for this Figure 3(a) in [157]) and concluded that retailoring atomic factors of the support material can elevate the $CO_2$ conversion value.

In the future, the presence of ligands on the $Au_{38}$ cluster is of potential interest to investigate the ligand dependence of the catalytic activity of $Au_{38}$ [157]. This open question can be transferred to any other gold nanocluster catalysts and needs to be answered if almost “perfect” gold nanocluster catalysts are to be engineered in the future.

# 6.Chapter: REFERENCES

# 7.Chapter: APPENDIX

## 8.1. INFRARED TABLE

*Table 8 Infrared Table including CO adsorbed species on gold and ceria.*

| | *Literature Wavenumber cm$^{-1}$* | *Classification* | *Reference* |
|---|---|---|---|
| **$CeO_2$** | 2150 | CO adsorption / slight thiolate removal | [48] |
| | 2117 | CO – $Au^{\delta+}$ | [105] |
| | 2150 -2117 | CO – $Au^{\delta+}$ | [48] |
| | | | |
| | 2158<br>2157 | CO – $Au^{+}$ | [48] |
| | 2177 | CO – $Ce^{4+}$ | [48] |
| | 2076<br>2078 | CO – $Au^{\delta-}$ | [48] |
| | 2160 | $Ce^{4+}$ - CO species | [146] |
| | 2180 | $Ce^{n+}$ - CO species | [105] |
| | 2133 | $Ce^{3+}$ - CO species | [148] |
| | 2160 | $Ce^{4+}$ - CO cations | [105] |
| **$Au_{38}(SR)_{24}/CeO_2$** | 2177 | $Ce^{4+}$ - CO species | [48] |
| | 2157 | CO – $Au^{+}$ | |
| | 2150 | | |
| | 2117 | CO – $Au^{\delta+}$ | |
| | 2076 | CO – $Au^{\delta-}$ | |
| | 2724 | Overtone band of C – H deformation mode | |
| | 3030 | C – H stretching | |
| | 3100 | C – H stretching | |
| | | | |
| | 2950<br>2850 | C – H modes of formats | |

*Table 9 Infrared Table including Carbonate and OH region.*

| | ***Literature Wavenumber*** | ***Classification*** | ***Reference*** |
|---|---|---|---|
| | ***cm$^{-1}$*** | | |
| **Carbonate Region** | 1620 - 1670 | Bridged bidentate carbonates | [161] |
| | 1730<br>1135<br>1395<br>1220 | Bridged carbonates | [105], [161], [162] |
| | 1590 – 1599<br>1612 | Hydrogen carbonates | |
| | 3617<br>1390 – 1413<br>1025 – 1045<br>1580 | Formats and bidentate carbonates | |
| | 1460 - 1570 | Monodentate carbonates | |
| | 1350 | Polydentate carbonates | |
| | 1060 | Polymeric carbonates | |
| | 1310 - 1316 | Carboxylate species, coordinated via oxygen atom to a surface Ce ion | |
| | 1590 – 1492<br>1297 – 1227 | Tridentate carbonates | |
| | | | |
| | 1580 - 1670 | Carboxy group [COOH]s | [161] |
| | 1480 - 1460<br>1380 - 1360 | Monodentate carbonates | [161] |
| | 1353 | Polydentate carbonates | [163] |
| | 1300 - 1370 | Monodentate carbonates | [164] |
| | 1700-1850 | C-O stretching mode | |
| | 1250-1300 | O-H bending | |
| | 1000-1100 | C-OH stretching | |
| **OH Region** | 3500 - 3600 | O-H stretching mode | [165] |
| | 3500 - 3600 | O-H stretching mode | |
| | 3649<br>3675 | bridged OH groups | [162], [165], [166] |
| | 3300-3500 | molecularly adsorbed water or hydrogen bonded OH groups on ceria | |
| | 3520<br>3710<br>3655 | Linearly adsorbed (monodentate) and tridentate OH groups on $CeO_2$ | |